\documentclass[hyper,12pt,letterpaper]{JHEP3}

\usepackage{amsmath,amssymb,multirow,array}
\newcommand{\half}{\frac{1}{2}} 
\newcommand{\Jsc}{J^{\mu}_{S\,\hbox{\tiny canon}}}
\newcommand{\Jsn}{J^{\mu}_{S\,\hbox{\tiny new}}}
\newcommand{\cd}{\!\cdot\!}
\newcommand{\str}{{*}}
\newcommand{\cl}[1]{{\vskip 0.5 cm}\noindent{\bf{#1}} \\}
\newcommand{\sone}{\sigma_1} %%Was previously y_1
\newcommand{\sthr}{\sigma_3} %%Was previously v_1
\newcommand{\sfou}{\sigma_4} %%Was previously v_2
\newcommand{\sfiv}{\sigma_5} %%Was previously v_3
\newcommand{\seig}{\sigma_8} %%Was previously y_2
\newcommand{\snin}{\sigma_9} %%Was previously C_{\omega}
\newcommand{\sten}{\sigma_{10}} %%Was previously C_B
\title{A theory of first order dissipative superfluid dynamics}
\author{Jyotirmoy Bhattacharya$^a$, Sayantani Bhattacharyya$^b$, Shiraz Minwalla$^a$ and Amos Yarom$^{c,d}$\\
$^a$Dept. of Theoretical Physics, Tata Institute of Fundamental Research \\ Homi Bhabha Rd, Mumbai 400005, India. \\
$^b$Harish-Chandra Research Institute, Chhatnag Road, Jhunsi, Allahabad-211019.\\
$^c$Joseph Henry Laboratories, Princeton University, Princeton, NJ 08544\\
$^d$Department of Physics, Technion, Haifa 32000, Israel \\
Email:\ \ {\bf jyotirmoy@theory.tifr.res.in, sayanta@hri.res.in, minwalla@theory.tifr.res.in, ayarom@princeton.edu}}
\abstract{We determine the most general form of the equations of relativistic superfluid hydrodynamics consistent with 
Lorentz invariance, time-reversal invariance, the Onsager principle and the second law of thermodynamics at first 
order in the derivative expansion. Once parity is violated, either because the $U(1)$ symmetry is anomalous or as a 
consequence of a different parity-breaking mechanism, our results deviate from the standard textbook analysis of superfluids. 
Our general equations require the specification of twenty parameters (such as the viscosity and conductivity). 
In the limit of small relative superfluid velocities we find a seven parameter set of equations.  
In the same limit, we have used the AdS/CFT correspondence to compute the parity odd contributions to the superfluid equations 
of motion for a generic holographic model and have verified that our results are consistent.}
\preprint{PUPT-2375}
\begin{document}
\maketitle
%
%********************************************************
\section{Introduction} \label{S:intro}
%********************************************************
%
In this paper we work out a theory of relativistic superfluid hydrodynamics including dissipative terms at first order in a gradient expansion, allowing for parity violation and the presence of triangle anomalies.
In other words we work out the most general form of the equations of $3+1$ 
dimensional 
$s$-wave superfluid hydrodynamics  consistent with Lorentz invariance, time-reversal invariance
 and the second law of thermodynamics.  
While we work in a relativistic context throughout this  paper, our final results admit a straightforward non relativistic limit, and are easily formulated in a non relativistic context. 

The theory of superfluid hydrodynamics has a long history. The equations of ideal superfluidity (i.e. superfluid dynamics in the absence of dissipative terms) were worked out over 60 years ago by Landau and Tisza \cite{Landau:1941, Tisza:1947zz} in a non relativistic setting. They were generalized to a 
relativistic superfluid in the early 80's by Israel and  Khalatnikov and 
Lebedev \cite{Israel1,LebedevKhalatnikov,Israel2} and later reformulated 
by Carter and Khalatnikov \cite{Carter:1992a,CarterKhalatnikov} and by 
Son \cite{Son:2000ht}. In most of this work we will use a formulation of 
superfluid dynamics close to that of \cite{Son:2000ht}. 
In a beautiful recent work, Sonner and 
Withers \cite{Sonner:2010yx} (see also \cite{Bhattacharya:2011ee}) have used the equations of Einstein gravity to rederive the Landau-Tisza equations for superfluids that admit a dual gravitational description via the AdS/CFT correspondence of string theory. This development yields independent evidence for the correctness and completeness of the Landau-Tisza theory of ideal superfluidity. 

The focus of the current paper is  on the one derivative (dissipative) 
corrections to the Landau-Tisza equations. Dissipative corrections to 
relativistic superfluids have been extensively studied in the literature 
\cite{Pujol:2002na,Gusakov:2006ga,Valle:2007xx,Mannarelli:2009ia,Gusakov:2007px,Herzog:2011ec,Bhattacharya:2011ee} and they generalize (and extend) the 
textbook  derivation of such corrections in the non-relativistic limit 
\cite{LLvol6,Clark,Putterman}. While it is straightforward to list the 
most general dissipative corrections to the stress tensor, charged current 
and Josephson relations allowed by Lorentz invariance, it turns out that 
such a listing also allows for many unphysical possibilities. It is a physical
requirement that any hydrodynamic flow be equipped with an entropy current 
$J^{\mu}_S$.
The second law of thermodynamics requires that the increase in the
entropy in any compact, spacelike, region be larger than the  incoming entropy flux 
through the surface of that region; this requires that at every point in 
spacetime and for every conceivable fluid flow
\begin{equation}
	\partial_{\mu} J_S^{\mu} \geq 0\,.
\end{equation}
An interesting and important fact about fluid dynamics, whether superfluid or not, is that the requirement that the divergence of the entropy current always be positive gives rise to important constraints on dissipative corrections to the equations of  motion.

As far as we are aware, all previous studies of dissipative corrections to 
the equations of superfluid hydrodynamics assume, on intuitive grounds, 
 that the entropy current of superfluid hydrodynamics takes a particular 
canonical form, $\Jsc$.  
These studies then determine the dissipative corrections to the energy momentum tensor and charged current (and Josephson condition) consistent with the positivity of the divergence of this canonical entropy current and with covariance under time reversal. The latter restriction is usually called the Onsager relations.

In \cite{ Son:2009tf}, Son and Sur\'owka observed that, in the presence of triangle anomalies, the entropy current of ordinary fluids (non superfluids) deviates from its canonical form. This observation makes clear that the intuition that
fixes the entropy current to its canonical form is not infallible. 
For this reason, the starting point of our analysis in this paper is the assumption that an entropy current with positive definite divergence exists. However, we make no assumption about the form of this current beyond the requirements of symmetry. 
 
More formally,  we allow the (postulated positive divergence)
entropy current to take the form 
\begin{equation}\label{ec}
J^\mu_S=\Jsc+ \sum_i v_i V_i^\mu
\end{equation}
where $\Jsc$ is the canonical entropy current referred to  above, 
$V_i^\mu$ is a basis of on-shell inequivalent one derivative vectors and 
$v_i$ are (initially) unconstrained coefficient functions.  
We then demand that the divergence of $J_S^{\mu}$ be positive semi-definite for any solution of the equations of superfluid hydrodynamics 
{\it on an  arbitrary background spacetime} 
and that the Onsager relations are satisfied.\footnote{Recall that the second
law of thermodynamics must apply in any conceivable consistent situation. 
In particular it must apply when the system is formulated on an arbitrary 
background spacetime provided the system is free of diffeomorphism anomalies. 
This condition is true of all experimental superfluids as well as 
all superfluids obtained via the AdS/CFT correspondence.}
These restrictions fix most of the $v_i$'s in 
\eqref{ec} and also restrict the possible transport coefficients of the theory.
 
As we discuss in detail throughout this work, the requirement that the entropy 
current be of positive divergence {\it in an arbitrary background spacetime}
provides powerful constraints on the form of the entropy current and through 
it, on the form of the possible dissipative corrections to the 
hydrodynamic constitutive relations {\it even in flat space}. For instance, 
the divergence of the entropy current could contain a term proportional to
\begin{equation}
	\nabla_{\mu} J_S^{\mu} \propto v_1 R_{\mu\nu}u^{\mu}u^{\nu} + \ldots
\end{equation}
where $u^\mu$ is the fluid velocity, $R_{\mu\nu}$ 
is the Ricci tensor, and $v_1$ is one of the coefficient functions in 
\eqref{ec}. The divergence of the entropy current may also contain many 
other terms proportional to $v_i$ and independent of curvatures. However, for any 
given fluid flow these other terms can be held fixed while 
$R_{\mu\nu}u^\mu u^\nu$ is made arbitrarily negative by tuning the curvature
tensor.\footnote{Note that curvature tensors do not contribute to the 
fluid equations at first order, so it is consistent to hold fluid flows 
fixed while taking curvatures to be very large.} It follows that the 
divergence of the entropy current is positive for an arbitrary fluid 
flow on an arbitrary spacetime only if $v_1=0$. Thus, we find a constraint 
on the entropy current for fluid motion in a flat space background, 
even though we needed to move to a curved spacetime in
order to obtain this constraint.

The result of our extended analysis is as follows. 
For superfluids which are parity and time-reversal invariant we find that  (see sections \ref{S:ParityInvariant} and \ref{S:SuperfluidPodd} for details) all the $v_i$'s in \eqref{ec} 
should be set to zero---the entropy current agrees with the canonical 
entropy  current.\footnote{As a warm up for this analysis we show in 
section \ref{cfm} that a similar result is true of parity invariant 
charged fluid dynamics
in the absence of superfluidity}
 This result provides a
check of the intuition reviewed in, say, \cite{LLvol6}, that the entropy current should take its canonical form. It follows that the most general structure 
of dissipative terms in parity and time-reversal preserving superfluids is given by 
the 14 parameter family described in \cite{Bhattacharya:2011ee} which generalized the
13 parameter construction of Clark and Putterman 
\cite{Clark,Putterman}). The generalization of this result to superfluids that preserve parity 
but not time-reversal invariance has been worked out in the paper \cite{Bhattacharyya:2012xi} and yields 
a 17 parameter set of independent transport coefficients.

Let us now turn to the case of superfluids that do not respect parity symmetry. In this case we find that the entropy current is not constrained to take 
the canonical form. Not only are the $v_i$'s non  zero, two of them remain 
undetermined (meaning that they are completely free functions of local 
thermodynamical variables). In additional there are 
two physically unimportant ambiguities which we describe in 
detail in sections \ref{S:ParityInvariant} and \ref{S:SuperfluidPodd}. 

It turns out that, under the assumption of time-reversal invariance,  the most general 
equations of parity odd superfluid hydrodynamics, at first order, are parameterized by 
14 +4 +2 =20 parameters, which  are consistent with 
the positivity of the divergence of any given choice of entropy current. 
14 of these parameters multiply parity invariant structures; these are the parameters 
that parameterize superfluid hydrodynamics in parity preserving systems.
The remaining 6 parameters multiply parity odd structures, and are completely new.  
2 out of these 6 parity odd parameters are the undetermined 
$v_i$s in the entropy current
\footnote{Note that these two parameters in the entropy current gets related to some of the parity odd transport coefficients 
in the constitutive relations, when we demand positivity of the divergence of entropy current.}. In the case of superfluids that 
are allowed to violate time reversal invariance it turns out that 
there is an additional three function parity even ambiguity in the entropy current and transport
coefficients \cite{Bhattacharyya:2012xi}, leading to a 23 parameter set of constitutive relations.

At this point it would be useful to clarify some of the terminology we use. 
We have stated that the equations of motion of superfluid dynamics require 
the specification of twenty unknown functions of the thermodynamic variables. This may be contrasted with the equations of motion of a normal, 
charged conformal fluid which requires the specification of two parameters, the shear viscosity and conductivity. For the normal charged conformal fluid, 
the shear viscosity and conductivity also control the amount of entropy produced by the fluid and so, these parameters are also called dissipative parameters. 
In the case of parity violating superfluids, six of the twenty parameters are not associated with entropy production so, only fourteen of the 
twenty parameters are dissipative in the sense described above. These fourteen parameters are precisely the ones that are present in parity and time-reversal preserving superfluids. 
In normal charged conformal fluids the conductivity 
controls not only the response of the current to an external electric field, but also the response of the current to 
changes in temperature and changes in chemical potential. The situation in superfluids is similar: the twenty parameters,
and in particular the two undetermined parameters associated with the entropy current, control the response of the system to various changes in the hydrodynamic variables.

When the superfluid velocity is too large superfluidity breaks down. For this 
reason, when considering experimentally accessible superfluids such as liquid Helium 
it is often particularly interesting to study dissipative 
corrections in the limit that the normal and superfluid 
velocities are collinear. In this limit the 14 parameter family of dissipative 
corrections of parity invariant superfluids reduce to the five parameter 
family described, for instance, in the classic 
text book of Landau and Lifshitz \cite{LLvol6}.\footnote{%
These reduce further to only three parameters in the case of a 
conformally invariant theory.} When parity is violated, the twenty parameter 
family of solutions reduces to seven.
The two additional parameters determine (in an appropriate frame) the response of the charged current to a magnetic 
field and to ``chiral vorticity'', i.e., changes in $\omega^{\mu} = \frac{1}{2}\epsilon^{\mu\nu \alpha\beta}u_\nu \partial_\alpha u_\beta$. 

When the superfluid density is set to zero we are in the normal phase. In this limit the two undetermined parameters of the superfluid theory reduce to an integration constant and the contributions 
to the charge current proportional to the magnetic field and 
chiral vorticity are completely determined by the triangle anomaly \cite{Son:2009tf}
(up to an integration constant). In such a limit, we recover the results 
of \cite{Erdmenger:2008rm, Banerjee:2008th, Son:2009tf} (see also 
\cite{Bhattacharyya:2007vs}). 

We emphasize that, as opposed to the configuration at zero superfluid density, 
the 6 new arbitrary parameters that appear in parity non invariant superfluid
hydrodynamics may be non vanishing even in the absence of an anomaly.
This difference may be of significance; 
practically speaking, anomalies are intrinsically relativistic phenomena 
and always vanish in a nonrelativistic setting. Consequently, the results of this paper suggest that 
a term in the particle number current proportional to the vorticity could 
possibly 
show up in non relativistic table top experiments involving superfluids or 
superconductors which violate parity (e.g. non centro symmetric
superconductors).
 
We were led to the study of dissipative effects in superfluids in order to understand the results of certain holographic computations using the AdS/CFT correspondence of string theory. We emphasize, however, that our eventual derivation of the general equations of superfluid hydrodynamics makes no use of the AdS/CFT correspondence and so, applies to all superfluids, not just those that admit a dual gravitational description.

In the second part of our paper we test the collinear limit of our general formalism 
using the AdS/CFT correspondence. In the case of parity preserving superfluids, the authors of \cite{Herzog:2011ec, Bhattacharya:2011ee}  showed that it is possible to use the fluid gravity map 
\cite{Bhattacharyya:2008jc,VanRaamsdonk:2008fp,Bhattacharyya:2008xc, Bhattacharyya:2008ji,Haack:2008cp,Erdmenger:2008rm,Banerjee:2008th,Bhattacharyya:2008mz} to derive dissipative 
corrections to the equations of superfluid dynamics by extending the analytically tractable superfluid model of \cite{Herzog:2010vz} which is valid close to the phase transition point.
The constitutive relations so obtained fit within the 14 parameter fluid 
dynamical framework spelt out in \cite{Bhattacharya:2011ee} which generalize 
the text book predictions of \cite{Clark,Putterman}). In particular, it was explicitly 
verified that the entropy current obtained by holographic methods 
matches the canonical entropy current in agreement with the intuition of 
\cite{LLvol6}. 

In this paper we consider a generic asymptotically $AdS_5$ gravitational system which involves a  Chern Simons term for a single bulk $U(1)$ field. Such a Chern Simons term signifies the presence of a $U(1)^3$ triangle anomaly in the dual field theory and so, 
in particular, introduces parity violating terms into the
effective superfluid dynamical description. Particular truncations of type IIB supergravity for which our analysis is valid can be found in 
\cite{Gubser:2009qm,Bhattacharyya:2010yg}. We find that, in the collinear limit, it is possible to obtain integral expressions for all the parity odd transport coefficients in terms of the background solution. Our results  are in perfect agreement with the 
collinear limit of the modified theory of parity violating superfluid dynamics 
which we present in section \ref{S:SuperfluidPodd}. We regard this agreement
 as a nontrivial (though 
as yet limited) check of the general results of section \ref{S:SuperfluidPodd}.
It should be possible to use the AdS/CFT correspondence away from the 
collinear limit to find much more extensive test our results; we leave this 
to future work.

{\it Note Added in v1: } After the work presented in this paper was completed, 
we received a paper \cite{Lin:2011mr} that has substantial overlap with section 
\ref{S:SuperfluidPodd} of this paper. While the approach of 
\cite{Lin:2011mr} is similar to the one 
adopted in this paper, our final results differ in several qualitative and quantitative
respects. We present a brief comparison with  \cite{Lin:2011mr} in section \ref{S:Summary}.  

{\it Note Added in v2: } The revised version of the paper \cite{Lin:2011mr} agrees better 
with our results. Our results have also been confirmed and generalized in \cite{Neiman:2011mj}, which 
also pointed out an error in the sign of one of the Onsager relations in the first version 
of this paper. The results of this paper have also been generalized to the study of 
superfluids that do not respect time reversal invariance in \cite{Bhattacharyya:2012xi}.

%*******************************************************************
\section{The theory of charged fluid dynamics}\label{cfm}
%*******************************************************************
%
In this pedagogical section we construct the most general equations of 
Lorentz invariant charged fluid dynamics consistent with the 
second law of thermodynamics. 
Our goal is to illustrate our method for determining the most general 
form of fluid-dynamical equations of motion in a simple and familiar context 
before tackling the slightly more complicated case of superfluids. 
The final results of this section are well known; the novelty of 
this section lies in our method of computation. 

The long-wavelength degrees of freedom of a locally equilibrated 
system with a single global $U(1)$ charge can be taken to be the velocity 
field $u_{\mu}(x)$ (normalized so that $u^{\mu}u_{\mu} = -1$), 
the temperature field $T(x)$ and a chemical potential field $\mu(x)$. Both the energy momentum tensor and the charged current can be written in terms of these five fields and their gradients.
The equations of motion of charged fluid dynamics are the conservation of 
the stress tensor and charge current 
\begin{equation} 
\label{backeqc}
\begin{split}
	\nabla_\mu T^{\mu\nu}&=F^{\nu \mu} J_\mu\\
	\nabla_\mu J^\mu &= -\frac{c}{8} \epsilon^{\mu\nu\rho \sigma}
F_{\mu\nu} F_{\rho \sigma}
\end{split}
\end{equation}
which provides the five equations for the five hydrodynamic fields. In these equations 
we have allowed for the possibility that the current in question 
has a $U(1)^3$ anomaly. We call the coefficient $c$ the anomaly coefficient. We have also allowed the current to be coupled to an external source with field strength $F_{\mu\nu}$.  To completely determine the equations of motion it remains to 
determine the dependence of $T^{\mu\nu}$ and $J^{\mu}$ on the fields $u^{\mu}(x)$, 
$T(x)$, $\mu(x)$ and their derivatives.

By considering a stationary fluid for which $u^{\mu} = (1,\,0,\,0,\,0)$ and using boost invariance one can argue that the stress tensor and charge current take the form
\begin{equation} \label{ltconst}
\begin{split}
T^{\mu\nu}&=(\rho+P) u^\mu u^\nu + P \eta^{\mu\nu} + T_{diss}^{\mu\nu} \\
J^\mu &=q u^\mu + J^\mu_{diss}\\
\end{split}
\end{equation}
where $T_{diss}^{\mu\nu}$ and $J^\mu_{diss}$ are the contributions to the 
stress tensor and charge current that involve derivatives of $\mu$, $T$ and $u^{\mu}$. The equations that express $T_{diss}^{\mu\nu}$ and $J^\mu_{diss}$
in terms of fluid dynamical fields and their derivatives are termed 
constitutive relations. 
In the long wavelength fluid dynamical limit it is sensible to expand the 
constitutive relations in powers of derivatives of the fluid dynamical 
fields $u^{\mu}$, $T$ and $\mu$. We will refer to such an expansion as a derivative expansion and refer to the terms which are linear in gradients as first order terms. 
In this paper we work only to first order in the 
derivative expansion. The electromagnetic source term $F^{\mu\nu}$ is taken
to be of first order in derivatives in this counting.

\cl{ Field Redefinitions and frame choices}
\noindent
Note that the fluid temperature $T$, chemical potential $\mu$ and 
velocity $u^\mu$ are thermodynamical concepts that are well defined in 
equilibrium but have no microscopic definitions 
in dynamical situations. In other words, we are always free to redefine the 
thermodynamic variables into primed ones according to the equations 
\begin{equation} \begin{split}\label{fo}
u^\mu&= u^{\prime\,\mu}  + \delta u^\mu\\
T&=T' + \delta T\\
\mu&=\mu'+ \delta \mu
\end{split}
\end{equation}
where $\delta u^\mu$ is an arbitrary one derivative 
vector that obeys $\delta u^\mu u_\mu=\delta u^{ \mu} u^{\prime}_{\mu} = 0$
and $\delta T$ and $\delta \mu$ are arbitrary one derivative scalars. 
The primed and unprimed fields are each equally good definitions of 
the velocity, temperature and chemical potential fields. 
Physically meaningful assertions, such as the 
constitutive relations for $T_{diss}^{\mu\nu}$ and $J^\mu_{diss}$, must
only involve field redefinition invariant quantities.

Thus, let us determine the field redefinition invariant combinations of 
$T^{\mu\nu}_{diss}$ and $J^\mu_{diss}$.
Under the field redefinition \eqref{fo},
\begin{equation} \begin{split} \label{channge}
\delta T_{diss}^{\mu\nu} & = 
(u^\mu \delta u^\nu + u^\nu \delta u^\mu) (P+\rho) + 
u^\mu u^\nu d(P+\rho) + \eta^{\mu\nu} dP\\
\delta J^{\mu}_{diss}&=q \delta u^\mu + d q u^\mu \\
\end{split}
\end{equation}
where 
$$\delta T_{diss}^{\mu \nu}=T_{diss}^{\prime\,\mu \nu}- T_{diss}^{\mu \nu}$$
$$\delta J^{\mu}_{diss} = J^{\prime\,\mu}_{diss} -J^{\mu}_{diss}$$
and $d f(\mu, T)$ represents the change in 
the function $f$ under the first order variable change \eqref{fo}. 
It useful to decompose $T^{\mu\nu}_{diss}$ and $J^\mu_{diss}$ into 
$SO(3)$ invariant tensors, vectors and scalars The $SO(3)$ that we are referring to is the group of 
rotations orthogonal to $u^\mu$. To this end we introduce the projection matrix
\begin{equation}
	P^{\mu\nu}= \eta^{\mu\nu} + u^{\mu}u^{\nu}\,.
\end{equation}
We find that there is one tensor, two vectors and three scalars.
The unique tensor 
\begin{equation} \label{unten} 
	P^\mu_{\phantom{\mu}\alpha} P^\nu_{\phantom{\nu}\beta} T^{\alpha\beta}_{diss}- 
\frac{P^{\mu \nu}}{3}  P_{\alpha \beta} T_{diss}^{\alpha \beta}
\end{equation}
is automatically field redefinition invariant. 
The two vectors $P^{\mu}_{\phantom{\mu}\alpha} T_{diss}^{\alpha\beta} u_{\beta}$ and $P^{\mu}_{\phantom{\mu}\alpha} J^{\alpha}$ transform under field redefinitions as 
\begin{equation}\label{vectransf} \begin{split}
\delta \left( P^{\mu}_{\phantom{\mu}\alpha} T_{diss}^{\alpha\beta} u_{\beta} \right)
&=-(P+\rho) \delta u^\mu\\
\delta \left( P^{\mu}_{\phantom{\mu}\alpha} J^{\alpha} \right)&= q \delta u^\mu
\end{split}
\end{equation}
so that the unique invariant combination of vectors is 
given by 
\begin{equation}\label{unvec}
P^\mu_{\phantom{\mu} \alpha} J^\alpha + 
\frac{q}{P+\rho} \left( P^\mu_{\phantom{\mu}\alpha} T_{diss}^{\alpha \beta} u_\beta \right) \,.
\end{equation}
The three scalars transform under field redefinitions as 
\begin{equation}\label{scaltransf} \begin{split}
\delta \left( P_{\alpha \beta} T_{diss}^{\alpha \beta} \right)
&=3 dP\\
\delta \left( u_\alpha T_{diss}^{\alpha \beta}u_\beta \right)&= d \rho\\
\delta \left( u_\alpha J^\alpha \right)&= -dq
\end{split}
\end{equation}
so that the unique invariant scalar is given by 
\begin{equation}\label{unscal}
\frac{1}{3}\left( P_{\alpha\beta} T_{diss}^{\alpha \beta}\right)
- \frac{\partial P}{\partial {\rho}} \left( u_\alpha T_{diss}^{\alpha \beta}u_\beta \right) + 
\frac{\partial P}{\partial {q}} \left( u_\alpha J^\alpha \right)
\end{equation}
where $\frac{\partial P}{\partial {\rho}}$ is taken at constant $q$ and 
$\frac{\partial P}{\partial {q}}$ is taken at constant $\rho$. 

Instead of working in a manifest field redefinition invariant manner, 
it is sometimes convenient to `fix' the field redefinition ambiguity 
by imposing five additional conditions on the thermodynamic fields so that 
they are well defined. Different choices of fixing the ambiguity are referred to as frames. One often used frame is the so called Landau frame, in which the velocity and temperature fields 
are defined to obey the conditions $T_{diss}^{\mu \nu}u_\nu=0$ and 
$ J^\mu_{diss}u_\nu=0$. This gives one vector and two scalar conditions, 
matching the field redefinition degrees of freedom. Another choice of frame is the Eckart frame which is defined by the conditions $J^\mu_{diss}=0$ together with 
$u_\mu T_{diss}^{\mu\nu}u_\nu=0$. The expressions for the invariant 
vector \eqref{unvec} and the invariant scalar \eqref{unscal} greatly 
simplify in either of these frames. In this paper we adopt no 
such `gauge' choice but work in a fully field redefinition invariant manner.

\cl{ {\bf The strategy for the rest of this section}}
\noindent
In order to complete our specification of the equations of charged 
fluid dynamics, we need to specify $T_{diss}^{\mu\nu}$ and $J^\mu_{diss}$ 
(or more precisely the field redefinition invariant parts 
\eqref{unten}, \eqref{unvec} and \eqref{unscal} of these 
expressions) as a function of first derivatives of fluid dynamical fields. 
Of course, in any particular dynamical system, the explicit form of the 
constitutive relations for $T_{diss}^{\mu\nu}$ and $J^\mu_{diss}$ can be 
determined only by a detailed dynamical computation. In this paper we will 
be interested not in computing the precise form of these quantities in any 
particular system, but in parameterizing the most general form that the 
constitutive relationship can take in any system.  
As we will see below, it will prove possible to completely determine the form 
of the first order constitutive relations up to three undetermined dissipative
parameters, each of which is an arbitrary function of $T$ and $\mu$.  

We proceed as follows. As in any effective field theory, 
we start by writing down all possible 
expressions which may contribute to $T_{diss}^{\mu\nu}$ and $J_{diss}^{\mu}$. We then eliminate those that do not satisfy the symmetries of the theory, 
Lorentz invariance in this case. In addition, since we are dealing with a 
hydrodynamic theory, we must ensure that the second law of thermodynamics is 
satisfied. As explained in the introduction, we demand the existence of 
an entropy current of positive semi-definite divergence even when the 
theory is formulated on a curved background. As alluded to in section \ref{S:intro}, the entropy current is defined to be a four vector $J_S^{\mu}$ satisfying two requirements. The first is that in a configuration where the fluid is 
in uniform motion,\footnote{Such a configuration is a stationary, dissipation 
free solution to the equations of fluid dynamics. Indeed it may be obtained 
by boosting a uniform fluid at rest (by which we mean a uniform fluid with velocity field 
$u^{\mu} = (1,\,0,\,0,\,0)$).}
\begin{equation}\label{entperf}
	J_S^{\mu} = s u^{\mu} \quad \hbox{(for a spacetime independent configuration)}
\end{equation}
with $s$ the entropy density which is related to $\rho$, $P$, $q$, $\mu$ and $T$ through
\begin{equation}
\label{E:gd}
	\rho + P = s T + \mu q \,.
\end{equation}
Our second requirement of the entropy current is that its divergence is positive semi-definite in an arbitrary curved background,
\begin{equation}
\label{E:dJs}
	\nabla_{\mu}J_S^{\mu} \geq 0
\end{equation}
implying that the entropy increase in any region is always greater 
than the entropy inflow into that region.  

For a perfect fluid level (i.e. a fluid in which all gradient terms have been neglected--- 
$T^{\mu\nu}_{diss}=J^\mu_{diss}=0$) the entropy current 
is given by \eqref{entperf}. At this order it is 
not difficult to verify that $\nabla_{\mu} (s u^{\mu}) = 0$ using 
\eqref{E:gd} and $dP = s dT + qd\mu$.

Once the gradients of $u^{\mu}$, $T$ and $\mu/T$ are non vanishing the divergence of the entropy current no longer vanishes. Indeed, the divergence of the entropy current at the one derivative level
will be the focus of much of the rest of this paper. 
 We will demand, on physical grounds, that it is possible 
to modify \eqref{entperf} by first order corrections so that 
\eqref{E:dJs} will be satisfied. This requirement will turn out to constrain the possible 
forms of $T^{\mu\nu}_{diss}$ and $J^\mu_{diss}$. 

We start our analysis in section \ref{S:NormalPEven} by considering  parity conserving charged fluids. In section \ref{S:NormalPOdd} we move on to describe parity-violating fluid dynamics.

\subsection{Parity Invariant Charged Fluid Dynamics}\label{S:NormalPEven}
%==========================================================================
%
Consider a hydrodynamic theory in the presence of external electromagnetic fields satisfying \eqref{backeqc} with $c=0$. Following the general prescription described at the beginning of this section, we would like to write the most general 
parity-invariant and Lorentz invariant contributions to $J_{diss}^{\mu}$, $T_{diss}^{\mu\nu}$ and $J_S^{\mu}$ which involves a single derivative of the hydrodynamic fields $u^{\mu}$, $T$ and $\mu$. This is carried out in section \ref{SS:classification}. We then work out the the restrictions on these terms by requiring that the entropy current has positive semi-definite divergence. This is described in section \ref{SS:positive}.

%-------------------------------------------------------------------------
\subsubsection{Classification of one and two derivative data}
\label{SS:classification}
%-------------------------------------------------------------------------
%
We begin our analysis on a technical point. The tangent space about
any point in our spacetime manifold has an $SO(3,1)$ rotational invariance. 
However, the fluid velocity vector, $u^\mu(x)$, takes a definite value at that 
point and breaks this rotational group down to $SO(3)$. It is  
useful to decompose all derivatives of fluid dynamical 
fields, at any given spacetime point, into representations of this 
residual $SO(3)$ rotational group. 

In the first column of table \ref{tb:PEoneD} we have classified all expressions 
formed from a single derivative of any of $u^\mu(x)$, $T(x)$ and $\mu(x)$ 
according to their $SO(3)$ and parity transformation properties. We refer to  
these expressions as one derivative fluid dynamical data. We have 
also classified one derivative expressions constructed out of the background 
electromagnetic fields according to their $SO(3)$ and parity
transformation properties. We will refer to these as background data. As fluid and background field data enter our 
analysis on an even footing, we have listed these expressions together 
in the first column of table \ref{tb:PEoneD}.\footnote{Since all curvature invariants built out 
of the background metric have at least two derivatives, there is no 
one derivative data associated with the metric.}

Not all the expressions in the first column of table \ref{tb:PEoneD} are independent under the equations of motion. The equations of motion can be used to solve for some pieces of data in terms of other data. The classification of the equations of motion according to their $SO(3)$ and parity transformation properties can be found in the middle column of table \ref{tb:PEoneD}. Note that 
there are no tensor equations of motion.

In the last column of table \ref{tb:PEoneD} we have listed a choice 
of independent data. By this we mean a choice of independent one derivative 
fluid dynamical expressions and one derivative field expressions in terms of which all others can be solved for. 

\TABLE{
\renewcommand\arraystretch{1.1}
\caption{One-derivative expressions classified according to their transformation laws under the $SO(3)$ residual symmetry and parity. The first column lists all one derivative data. The second column lists the equations of motion. 
The last column lists a choice of independent data. See 
\eqref{E:sigmadef} and \eqref{pdef}  for the definition of 
$\sigma_{\mu\nu}$ and $P_{\mu\nu}$ respectively.}
\label{tb:PEoneD}
\begin{tabular}{| p{2.6 cm} | c | c | l |}
\hline        
	 \raggedright $SO(3)$ and $P$ classification & All  data & Equations of motion & Independent  data\\
\hline
\hline
	\multirow{3}{*}{Scalars} & 
	$ u^{\mu}\partial_{\mu} T$ & 
	\multirow{3}{*}{
		\begin{minipage}{2.5 cm}
			\begin{flushright} 
				$u_{\mu}\nabla_{\nu} T^{\mu \nu} = 0$  \\  $\nabla_{\mu} J^{\mu}=0$ 
			\end{flushright}
		\end{minipage} 
	} &
	\ \multirow{3}{*}{$S_1 = \partial_{\mu} u^{\mu}$} \\
	& $u^{\mu}\nabla_{\mu} \frac{\mu}{T}$ & & \\
	& $\partial_{\mu} u^{\mu}$ & & \\
\hline
	\multirow{4}{*}{Vectors} & 
	$P^{\mu \nu} \partial_{\nu} T$ & 
	\multirow{4}{*}{$P^{\mu}_{\phantom{\mu}\nu}\partial_{\rho}T^{\rho \nu} =0$} &
	\multirow{4}{*}{
		\begin{minipage}{4 cm}
			\begin{flushleft} 
				$V_1 = -P^{\mu \nu} \partial_{\nu}\frac{\mu}{T} + \frac{F^{\mu\nu}u_{\nu}}{T}$ \\ 
                                $V_2 = u^{\nu}\nabla_{\nu} u^{\mu}$ \\  $V_3 = F^{\mu\nu}u_{\nu}$ 
			\end{flushleft}
		\end{minipage} 
	} \\
	&  $u^{\nu}\partial_{\nu} u^{\mu}$ & & \\
	& $P^{\mu \nu} \partial_{\nu}\frac{\mu}{T}$ & & \\
	&  $F^{\mu\nu}u_{\nu}$ & & \\
\hline
	Tensors &$\sigma_{\mu\nu}$ & -- &\ $T_1 = \sigma_{\mu\nu}$ \\
\hline
	\multirow{2}{*}{Pseudo vectors} & $\frac{1}{2}\epsilon^{\mu\nu \alpha \beta} u_\nu \partial_{\alpha} u_{\beta} $ & \multirow{2}{*}{--} & \ $\omega^{\mu} = \frac{1}{2}\epsilon^{\mu\nu \alpha \beta} u_\nu \partial_{\alpha} u_{\beta} $ \\
	& $\frac{1}{2}\epsilon^{\mu\nu \alpha \beta} u_\nu F_{\alpha\beta}  $ & & \ $B^{\mu} = \frac{1}{2}\epsilon^{\mu\nu \alpha \beta} u_\nu F_{\alpha\beta}$\\
\hline
\end{tabular}
}

While some of the expressions used in table \ref{tb:PEoneD} such as
\begin{equation}
\label{E:sigmadef}
	\sigma_{\mu\nu} =\half P^{\mu \alpha} P^{\nu \beta} 
	\left( \nabla_{\alpha} u_{\beta} + \nabla_{\beta} u_{\alpha} - 
	P_{\alpha \beta} \left( \nabla_{\lambda} u^{\lambda}\right)\right)
\end{equation}
and
\begin{equation}\label{pdef}
	P^{\mu\nu} = u^{\mu}u^{\nu} + \eta^{\mu\nu}
\end{equation}
are standard, some of our notation isn't. The new notation has been 
introduced in order to prepare the reader for later sections. In particular
$V_3$ is the electric field in the rest frame of a fluid element. In the 
conventions of Son and Sur\'owka \cite{ Son:2009tf} we have 
\begin{equation}\label{ebdef} 
V_3= E_\mu\,.
\end{equation}

We will soon construct an entropy current that includes terms which are  first 
order in derivatives. The divergence of such an entropy current is 
of second order in derivatives and includes terms quadratic in first order 
fluid (and background field) data plus expressions built out 
of two derivatives acting on fluid fields or single derivatives of electromagnetic 
field strengths. We refer to the second class of expressions as two derivative scalar data.
When studying the divergence of the entropy current it is useful to have a listing of 
independent scalar two derivative data.

 In the first column of table \ref{tb:PEtwoD} we list the most 
general fluid and background field (but not curvature related) two derivative 
data that transforms as an $SO(3)$ scalar. More explicitly, we list all 
scalar expressions formed by acting with two derivatives on $u^\mu(x)$, 
$T(x)$ and $\mu(x)$ together with all scalars formed from the action of 
a single derivative on electromagnetic field strengths.\footnote{It is also 
easy to list 
two derivative fluid  data in the 3, 5 and 7 dimensional representations of 
$SO(3)$, but that will not be required in what follows, so we do not present
such a listing.} In the second column of the same table we list all scalar two derivative equations of motion. In the last column of the same table 
we list our choice of independent two derivative scalar data (in terms of which
we have solved for all the other two derivative scalars). 

\TABLE{
\renewcommand\arraystretch{1.1}
\caption{Parity even two derivative scalar data for charged fluids. 
The first column lists all six second order scalars constructed from two 
derivatives of the hydrodynamic variables and background field strengths. 
The second 
column lists the three scalar two derivative equations of motion.
The last column lists one choice of a $6-3=3$ dimensional basis for  the
independent two derivative scalar data.}
\label{tb:PEtwoD}
\begin{tabular}{|c|c|c|}
\hline        
  All  data & Equations of motion & Independent  data\\
\hline
\hline
	$ u^{\mu} u^{\nu} \partial_{\mu} \partial_{\nu} T$ & 
	\multirow{6}{*}{
		\begin{minipage}{3.3 cm}
			\begin{flushright}
				$ \nabla_{\mu} \nabla_{\nu} T^{\mu \nu} =0$ \vskip 0.2cm
				$P^{\rho}_{~\nu} \nabla_{\rho}\nabla^{\mu} T^{\mu \nu} =0$ \vskip 0.2cm
				$u^{\nu} \nabla_{\nu} \nabla^{\mu} J^{\mu} =0$
			\end{flushright}
		\end{minipage}
	} &
	\multirow{6}{*}{
		\begin{minipage}{2.5 cm}
			\begin{center}
				$P^{\mu \nu} \nabla_{\mu} \partial_{\nu} \frac{\mu}{T}$ \vskip 0.2cm
				$u^{\mu}\nabla_{\mu}\partial_{\nu}u^{\nu}$ \vskip 0.2cm
				$ \nabla_{\mu}(F^{\mu\nu}u_{\nu})$
			\end{center}
		\end{minipage}
	} \\
	$ P^{\mu \nu} \partial_{\mu} \partial_{\nu} T$ && \\
	$ u^{\mu} u^{\nu} \partial_{\mu} \partial_{\nu} \left( \frac{\mu}{T} \right)$ && \\
	$P^{\mu \nu} \partial_{\mu} \partial_{\nu} \left(\frac{\mu}{T} \right)$ && \\
	$u^{\mu}\partial_{\mu}\partial_{\nu}u^{\nu}$ && \\
	$\partial_{\mu}(F^{\mu\nu}u_{\nu})$ && \\
\hline
\end{tabular}
}
%
%

%---------------------------------------------------------------------
\subsubsection{The general entropy current and its divergence}
\label{SS:positive}
%---------------------------------------------------------------------
%
Armed with the listings in tables \ref{tb:PEoneD} and \ref{tb:PEtwoD} we now proceed with our analysis. Traditional studies of first order charged fluid dynamics (see, for example, \cite{LLvol6}) assume that the entropy current takes a canonical form,\footnote{As explained in \cite{Bhattacharya:2011ee} the expression in \eqref{canent} is frame invariant, 
i.e. invariant under a first order field redefinition of $T$, $u^\mu$ and $\mu$. 
Note that the second term on the right hand side vanishes in the Landau 
frame while the third term vanishes in the Eckart frame.} 
\begin{equation}\label{canent} 
\Jsc = s u^{\mu} - \frac{1}{T} u_{\mu} T_{diss}^{\mu \nu} - \frac{\mu}{T} J^{\mu}_{diss} \,.
\end{equation}
As we explained in the introduction, in this work we 
will not make any prior assumption about the form of the entropy current. According 
to the analysis of section \ref{SS:classification} the most general parity even 
first order entropy current is given by  
\begin{equation}\label{entcurf}
 J^{\mu}_{S} = \Jsc + s_1 \, S_1 u^{\mu} 
+ \sum_{i=1}^3 v_i V_i^\mu 
\end{equation} 
where $S_1$ and $V_i$ are defined in the last column of table \ref{tb:PEoneD},
and $s_1$ and the $v_i$'s are arbitrary functions of $\frac{\mu}{T}$ and 
$T$. 

We now explore the constraints obtained by enforcing the positivity of 
the divergence of the entropy current \eqref{entcurf}. It is easily 
demonstrated (see, for instance, \cite{LLvol6,Bhattacharya:2011ee}) that the divergence of the 
canonical part of the entropy current is given by
\begin{equation}\label{diventcan}
 \nabla_{\mu} \Jsc = -\nabla_{\mu} \left( \frac{u_{\nu}}{T} \right) T_{diss}^{\mu \nu} 
- \left( \partial_{\mu} \left( \frac{\mu}{T}\right) - \frac{F_{\mu\nu}u^{\nu}}{T}\right) J^{\mu}_{diss}.
\end{equation}
The right hand side of \eqref{diventcan} is a quadratic form in one derivative fluid and 
background electromagnetic field data. The divergence of the non canonical part of the  
entropy current in \eqref{entcurf} is also a two derivative expression but 
is composed of two kinds of terms. The first set of terms are linear in independent
two  derivative and curvature data. Such terms are always inconsistent with 
the positivity of the entropy current, and so we must choose $s_1$ and 
$v_i$ so that these terms vanish. The second set of terms contains  
products of one derivative terms. Such terms would modify the quadratic form on the right hand side of \eqref{diventcan} and do not necessarily vanish. Schematically, we have
\begin{equation}
\label{E:schematic}
	\partial_{\mu} J_S^{\mu} = \left( \substack{ \hbox{\small independent two} \\  
\hbox{\small derivative and curvature data} } \right) +  \left( \substack{ \hbox{\small quadratic form in} \\  \hbox{\small first order data} } \right)\,.
\end{equation}
The first term on the right hand side of \eqref{E:schematic} must vanish while the 
second term must be tuned to be positive.

%--------------------------------------------------------------------------------------
\subsubsection{Constraints from positivity of the divergence of the entropy current}
%-------------------------------------------------------------------------------------
%
We will first explore the constraints that follows from the requirement that 
no two derivative data appears in the divergence of the entropy current. As explained previously, this implies that the first set of terms on the right hand 
side of \eqref{E:schematic} must vanish. We will implement this condition 
separately for two derivative and curvature terms.

\cl{\bf Constraints from the vanishing of 2 derivative terms}
\noindent
The two derivative part of the divergence of the entropy is given by 
$$ 
	-v_1 P^{\mu \nu} \nabla_{\mu} \partial_{\nu} \frac{\mu}{T}
+ (s_1+v_2)u^{\mu}\nabla_{\mu}\partial_{\nu}u^{\nu} + \left(v_3 + \frac{v_1}{T}\right) \nabla_{\mu}(F^{\mu\nu}u_{\nu})\,.
$$
This expression is a linear combination of the three independent 
two derivative pieces of data (see Table \ref{tb:PEtwoD}). 
It follows that the vanishing of two 
derivative terms requires us to set the coefficients of each of these 
terms to zero, i.e. to set $v_1=v_3=0$ and $v_2=-s_1$. Thus the vanishing of two derivative terms in the divergence of the entropy current restricts the  entropy current 
\eqref{entcurf} to take the form
\begin{equation}\label{entcurf2}
 J^{\mu}_{S} = \Jsc + s_1 \left( S_1 u^{\mu} 
- V_2^\mu \right) \,.
\end{equation} 
where $s_1$ is still an arbitrary function of $T$ and $\mu$.

\cl{\bf Constraints from vanishing of curvature terms}
\noindent
According to \eqref{entcurf2} the
entropy current has a one parameter ambiguity, $s_1$. Were we to restrict our attention 
to a flat space background we would not have been able resolve this ambiguity. Consider a charged fluid propagating on 
an arbitrary curved background. The cancellation of two derivative terms 
proportional to $s_1$ is now incomplete; it is not difficult to 
check that there is an additional, curvature dependent term in the divergence
of the entropy current proportional to 
$s_1R_{\alpha \beta} u^{\alpha}u^{\beta}$ with $R_{\alpha\beta}$ the Ricci tensor.
This term is inconsistent with positivity of the divergence of $J_S^{\mu}$. Thus, we are forced to set
$s_1=0$.

We conclude that the requirement that the divergence of the entropy current 
is positive {\it in an arbitrary curved background} forces the entropy current 
to take its canonical form, justifying the assumptions of standard 
treatments of fluid dynamics e.g. \cite{LLvol6}.

%---------------------------------------------------------------------
\subsubsection{Constraints on dissipative terms}
\label{SS:Constraints}
%---------------------------------------------------------------------
%
We have demonstrated that the entropy current takes its canonical form 
and consequently that its divergence is given by \eqref{diventcan}. 
It is now not difficult to work out the 
constraints on dissipative terms that ensure the positivity of 
the quadratic form on the right hand side of \eqref{diventcan}. 
We outline the calculation here. 

Consider the expansion of $\nabla_{\mu} \left(\frac{u_{\nu}}{T}\right)$ and $-V_1 = \partial_{\mu} \frac{\mu}{T} - \frac{E_{\mu}}{T}$ which appear on the right hand side of \eqref{diventcan} into $SO(3)$ invariant tensors vectors and scalars. We find a single tensor, $\sigma_{\mu\nu}$, 
two vectors,
$$V_{1}^{\mu}, \quad \hbox{and} \quad \left(P^{\mu\nu}\frac{\partial_\nu T}{T} + (u.\partial)u^\mu \right)$$
(see table \ref{tb:PEoneD} for a definition of the vector $V_1^\mu$) 
and three scalars,
$$\frac{(u \cd \partial)T}{T}, \quad (u \cd \partial)\nu,  \quad \hbox{and} \quad (\nabla \cd u)\,.$$
While the two vectors are completely distinct off-shell, it turns out that 
the equations of motion imply that they are proportional to each other on-shell.
Similarly, the equations of motion imply that the three scalars are also 
proportional to each other on-shell. As we demonstrate in Appendix \ref{na} 
the explicit relations are
\begin{equation}\label{simpeom}
\begin{split}
&\frac{(u.\partial)T}{T} = -\left[\frac{\partial P}{\partial \rho}\right]_q (\nabla.u)\\
&(u.\partial)\nu = -\frac{1}{T}\left[\frac{\partial P}{\partial q}\right]_\rho (\nabla.u)\\
&P^{\mu\nu}\frac{\partial_\nu T}{T} + (u.\partial)u^\mu = 
\frac{q T}{\rho +P} V_1^\mu\,.
\end{split}
\end{equation}
Plugging these relations into \eqref{diventcan}, we can rewrite the 
divergence of the entropy current in the form 
\begin{equation}\label{divergsimp}
\begin{split}
\nabla_{\mu} J_S^{\mu}  =&- \frac{\left(\nabla_\mu u^\mu\right)}{T}\left[\frac{(T_{diss})_{ab} P^{ab}}{3} - \frac{\partial P}{\partial\rho} (u_\mu u_\nu T_{diss}^{\mu\nu}) +\frac{\partial P}{\partial q} (u_\mu J_{diss}^\mu )\right]\\
&+ V_{1\,\mu} \left[J^\mu_{diss} + \frac{q}{\rho + P} (u_\nu T_{diss}^{\mu\nu})\right]-
\frac{T_{diss}^{\mu\nu}\sigma_{\mu\nu}}{T}\\
\end{split}
\end{equation}
where $V_1^{\mu}$, $B^{\mu}$ and $E^{\mu}$ were defined in Table \ref{tb:PEoneD} and \eqref{ebdef}. We collect their definitions here for convenience:
\begin{align*}
	E_\mu &=F_{\mu\nu} u^\nu \\
	B_\mu &=\frac{1}{2}\epsilon_{\mu\nu \alpha \beta} u^\nu F^{\alpha \beta}\\
	V_{1\,\mu} &= \frac{E_\mu}{T} - P_\mu ^\theta \partial_\theta \nu\,.
\end{align*}

We will now use \eqref{divergsimp} to constrain the constitutive 
relations of fluid dynamics, i.e. the expressions for  
$T^{\mu\nu}_{diss}$ and $J^\mu_{diss}$ as a linear expansion in 
first order scalars, vectors and tensors. To first order in gradients there is only one independent scalar data so the scalar parts of $T^{\mu\nu}_{diss}$ and $J^\mu_{diss}$ are necessarily proportional to $\nabla \cd u$. The vector parts of 
$T^{\mu\nu}_{diss}$ and $J^\mu_{diss}$ must each be expanded as a linear sum of 
the three independent vectors listed in Table \ref{tb:PEoneD}. The tensor 
in Table \ref{tb:PEoneD} is proportional to $\sigma_{\mu\nu}$ since there is only one $SO(3)$ invariant tensor. 
It follows from group theory that positivity of the divergence of the entropy current implies positivity of the scalar, vector and tensor components separately. Thus, we have
\begin{equation}\begin{split}\label{constraints}
P^\mu_\alpha P^\nu_\beta T_{diss}^{\alpha \beta}  - \frac{P^{\mu\nu}}{3}P_{\alpha\beta}T_{diss}^{\alpha\beta} &
= -\eta \sigma^{\mu\nu}\\
P_\alpha^\mu 
\left( J^\alpha_{diss} + \frac{q}{\rho + P} (u_\nu T_{diss}^{\alpha\nu}) \right) 
&= \kappa V_1^\mu \\
\frac{(T_{diss})_{ab} P^{ab}}{3} - \frac{\partial P}{\partial\rho} (u_\mu u_\nu T_{diss}^{\mu\nu}) +\frac{\partial P}{\partial q} (u_\mu j_{diss}^\mu )& 
= -\beta \partial_\alpha u^\alpha 
\end{split}
\end{equation}
where 
$$ \eta \geq 0, ~~~ \kappa \geq 0, ~~~ \beta \geq 0 .$$
These three coefficients are the shear viscosity, $\eta$, the heat conductivity, $\kappa$, and the bulk viscosity, $\beta$. The bulk viscosity is traditionally denoted by $\zeta$ but in this work we reserve $\zeta$ for different use.\footnote{Note that the speed of sound, $c_s$, is related to the variation of the pressure with respect to energy density through $\frac{\partial P}{\partial \rho} = c_s^2$. Using dimensional analysis one can conclude that $\frac{\partial P}{\partial q}=0$ in a scale invariant theory. It then becomes clear that in a conformal theory the left hand side of the last equality in \eqref{constraints} vanishes as it should.}

Several aspects of \eqref{constraints} deserve comment. First, the requirement 
of positivity does not individually constrain the three scalar and two 
vector pieces in $T_{diss}^{\mu\nu}$ and $J_{diss}^\mu$, but only constrains 
the combinations that appear in \eqref{unscal} and \eqref{unvec}. This is 
exactly as we would expect: only field redefinition invariant data 
can be constrained in a physical way. The vectors and scalars 
that are left undetermined are unphysical; they can be changed, or chosen arbitrarily, by a field redefinition. Despite 
appearances, \eqref{constraints} constitutes a complete determination 
of the constitutive relations of our system. 

We also note that we could have used the fact that the divergence of the entropy current is frame invariant (see \cite{Bhattacharya:2011ee}) to determine the frame invariant scalar, vector and tensor combinations in \eqref{unten}, \eqref{unvec} and \eqref{unscal}; the expression on the right hand side of \eqref{divergsimp} must arrange itself into such frame invariant combinations.

The third aspect to note is that the constraints of positivity are relatively mild in the 
scalar and tensor sector. The expansion of scalars and tensors is the most general one permitted 
by symmetry; the requirement of positivity merely imposes inequalities in the coefficients of this 
expansion. However, the constraint on vectors is much stronger. Symmetry alone would have allowed
the expansion of the second line in \eqref{constraints} as an arbitrary linear combination of the 
3 vectors $V_1$, $V_2$ and $V_3$. However the requirement of positivity sets the coefficients $V_2$ and $V_3$ to zero,\footnote{The origin of this constraint is the observation that the 
quadratic form $ax^2 +bxy +c x z$ is positive only when $b=c=0$ and $a \geq 0$. The role of $x$ 
is played by the vector $V$, while the roles of $y$ and $z$ are played by the other two vectors} apart 
from imposing an inequality on the coefficient of the third. We will see this pattern repeated and magnified in the study of superfluid dynamics in sections \ref{S:ParityInvariant} and \ref{S:SuperfluidPodd} in the scalar, vector and tensor sector.

%==========================================================================
\subsection{Parity non invariant charged fluid dynamics}
\label{S:NormalPOdd}
%==========================================================================
%
Let us now turn to the dynamics of fluids that are not invariant under parity transformations. According to table \ref{tb:PEoneD} we should allow the entropy current to depend on an additional arbitrary pseudo vector. Thus, the most general entropy current for such a fluid takes the form 
\begin{equation}\label{entcurfp}
 J^{\mu}_{S} = \Jsc + s_1 \, S_1 u^{\mu} 
+ \sum_{i=1}^3 v_i V_i^\mu + \sigma_{\omega}\omega^{\mu} + \sigma_B B^{\mu}\,.
\end{equation} 
In the parity even sector the divergence of this entropy current is identical to the one discussed in subsection \ref{S:NormalPEven}; the arguments in 
\ref{S:NormalPEven} go through unchanged and in particular 
the cancellation of two derivative and scalar terms 
set $s_1 =v_i=0$. In the parity odd sector the divergence of 
the entropy current receives contributions involving the dot product of the pseudo vectors 
$\omega^{\mu}$ and $B^{\mu}$ with ordinary vectors. Positivity of the 
divergence of the entropy current implies that such products vanish.\footnote{We will see later that products of vectors and pseudo vectors do not necessarily need to vanish in the case of superfluid dynamics. 
In the current setup vanishing of such bilinear terms follows from the fact that the divergence has no squares of pseudo vectors and contains only a single squared vector.} This restriction was analyzed in detail by Son and Sur\'owka \cite{Son:2009tf} who found that it leads to 
\begin{equation}\begin{split}\label{constraints2}
P^\mu_\alpha P^\nu_\beta T_{diss}^{\alpha \beta}  - \frac{P^{\mu\nu}}{3} 
P_{\alpha \beta} T^{\alpha \beta}_{diss} & = -\eta \sigma^{\mu\nu}\\
P_\alpha^\mu \left( J^\alpha_{diss} + \frac{q}{\rho + P} (u_\nu T_{diss}^{\alpha\nu}) \right) 
&= \kappa V^\mu + \tilde{\kappa}_{\omega} \omega^\mu 
+\tilde{\kappa}_B B^\mu\\
\frac{(T_{diss})_{ab} P^{ab}}{3} - \frac{\partial P}{\partial\rho} (u_\mu u_\nu T_{diss}^{\mu\nu}) +\frac{\partial P}{\partial q} (u_\mu j_{diss}^\mu )& 
= -\beta \partial_\alpha u^\alpha 
\end{split}
\end{equation}
where 
\begin{equation}\label{res}\begin{split}
\sigma_{\omega}&=c \frac{\mu^3}{3T} + T \mu k_2 + T^2 k_1\\
\sigma_B&=c \frac{\mu^2}{2T} + \frac{T}{2} k_2\\
\tilde{\kappa}_{\omega}&=c \left( \mu^2 -\frac{2}{3} \frac{q}{\rho +P} \mu^3 \right)  + T^2 \left(1-\frac{2q}{\rho +P} \mu \right) k_2 - \frac{2q}{\rho +P} k_1   \\
\tilde{\kappa}_{B}&=c\left( \mu -\frac{1}{2}\frac{q}{\rho_n+P} \mu^2 \right) - \frac{T^2}{2}\frac{q}{\rho+P} k_2 \\
\end{split}
\end{equation}
and $k_1$ and $k_2$ are integration constants. We will now argue that
the requirement of CPT invariance forces $k_2$ to vanish. \footnote{
We thank D. Son for pointing this out to us.} The argument goes as follows. 
Consider the CPT transformation $x^\mu \rightarrow -x^\mu$ $q \rightarrow 
-q$ (and so  $\mu \rightarrow -\mu$). Under this transformation 
$T^{\mu\nu}_{diss} \rightarrow T^{\mu\nu}_{diss}$ and 
$J^\mu_{diss} \rightarrow -J^\mu_{diss}$. Also $u^\mu \rightarrow u^\mu$ so 
that $\omega^\mu \rightarrow -\omega^\mu$ and $B^\mu \rightarrow B^\mu$. 
Thus under a CPT transformation it must be that 
$\tilde{\kappa}_\omega \rightarrow \tilde{\kappa}_\omega$ while 
$\tilde{\kappa_B} \rightarrow - \tilde{\kappa_B}$. Consistency of 
this requirement with \eqref{res} sets $k_2=0$. Nothing in our 
argument requires that $k_1$ vanish (although it would be interesting to 
find a specific system with $k_1 \neq 0$; $k_1$ vanishes 
in all AdS/CFT computations performed so far).  

The results \eqref{constraints2} and \eqref{res} have several interesting features. First, the presence of an anomaly 
forces the entropy current to depart from the canonical form (i.e. $\sigma_B$ and 
$\sigma_{\omega}$ are never zero if $c$ is nonzero). Second, it induces new terms in the 
vector part of the constitutive relations, proportional to the vorticity and the magnetic field. 
Third, the new contributions to both the entropy current and the vector part of the constitutive 
relations are completely determined (up to an integration constant that is
 independent of $T$ and $\mu$) by the 
anomaly. In other words, although the constitutive relations take a different form from the parity 
even case, this change in form is completely determined by the anomaly, and we have no new free parameters apart from the integration constant $k_2$. 
%

%*****************************************************************
\section{Parity invariant Superfluid hydrodynamics}
\label{S:ParityInvariant}
%*****************************************************************
%
By definition, a superfluid is a fluid phase of a system with a spontaneously broken global symmetry. When discussing superfluids this forces us to consider the gradient of the Goldstone boson as an extra hydrodynamical degrees of freedom in addition to the standard variables $u^{\mu}$, $T$ and $\mu$. More precisely, if we denote the Goldstone Boson by $\psi$ ($\psi$ is the phase of the condensate of the charged scalar operator) and we also wish turn on a 
background gauge field $A_{\mu}$ then
\begin{equation}
\label{E:xidef}
	\xi_\mu= -\partial_\mu \psi + A_\mu
\end{equation}
represents the covariant derivative of the Goldstone Boson and is an extra hydrodynamic degree of freedom.\footnote{In \cite{Herzog:2008he,Herzog:2011ec} $\xi$ was defined with an opposite sign.} According to the Landau-Tisza two fluid model the superfluid should be thought of as a two component fluid: a condensed component and a non condensed or normal component. The velocity field of the normal fluid 
is given by $u_{\mu}$ and the velocity of the condensed phase is proportional to $\xi_{\mu}$. It is often convenient to define the component of $\xi$ orthogonal to $u$,
\begin{equation}
\label{E:zetadef}
	\zeta^{\mu} = P^{\mu\nu}\xi_{\nu}\,.
\end{equation}

The equations of motion of the superfluid are given by 
\begin{equation} \label{backceq}\begin{split}
\partial_\mu T^{\mu\nu}&=F^{\nu \mu} J_\mu\\
\partial_\mu J^\mu &= c E_\mu B^\mu\\
\partial_\mu \xi_\nu-\partial_\nu \xi_\mu &= F_{\mu\nu}\\
\end{split}
\end{equation}
together with the constitutive relations
\begin{equation} \label{ltconsts}
\begin{split}
T^{\mu\nu}&=(\rho+P) u^\mu u^\nu + P \eta^{\mu\nu} + 
f \xi^\mu \xi^\nu + T_{diss}^{\mu\nu} \\
J^\mu &=q u^\mu - f{\xi}^{\mu} + J^\mu_{diss}\\
u \cd \xi & = \mu + \mu_{diss}\,
\end{split}
\end{equation}
where we have chosen to work in an arbitrary `fluid frame' (see \cite{Bhattacharya:2011ee} for an explanation
of this terminology and  a fuller introduction to dissipative superfluid dynamics).

As was the case for the theory of charged fluids which we described in the previous section, superfluids also allow for a simple `canonical' entropy current \cite{Bhattacharya:2011ee}
\begin{equation}\label{entcur} 
\Jsc= s u^\mu - \frac{\mu}{T} J^\mu_{diss} - \frac{u_\nu T_{diss}^{\mu\nu}}{T}
\end{equation}
where $s$ is the thermodynamical entropy density of our fluid and is related to $\rho$ and $P$ through the Gibbs-Duhem relation
\begin{equation}
\label{E:gdrelation}
	\rho+ P = s T + \mu q
\end{equation}
and
\begin{equation}
	dP = s dT + q d\mu + \frac{1}{2}fd\xi^2
\end{equation}
where
\begin{equation}
\label{xidef}
	\xi = \sqrt{-\xi^{\mu}\xi_{\mu}}\,.
\end{equation}
It has been 
demonstrated in \cite{Bhattacharya:2011ee} that the entropy current \eqref{entcur} is invariant under 
field redefinitions. It was also shown in \cite{Bhattacharya:2011ee}  that 
the divergence of this entropy current is given by
\begin{equation}\label{diventcur}
\partial_\mu  J^\mu_s = -\partial_\mu \left(\frac{u_\nu}{T} \right)T_{diss}^{\mu\nu}  
-\left( \partial_\mu\left(\frac{\mu}{T} \right) - \frac{E_{\mu}}{T} \right)J_{diss}^{\mu} 
+\frac{\mu_{diss}}{T}\partial_\mu\left(f \xi^\mu\right)
\end{equation}

The rest of this section closely follows section \ref{cfm}. In \ref{S:SFdata} we list the independent first order data and second order scalar data, in section \ref{S:Js} we construct the most general positive divergence parity conserving entropy current consistent with Lorentz invariance. We find that up to a certain ambiguous term which is physically trivial, the entropy current agrees with its canonical form \eqref{entcur} and therefore the analysis of \cite{Bhattacharya:2011ee} follows. 

%====================================================================
\subsection{Onshell inequivalent First order independent data}
\label{S:SFdata}
%====================================================================
%
In the case of superfluid dynamics, the $SO(3,1)$ tangent space symmetry at any 
point is generically broken down to $SO(2)$ by the nonzero velocity fields $u^\mu$ and 
$\xi^\mu$. In the special case that $u^\mu$ and $\xi^\mu$ are 
collinear, $SO(2)$ is enhanced to $SO(3)$. This special case is physically interesting since it implies that the superfluid component is motionless relative to the normal component---once the superfluid velocity is too large superfluidity breaks down. We will find it convenient 
to decompose all first order fluid dynamical data into representations 
of $SO(2)$ and treat the collinear limit as a special point in parameter space.

Representations of $SO(2)$ are all one dimensional. We refer to 
fluid dynamical data that is invariant under $SO(2)$ as scalar data. All other
fluid data has charge $\pm m$ under $SO(2)$, where $m$ is an integer. There
is always as much $+ m$ as $-m$ data. We will find it useful to group 
together $+1$ and $-1$ charge data into a two column which we refer to as 
vector data; similarly we group $+2$ and $-2$ data together into tensor
data. 

Now consider a vector $\mathcal{V}^\mu$ whose $m=1$ and $m=-1$ components are 
$(a, b)$. The vector 
\begin{subequations}
\label{E:defstar}
\begin{equation}
	\tilde{\mathcal{V}}^\mu= \epsilon^{\mu\nu \alpha \beta} u_\nu \xi_\alpha 
\mathcal{V}_\beta \equiv \str \mathcal{V}_{\mu}
\label{E:defstarV}
\end{equation}
is a pseudo vector. Its components are proportional to $(a, -b)$. 
Thus, when considering representations of $SO(2)$, the same data can be packaged into either 
a vector or into a pseudo vector. The story is similar for all non-scalar 
representations. For instance, a traceless symmetric tensor $\mathcal{T}_{\mu\nu}$, 
whose $m=2$ and $m=-2$ components are $(a,b)$ is simply related to 
a traceless symmetric pseudo tensor 
\begin{equation}
	\tilde{\mathcal{T}}_{\mu\nu}=\epsilon_{\mu \alpha \beta \gamma} u^\alpha \xi^\beta 
\mathcal{T}^{\gamma \nu} \equiv \str \mathcal{T}_{\mu\nu}
\end{equation}
\end{subequations}
with components proportional to $(a, -b)$. Hence all tensor data can be packaged into
pseudo tensors. 

We now turn to a listing of the one derivative fluid dynamical and field 
data for superfluids. In Table 
\ref{tb:PESoneD}
we explicitly list 
all one derivative data, one derivative equations of motion, and then 
eventually independent one derivative data. 
The scalar $\xi$ used in this table is given by  
\eqref{xidef}.
We do not list pseudo vectors 
and pseudo tensors independently from vectors and tensors as they are 
isomorphic and contain the same data. 
In Table \ref{tb:fdsv} we assign labels to our independent data. In the same table we also present a second 
listing of a basis for independent scalar data which will be more convenient at places.
In Appendix \ref{linind} we 
demonstrate that both sets of seven scalars and the seven vectors listed are 
independent data, i.e. that we can solve for all other scalars and all other vectors 
in terms of the chosen basis.

{\small
\TABLE{
\renewcommand\arraystretch{1.1}
\caption{One derivative data for superfluids. The first column lists all quantities formed from the 
action of a single derivative on fluid and background fields. 
The second column lists all one derivative equations of motion. The last columns lists a choice 
of independent data. The tensors $\sigma^{\xi}_{\mu\nu}$ and $\sigma^{u}_{\mu\nu}$ are defined in \eqref{E:sigmaxidef}-\eqref{E:sigmaudef}. We also used $e^{\alpha\beta} = \epsilon^{\mu\nu\alpha\beta}u_{\mu}\xi_{\nu}$. }
\label{tb:PESoneD}
\begin{tabular}{| c | c c | r @{$=$} l | p{2.87 cm}|}
\hline        
 	Classification & \multicolumn{2}{|c|}{All  data} & \multicolumn{2}{|c|}{Equations of motion} & Independent  data\\
\hline
\hline
	 \multirow{7}{*}{Scalars (set 1)}
	 &$  \partial_{\mu} u^{\mu}$ &  $\partial_{\mu} (T \xi^{\mu})$ 
	 &\multicolumn{2}{|c|}{} 
	 &$  \tilde P^{\mu \nu} \partial_{\mu} u_{\nu}$ \\
	&$\xi^{\mu} \xi^{\nu} \partial_{\mu} u_{\nu}$&$\xi^{\mu}  \partial_{\mu} \left(\frac{\xi}{T} \right)$
	&\multicolumn{2}{|c|}{}
	&$  \tilde P^{\mu \nu} \partial_{\mu} (T \xi_{\nu})$ \\
	&$\xi^{\mu} \partial_{\mu} \left(\frac{\mu}{T}\right)$&$\xi^{\mu} \partial_{\mu} T$
	&$\partial_{\mu} J^{\mu}$&$c E^{\mu} B_{\mu}$
	&$ \xi^{\mu} \xi^{\nu} \partial_{\mu} u_{\nu}$ \\
	&$E \cd \xi$& $u^{\mu} \partial_{\mu} \left(\frac{\mu}{T}\right)$
	&$\xi^{\mu}u^{\nu} \left( \partial_{\mu} \xi_{\nu}-\partial_{\nu} \xi_{\mu}\right)$&$\xi \cd E$
	&$\xi^{\mu} \partial_{\mu} \left(\frac{\mu}{T}\right)$ \\
	&$u^{\mu} \partial_{\mu} T$&$u^{\mu}  \partial_{\mu} \left(\frac{\xi}{T} \right)$
	&$\xi_{\mu} \partial_{\nu} T^{\mu \nu} $&$ \xi^{\mu} F_{\mu \nu} J^{\nu}$
	&$ \xi^{\mu} \partial_{\mu} \left(\frac{\mu}{T}\right)$ \\
	&$\xi^{\mu}u^{\nu}\partial_{\nu} u_{\nu}$&
	&$u_{\mu} \partial_{\nu} T^{\mu \nu} $&$ -E_{\mu} J^{\mu}$
	&$ \xi^{\mu}\partial_{\mu} T$ \\
	&&
	&\multicolumn{2}{|c|}{}
	&$ E \cd \xi$ \\
\hline
	 \multirow{7}{*}{Scalars (set 2)}
	 &$  \partial_{\mu} u^{\mu}$ &  $\partial_{\mu} (T \xi^{\mu})$ 
	 &\multicolumn{2}{|c|}{} 
	 &$  \tilde P^{\mu \nu} \partial_{\mu} u_{\nu} $\\
	&$\xi^{\mu} \xi^{\nu} \partial_{\mu} u_{\nu}$&$\xi^{\mu}  \partial_{\mu} \left(\frac{\xi}{T} \right)$
	&\multicolumn{2}{|c|}{} 
	&$  \tilde P^{\mu \nu} \partial_{\mu} (T \xi_{\nu})$ \\
	&$\xi^{\mu} \partial_{\mu} \left(\frac{\mu}{T}\right)$&$\xi^{\mu} \partial_{\mu} T$
	&$\partial_{\mu} J^{\mu}$&$c E^{\mu} B_{\mu}$
	&$ u^{\mu} \xi^{\nu} \partial_{\mu} u_{\nu}$ \\
	&$E \cd \xi$& $u^{\mu} \partial_{\mu} \left(\frac{\mu}{T}\right)$
	&$\xi^{\mu}u^{\nu} \left( \partial_{\mu} \xi_{\nu}-\partial_{\nu} \xi_{\mu}\right)$&$\xi \cd E$
	&$u^{\mu} \partial_{\mu} \left(\frac{\mu}{T}\right)$ \\
	&$u^{\mu} \partial_{\mu} T$&$u^{\mu}  \partial_{\mu} \left(\frac{\xi}{T} \right)$
	&$\xi_{\mu} \partial_{\nu} T^{\mu \nu}$&$\xi^{\mu} F_{\mu \nu} J^{\nu}$
	&$ u^{\mu} \partial_{\mu} \left(\frac{\mu}{T}\right)$ \\
	&$\xi^{\mu}u^{\nu}\partial_{\nu} u_{\nu}$&
	&$u_{\mu} \partial_{\nu} T^{\mu \nu}$&$-E_{\mu} J^{\mu}$
	&$ u^{\mu}\partial_{\mu} T$ \\
	&&
	&\multicolumn{2}{|c|}{} 
	&$ E \cd \xi$ \\
\hline
	\multirow{3}{*}{Pseudo scalars}
	& \multicolumn{2}{|c|}{$\omega \cd \xi $}
	& \multicolumn{2}{|c|}{}  
	& $  \omega \cd \xi $ \\
	&  \multicolumn{2}{|c|}{$B \cd \xi $}
	& $e^{\alpha \beta} (\partial_{\alpha} \xi_{\beta}
- \partial_{\beta} \xi_\alpha) $&$ e^{\alpha \beta}F_{\alpha \beta}$
	& $B \cd \xi $\\
	&  \multicolumn{2}{|c|}{$\epsilon^{\mu \nu \alpha \beta} u_{\mu} \xi_{\nu} \partial_{\alpha} \xi_{\beta}$}
	& \multicolumn{2}{|c|}{}  
	&  \\
\hline
	\multirow{7}{*}{Vectors} 
	&&
	&\multicolumn{2}{|c|}{} 
	& $ \tilde P^{\mu \nu} u^{\rho} \partial_{\rho} u_{\nu}$\\
	&$ \tilde P^{\mu \nu} \xi^{\rho} \partial_{\rho} u_{\nu}$&$\tilde P^{\mu \nu} \xi^{\rho} \partial_{\rho} \xi_{\nu}$
	&$\tilde P^{\mu \nu} \partial_{\beta} T^{\beta}_{~\nu} $&$ \tilde P^{\mu \nu} F_{\nu \beta} J^{\beta} $
	&$\tilde P^{\mu \nu} u^{\rho} \partial_{\rho} \xi_{\nu}$ \\
	&$\tilde P^{\mu \nu}  E_\nu$ & $ \tilde P ^{\mu \nu} F_{\nu \beta} \xi^{\beta}$
	&$\tilde P^{\alpha \mu}u^{\nu}\left( \partial_{\mu} \xi_{\nu}-\partial_{\nu} \xi_{\mu}\right) $&$ \tilde P^{\alpha \mu}E_{\mu}$
	&$ \tilde P^{\mu \nu} \xi^{\rho} \partial_{\rho} u_{\nu}$ \\
	&$ \tilde P^{\mu \nu} \partial_{\nu} \frac{\mu}{T}$ & $\tilde P^{\mu \nu} \partial_{\nu} \frac{\xi}{T}$
	&$\tilde P^{\alpha \mu}\xi^{\nu}\left( \partial_{\mu} \xi_{\nu}-\partial_{\nu} \xi_{\mu}\right) $&$ \tilde P^{\alpha \mu}F_{\mu \nu} \xi^{\nu}$
	& $\tilde P^{\mu \nu} \xi^{\rho} \partial_{\rho} \xi_{\nu}$\\
	&$ \tilde P^{\mu \nu} \partial_{\nu} T $ & $\tilde P^{\mu \nu} \xi^{\alpha} \partial_{\nu}u_{\alpha}$
	& \multicolumn{2}{|c|}{} 
	& $\tilde P^{\mu \nu}  E_\nu$\\
	&$  \tilde P^{\mu \nu} u^{\rho} \partial_{\rho} u_{\nu}$ & $\tilde P^{\mu \nu} u^{\rho} \partial_{\rho} \xi_{\nu}$
	&\multicolumn{2}{|c|}{} 
	& $ \tilde P ^{\mu \nu} F_{\nu \beta} \xi^{\beta}$\\
	&&
	&\multicolumn{2}{|c|}{} 
	& $\tilde P^{\mu \nu} \partial_{\nu} \frac{\mu}{T}$\\
\hline
	\multirow{2}{*}{Tensors}
	& \multicolumn{2}{|c|}{$\sigma^u_{\mu\nu}$}
	& \multicolumn{2}{|c|}{--}  
	& $\sigma^u_{\mu\nu} $ \\
	&  \multicolumn{2}{|c|}{$\sigma^{\xi}_{\mu\nu} $}
	& \multicolumn{2}{|c|}{--}  
	&$\sigma^{\xi}_{\mu\nu} $\\
\hline
\end{tabular}
}
}
\TABLE{
\renewcommand\arraystretch{1.1}
\caption{Labels for the two sets of independent one derivative scalars  and one set of independent vectors.}
\label{tb:fdsv}
\begin{tabular}{|c|c|c|c|c|c|c|c|}
\hline        
i  & 1 & 2 & 3 & 4 & 5 & 6 & 7\\
\hline
	$\mathcal{S}^a_i$ &  
	$  \tilde P^{\mu \nu} \partial_{\mu} u_{\nu}$ & 
	$ \tilde P^{\mu \nu} \partial_{\mu} (T \xi_{\nu})$ &
	$\xi^{\mu} \xi^{\nu} \partial_{\mu} u_{\nu}$ & 
	$\xi^{\mu}  \partial_{\mu} \left(\frac{\xi}{T} \right)$ & 
	$\xi^{\mu} \partial_{\mu} \left(\frac{\mu}{T}\right)$& 
	$\xi^{\mu} \partial_{\mu} T$ & 
	$E. \xi$\\ 
\hline
	$\mathcal{S}^b_i$ & 
	$ \tilde P^{\mu \nu} \partial_{\mu} u_{\nu}$ & 
	$ \tilde P^{\mu \nu} \partial_{\mu} (T \xi_{\nu})$ &
	$u^{\mu} \xi^{\nu} \partial_{\mu} u_{\nu}$ &  
	$ u^{\mu}  \partial_{\mu} \left(\frac{\xi}{T} \right)$ &
	$u^{\mu} \partial_{\mu} \left(\frac{\mu}{T}\right)$ & 
	$ u^{\mu} \partial_{\mu} T$ &
	$E. \xi$ \\ 
\hline
	$\mathcal{V}^a_i$ & 
	$  \tilde P^{\mu \nu} u^{\rho} \partial_{\rho} u_{\nu}$ & 
	$ \tilde P^{\mu \nu} u^{\rho} \partial_{\rho} \xi_{\nu}$ & 
	$ \tilde P^{\mu \nu} \xi^{\rho} \partial_{\rho} u_{\nu}$ &
	$\tilde P^{\mu \nu} \xi^{\rho} \partial_{\rho} \xi_{\nu}$ &
	$\tilde P^{\mu \nu} \partial_{\nu} \frac{\mu}{T}$ & 
	$ \tilde P^{\mu \nu} \partial_{\nu} E_\mu$ &
	$ \tilde P^{\mu \nu} F_{\nu \beta} \xi^{\beta}$\\ 
\hline
\end{tabular}
}

As can be seen from Tables \ref{tb:PESoneD} and  \ref{tb:fdsv}, after imposing the equations of motion  
we have six  first order scalars and one first order pseudo scalar built 
out of fluid data, one first order scalar and one first order pseudo scalar  
built out of background field strengths, five first order vectors 
built out of fluid data, two first order vectors built 
from background fields and two independent tensors. 
The first tensor is simply the usual shear tensor $\sigma^{\mu\nu}$ projected orthogonal to the plane 
formed by the two fluid velocities. 
\begin{equation}
\label{E:sigmaudef}
	\sigma^u_{\mu\nu} =  \tilde P^{\mu \alpha} \tilde P^{\nu \beta} 
\left( \sigma_{\alpha\beta}  - \frac{1}{2} 
\eta_{\alpha\beta} \tilde P_{\gamma \delta} \sigma^{\gamma\delta} \right)\,.
\end{equation}
The second tensor $\sigma^\xi_{\mu\nu}$ is defined by 
\begin{equation}
\label{E:sigmaxidef}
	\sigma^\xi_{\mu\nu} =\half \tilde P^{\mu \alpha} \tilde P^{\nu \beta} 
\left( \partial_{\alpha} \xi_{\beta} + \partial_{\beta} \xi_{\alpha} - 
\tilde P_{\alpha \beta} \tilde{P}^{\gamma\delta}\partial_{\gamma}\xi_{\delta}\right)\,.
\end{equation} 
The counting of data in the absence of background fields agrees with \cite{Bhattacharya:2011ee}.

As was the case for normal fluids,  the divergence of the first order superfluid entropy current is a sum over 
quadratic one derivative terms and two derivative pieces of data.  In order to assist the analysis of the 
positivity of the divergence of the entropy current we list 
all the scalar two derivative data, the two derivative equations of motion and a basis for onshell independent 
two derivative scalars in Table \ref{tb:std}. Note that we have nine independent pieces of two derivative fluid dynamical 
data (as reported in \cite{Bhattacharya:2011ee} ) together with four additional pieces 
of two derivative data from background field strengths. In Appendix \ref{linind} demonstrate that the scalars listed in the last column of table \ref{tb:std} form a basis of onshell independent scalars.

{\small
\TABLE{
\renewcommand\arraystretch{1.1}
\caption{Two derivative scalar data. The first row gives all two derivative scalar data, the second row lists all the equations of motion. The third row represents a particular choice of independent second order data. }
\label{tb:std}
\begin{tabular}{|p{2 cm}| l l l l |}
\hline  
\multirow{6}{*}{All data}
	& $ \tilde P^{\mu \nu} u^{\rho} \partial_{\rho} \partial_{\mu} u_{\nu}$ 
	& $\tilde P^{\mu \nu} \xi^{\rho} \partial_{\rho} \partial_{\mu} u_{\mu}$ 
	& $ \tilde P^{\mu \nu}  u^{\rho} \partial_{\rho} \partial_{\mu} (T \xi_{\nu})$
	& $\tilde P^{\mu \nu} \xi^{\rho} \partial_{\rho} \partial_{\mu} (T \xi_{\nu})$\\
	&$ \xi^{\mu} \xi^{\nu} u^{\rho} \partial_{\rho} \partial_{\mu} u_{\nu}$
	&$u^{\mu} \xi^{\rho} \partial_{\rho} \partial_{\mu} \left( \frac{\xi}{T} \right)$
	&$u^{\mu} \xi^{\rho} \partial_{\rho} \partial_{\mu}\left( \frac{\mu}{T}\right)$
	&$u^{\mu}\xi^{\rho} \partial_{\rho}  \partial_{\mu}T$\\
	&$\tilde P^{\mu \nu} \partial_{\mu}  \partial_{\nu}\left( \frac{\mu}{T}\right)$
	&$\tilde P^{\mu \nu} \partial_{\mu}  \partial_{\nu}\left( \frac{\xi}{T}\right)$ 
	&$\tilde P^{\mu \nu} \partial_{\mu}  \partial_{\nu}T$
	&$u^{\mu} u^{\rho} \partial_{\rho} \partial_{\mu} \left( \frac{\xi}{T} \right)$\\
	& $\xi^{\mu} \xi^{\rho} \partial_{\rho} \partial_{\mu} \left( \frac{\xi}{T} \right)$
	&$\xi^{\mu} \xi^{\rho} \partial_{\rho} \partial_{\mu} \left( \frac{\mu}{T} \right)$
	&$\xi^{\mu} \xi^{\rho} \partial_{\rho} \partial_{\mu} T$
	&$u^{\mu} u^{\rho} \partial_{\rho} \partial_{\mu} \left( \frac{\mu}{T} \right)$\\
	&$u^{\mu} u^{\rho} \partial_{\rho} \partial_{\mu} T$
	&$\xi^{\mu} \xi^{\nu} \xi^{\rho} \partial_{\rho} \partial_{\mu} u_{\nu}$ 
	&$u^{\mu} \xi^{\nu} u^{\rho} \partial_{\rho} \partial_{\mu} u_{\nu}$
	&$\xi^{\nu} \tilde P^{\rho \mu}\partial_{\rho} \partial_{\mu} u_{\nu}$\\
	& $u^{\mu}\xi^{\nu}\partial_{\mu}E_{\nu}$
	&$\xi^{\mu}\xi^{\nu}\partial_{\mu}E_{\nu}$
	&$\tilde P^{\mu \nu}\partial_{\mu}E_{\nu}$
	&$\epsilon^{\mu \nu \lambda \sigma} \xi_{\mu}u_{\lambda}\partial_{\mu}B_{\sigma}$\\
\hline
\multirow{6}{2 cm}{Equations of motion}
	& \multicolumn{2}{l}{$u^{\beta}\partial_{\beta}\left( \partial_{\mu} 
J^{\mu} \right)=u^{\beta}\partial_{\beta}\left( c E^{\mu} B_{\mu}\right)$}
	& \multicolumn{2}{l |}{$\xi^{\beta}\partial_{\beta}\left( \partial_{\mu} 
J^{\mu} \right)=\xi^{\beta}\partial_{\beta}\left( c E^{\mu} B_{\mu}\right)$}\\
	&  \multicolumn{2}{l}{$u^{\beta}\partial_{\beta}\left(\xi_{\mu} \partial_{\nu} T^{\mu \nu} \right)
= u^{\beta}\partial_{\beta}\left(\xi^{\mu} F_{\mu \nu} J^{\nu}\right)$}
	& \multicolumn{2}{l |}{$\xi^{\beta}\partial_{\beta}\left(\xi_{\mu} \partial_{\nu} T^{\mu \nu} \right)
= \xi^{\beta}\partial_{\beta}\left(\xi^{\mu} F_{\mu \nu} J^{\nu}\right)$}\\
	&  \multicolumn{2}{l}{$u^{\beta}\partial_{\beta}\left( u_{\mu} \partial_{\nu} T^{\mu \nu} \right)
= u^{\beta}\partial_{\beta}\left(-E_{\mu} J^{\mu}\right)$}
	&  \multicolumn{2}{l |}{$\xi^{\beta}\partial_{\beta}\left( u_{\mu} \partial_{\nu} T^{\mu \nu} \right)
= \xi^{\beta}\partial_{\beta}\left(-E_{\mu} J^{\mu}\right)$}\\
	&  \multicolumn{2}{l}{$u^{\beta}\partial_{\beta}\left(\xi^{\mu}u^{\nu} \left( \partial_{\mu}
 \xi_{\nu}-\partial_{\nu} \xi_{\mu}\right) \right)=u^{\beta}\partial_{\beta}\left(\xi^{\mu} E_{\mu} \right)$}
	& \multicolumn{2}{l |}{$\xi^{\beta}\partial_{\beta}\left(\xi^{\mu}u^{\nu} \left( \partial_{\mu}
 \xi_{\nu}-\partial_{\nu} \xi_{\mu}\right) \right)=\xi^{\beta}\partial_{\beta}\left(\xi^{\mu} E_{\mu} \right)$}\\
	& \multicolumn{2}{l}{$\partial_{\alpha}\left(\tilde P^{\alpha \mu}u^{\nu}\left( \partial_{\mu}
 \xi_{\nu}-\partial_{\nu} \xi_{\mu}\right) \right)= \partial_{\alpha}\left(\tilde P^{\alpha \mu}E_{\mu} \right)$}
	& \multicolumn{2}{l |}{$\partial_{\mu}\left(\tilde P^{\mu \nu} \partial_{\beta} T^{\beta}_{~\nu} \right)
= \partial_{\mu}\left(\tilde P^{\mu \nu} F_{\nu \beta} J^{\beta} \right)$}\\
	&  \multicolumn{2}{l}{$\partial_{\alpha}\left(\tilde P^{\alpha \mu}\xi^{\nu}\left( \partial_{\mu}
 \xi_{\nu}-\partial_{\nu} \xi_{\mu}\right) \right)= \partial_{\alpha}\left( \tilde P^{\alpha \mu}F_{\mu \nu}
\xi^{\nu}\right)$}
	&&\\
	\hline
	\multirow{3}{2 cm}{Independent data}
	&$ \tilde P^{\mu \nu} u^{\rho} \partial_{\rho} \partial_{\mu} u_{\nu}$
	&$\tilde P^{\mu \nu} \xi^{\rho} \partial_{\rho} \partial_{\mu} u_{\mu}$ 
	&$ \tilde P^{\mu \nu}  u^{\rho} \partial_{\rho} \partial_{\mu} (T \xi_{\nu})$
	&$  \tilde P^{\mu \nu} \xi^{\rho} \partial_{\rho} \partial_{\mu} (T \xi_{\nu})$\\
	& $ \xi^{\mu} \xi^{\nu} u^{\rho} \partial_{\rho} \partial_{\mu} u_{\nu}$
	& $u^{\mu} \xi^{\rho} \partial_{\rho} \partial_{\mu} \left( \frac{\xi}{T} \right)$
	& $u^{\mu} \xi^{\rho} \partial_{\rho} \partial_{\mu}\left( \frac{\mu}{T}\right)$
	& $ u^{\mu}\xi^{\rho} \partial_{\rho}  \partial_{\mu}T$\\
	& $\tilde P^{\mu \nu} \partial_{\mu}  \partial_{\nu}\left( \frac{\mu}{T}\right)$
	& $u^{\mu}\xi^{\nu}\partial_{\mu}E_{\nu}$
	&$\xi^{\mu}\xi^{\nu}\partial_{\mu}E_{\nu}$
	& $\tilde P^{\mu \nu}\partial_{\mu}E_{\nu}$ \\
	&$ \tilde P^{\mu \nu} \left( \partial_{\mu} F_{\nu \beta}\right) \xi^{\beta}$ & & & \\
\hline
\end{tabular}
}
}
%
%====================================================================
\subsection{Constructing the entropy current}
\label{S:Js}
%====================================================================
%
With the independent data at hand we proceed with our analysis. 
The most general entropy current allowed by symmetries takes the form  
\begin{equation}\label{encurform}
 J^{\mu}_{S} = \Jsc + u^{\mu} \sum_{i=1}^7  s^a_i \mathcal{S}^a_i 
+ \xi^{\mu} \sum_{i=1}^7  s^b_i  \mathcal{S}^b_i + \sum_{i=1}^7 v_i \mathcal{V}^{a\,\mu}_i,
\end{equation}
where the coefficients $s^a_i$, $s^b_i$, and 
$v_i$ are, at the moment, arbitrary functions of  $\frac{\mu}{T}$, 
$T$, and $\xi^2$. Note that we have chosen to expand the terms proportional 
to $u^\mu$ in the basis $\mathcal{S}^a_i$ while terms proportional to $\xi^\mu$ 
are expanded in the basis $\mathcal{S}^b_i$. This choice will prove
algebraically convenient below. In total we start with twenty one free parameters in the
entropy current. 

The two derivative terms in the divergence of the entropy current  \eqref{encurform}
are given by
\begin{equation}\label{2decdiv}
\begin{split}
  \partial_\mu J^{\mu}_{S} =& \left( s_1^a + v_1\right) \tilde P^{\mu \nu} u^{\rho} \partial_{\rho} \partial_{\mu} u_{\nu}
+ \left( s_2^a + v_2\right) \tilde P^{\mu \nu}  u^{\rho} \partial_{\rho} \partial_{\mu} (T \xi_{\nu})
+\left( s_1^b + v_3\right) \tilde P^{\mu \nu} \xi^{\rho} \partial_{\rho} \partial_{\mu} u_{\nu}
\\&+ \left( s_2^b + v_4\right)\tilde P^{\mu \nu} \xi^{\rho} \partial_{\rho} \partial_{\mu} (T \xi_{\nu})
+\left( s_3^a + s_3^b\right)\xi^{\mu} \xi^{\nu} u^{\rho} 
\partial_{\rho} \partial_{\mu} u_{\nu}
\\&+\left( s_3^a + s_3^b\right) u^{\mu} \xi^{\rho} \partial_{\rho} \partial_{\mu} 
\left( \frac{\xi}{T} \right)
+\left( s_3^a + s_3^b\right) u^{\mu} \xi^{\rho} \partial_{\rho} \partial_{\mu}\left( \frac{\mu}{T}\right)
\\&+\left( s_3^a + s_3^b\right) u^{\mu}\xi^{\rho} \partial_{\rho}  \partial_{\mu}T
+ v_5 ~\tilde P^{\mu \nu} \partial_{\mu}  \partial_{\nu}\left( \frac{\mu}{T}\right)
+s_7^a ~u^{\mu}\xi^{\nu}\partial_{\mu}E_{\nu} 
\\&+s_7^b ~\xi^{\mu}\xi^{\nu}\partial_{\mu}E_{\nu}
+v_6 ~\tilde P^{\mu \nu}\partial_{\mu}E_{\nu}
+v_7 ~\tilde P^{\mu \nu} \left( \partial_{\mu} F_{\nu \beta}\right) \xi^{\beta} + \dots\,.
\end{split}
\end{equation}
Following the algorithm of the previous section, we first set the coefficient
of each of the thirteen independent two derivative terms listed in Table 
\ref{tb:std}, which appear in the divergence of the entropy current, to zero. The vanishing of the nine fluid dynamical two derivative 
terms yields the following nine relations between the $s^a$'s $s^b$'s and $v$'s
\begin{align} \label{fr} 
	 v_1 &=-s^a_1\, &
	 v_2 &=-s^a_2\, &
	 v_3 &=-s^b_1\,&
	 v_4 &=-s^b_2\,&
	s^b_3 &= -s^a_3\, \\
\notag
	s^b_4 &= -s^a_4\,&
	s^b_5 &= -s^a_5\,&
	s^b_6 &= -s^a_6\, &
	v_5 &=0\,. &
	&&
\end{align} 
The vanishing of the four electromagnetic field related two derivative scalars yields 
the additional four relations
\begin{equation}\label{sr}
s^a_7=s^b_7=v_6=v_7=0\,.
\end{equation}

Apart from the two derivative fluid dynamical and background 
electromagnetic field data, there are four nontrivial curvature invariants
one can form out of the contractions of $u^\mu, \xi^\mu$ and $g^{\mu\nu}
$ with the the Reimann tensor $R_{\alpha \beta\mu\nu}$.\footnote{We omit the curvature scalar $R$ in 
this listing since it is a pure gravitational term and therefore never appears in the divergence of 
fluid dynamical entropy current.} After plugging in the constraints in \eqref{fr} and \eqref{sr}
into the expression for the entropy current \eqref{encurform} we find 
\begin{equation}\label{E:cterms}
 \begin{split}
  \partial_\mu J^{\mu}_{S} =& s_1^a \tilde P^{\alpha \beta} u^{\mu} u^{\lambda} R_{\lambda \beta \alpha \mu}
+ s_2^b \tilde P^{\alpha \beta} \xi^{\mu} \xi^{\lambda} R_{\lambda \beta \alpha \mu}
\\&+ \left( T s_2^a + s_1^b \right) \xi^{\mu} u^{\lambda} R_{\lambda \beta \alpha \mu}
+ s_3^a u^{\alpha} \xi^{\beta} u^{\gamma} \xi^{\delta} R_{\alpha \beta \gamma \delta} + \dots.
 \end{split}
\end{equation}
Each of the terms in \eqref{E:cterms} is of indefinite sign. Thus, the coefficients of these four terms must vanish. This implies
\begin{equation} \label{tr} 
	s^a_1 = 0\,,\quad
	s_1^b = -T s_2^a\,,\quad
	s^b_2 =0\,,\quad
	s_3 =0.
\end{equation}

To summarize, by setting the two derivative and curvature terms that appear in the divergence of the entropy current to zero we have eliminated $9+4+4=17$ of the original 21 coefficients and are left with an entropy current with four undetermined coefficients,
\begin{equation}\label{fulentcur}
\begin{split}
 J^{\mu}_S &= \Jsc
+ s_2^a \left(
u^{\mu} \tilde P^{\alpha \beta} \partial_{\alpha} ( T \xi_{\beta}) 
- \tilde P^{\mu \beta} u^{\nu} \partial_{\nu} (T\xi_{\beta}) - 
 T \xi^{\mu} \tilde P^{\alpha \beta} \partial_{\alpha} u_{\beta} 
+ T \tilde P^{\mu \beta} \xi^{\nu} \partial_{\nu} u_{\beta} \right) \\&
+ s_4^a \left( u^{\mu} \xi^{\nu}  \partial_{\nu} \frac{\xi}{T} 
- \xi^{\mu} u^{\nu}  \partial_{\nu} \frac{\xi}{T}\right)
+ s_5^a \left( u^{\mu} \xi^{\nu} \partial_{\nu} \frac{\mu}{T} - \xi^{\mu} u^{\nu} \partial_{\nu} \frac{\mu}{T} \right)
+ s_6^a \left( u^{\mu} \xi^{\nu} \partial_{\nu} T - \xi^{\mu} u^{\mu} \partial_{\nu} T \right).
\end{split}
\end{equation}

The entropy current \eqref{fulentcur} can be rewritten in a simpler form by introducing the antisymmetric tensor
\begin{equation}\label{qdef}
{\cal Q}_{\mu \nu} = T(\xi_{\mu} u_{\nu}-\xi_{\nu}u_{\mu})\,,
\end{equation} 
introducing a unified notation for the three thermodynamical scalar fields
$$ 
	\Sigma_{i} = \left\{ \frac{\mu}{T}, \frac{\xi}{T},T \right\} \qquad i=1,2,3\,,
$$
and also redefining our coefficient functions 
$$
	 c_0=s_2^a\,\quad
	 c_1=s_5 -\frac{T \mu}{\mu^2 - \xi^2}\,\quad
	 c_2=s_4 + \frac{T \xi}{\mu^2 - \xi^2}\,\quad
	 c_3=s_6 -\frac{2}{T}.
$$ 
Then \eqref{fulentcur} takes the form 
\begin{equation}\label{fulentcursimp}
\begin{split}
 J^{\mu}_{S} &= \Jsc 
+ c_0 \partial_{\nu} {\cal Q}^{\nu \mu}
+ \sum_{i=1}^3 c_i {\cal Q}^{\mu \nu} \partial_{\nu} \Sigma_{i} \,.
\end{split}
\end{equation}
The one parameter subclass of this parameter set of entropy currents 
\begin{equation}
\label{E:cis}
	c_i = - \partial_{\Sigma_i} c_0 \,.
\end{equation}
is trivial as (inserting \eqref{E:cis} into \eqref{fulentcursimp}) it yields
\begin{equation}\label{fecf}
 J^{\mu}_{S} = \Jsc 
+ \partial_{\nu} \left( c_0  {\cal Q}^{\nu \mu} \right)\,
\end{equation}
i.e an entropy current whose divergence vanishes identically 
\footnote{The parameter $c_0$ is essentially trivial and is related to a pullback ambiguity as we explain in appendix \S\ref{pullamb}.}. 
The remaining three parameters are nontrivial, and in general lead to physical effects 
(see \cite{Bhattacharyya:2012xi} for a thorough analysis). In the current paper, 
however, we focus attention on superfluids that preserve invariance under 
time reversal, (or, equivalently, by the CPT theorem,  CP invariance). 
The terms multiplying $c_0$ and $c_i$ ($i= 1 \ldots 3$) are all odd under this 
symmetry (this is a consequence of the fact that $\xi_i$ is odd under time 
reversal) and so must vanish in a time reversal invariant theory. 
\footnote{The divergence of the entropy current \eqref{fulentcursimp} is given by 
\begin{equation}\label{fulentcurdiv}
\begin{split}
  \partial_{\mu} J^{\mu}_{S} =& - \partial_{\mu}\left( \frac{u_{\nu}}{T}
 \right) 
T_{diss}^{\mu \nu}+
V_{1\,\mu} J^{\mu}_{diss} 
+ \frac{\mu_{diss}}{T} \partial_{\mu}\left(f \xi_{\nu}\right)\\
&+ \left(\partial_{\Sigma_i} c_0\right) \left( \partial_{\mu} \Sigma_i\right) \partial_{\nu} {\cal Q}^{\nu \mu}
+c_i \partial_{\mu}{\cal Q}^{\mu \nu} \partial_{\nu} \Sigma_{i} 
+ \left(\partial_{\Sigma_j}c_i\right) \left( \partial_{\mu} \Sigma_j\right) {\cal Q}^{\mu \nu} 
\left( \partial_{\nu} \Sigma_{i} \right)\,.
\end{split}
\end{equation}
In an earlier version of this paper we had used \eqref{fulentcurdiv} together with the assumption that 
the six vectors
$$ 
\tilde P^{\alpha \nu} \xi^{\mu}\sigma_{\mu \nu}\,,\quad
\tilde P^{\alpha}_{~\mu} \partial_{\nu}{\cal Q}^{\mu \nu}\,,\quad
	\tilde P^{\alpha}_{\phantom{\alpha}\mu} V_1^\mu\,,\quad
	\tilde P^{\alpha \nu} \left( \partial_{\nu} \Sigma_i \right)\,\ (i=1,\ldots,3)\,.
$$
are linearly independent to conclude that positivity of the \eqref{fulentcurdiv} forces the 
entropy current to take the form \eqref{fecf}, independent of the assumption of time-reversal invariance. 
This result is incorrect as the six vectors above are not linearly independent as may be seen using the 
equation \eqref{step3}. We thank the authors of \cite{Bhattacharyya:2012xi} for pointing this out to us.}
It follows that the entropy current is forced to take the canonical form. 

It follows that the constraints on the dissipative terms $T_{diss}^{\mu\nu}$, $J_{diss}^{\mu}$ and $\mu_{diss}$ from demanding the 
positivity of the divergence of $J_S$ are identical 
to the constraints from the positivity of the divergence of the 
canonical entropy current $\Jsc$. Such an analysis has been carried out in \cite{Bhattacharya:2011ee} 
and we will be carried out in more generality in the next section. Note also that the fluid
dynamical analysis for the most general form of the superfluid constitutive relations presented in 
\cite{Bhattacharya:2011ee} applies only under the assumption that the superfluid preserves both 
parity as well as time-reversal invariance. The agreement between 
the fluid dynamical analysis and gravitational results in \cite{Bhattacharya:2011ee} is explained by the 
fact that the gravitation theory analyzed in this preserve both these symmetries.

%*************************************************************************
 \section{Superfluid dynamics without parity invariance}
\label{S:SuperfluidPodd}
%*************************************************************************
%
The main goal of this work is to determine the most general expression for the constitutive relations for a Lorentz invariant but parity breaking superfluid. We tackle this problem using the same algorithm which has been spelt out in the previous sections: we determine the most general entropy current which is compatible with the symmetries. The restrictions on the constitutive relations are then determined by demanding positivity of the entropy current.

%==========================================================
\subsection{Onshell inequivalent data}
%==========================================================
%
As we've discussed in section \ref{S:SFdata}, the listing of one derivative pseudo vectors and pseudo-tensors of $SO(2)$ can be constructed from the tensors of $SO(2)$ using the $\epsilon$ tensor. If $\mathcal{V}_{\mu}$ and $\mathcal{T}_{\mu\nu}$ are vectors and tensors of $SO(2)$ then we define the associated pseudo vectors $\tilde{\mathcal{V}}^{\mu}$ and pseudo-tensors $\tilde{\mathcal{T}}^{\mu\nu}$ through: 
\begin{align}
\begin{split}
\label{E:starop}
	\tilde{\mathcal{V}}^{\mu} &= \str \mathcal{V}_{\mu} 
= \epsilon^{\mu\nu\alpha\beta}u_{\nu} \xi_{\alpha} \mathcal{V}_{\beta} \\
	\tilde{\mathcal{T}}^{\mu\nu} &= \str \mathcal{T}_{\mu\nu} 
=  \epsilon^{\mu\rho\alpha\beta}u_{\rho}\xi_{\alpha} \mathcal{T}_{\beta}^{\phantom{\beta}\nu} +  \epsilon^{\nu\rho\alpha\beta}u_{\rho}\xi_{\alpha} \mathcal{T}_{\beta}^{\phantom{\beta}\mu}\,.
\end{split}
\end{align}
We introduce the following terminology:  $\str \mathcal{V}_{\mu}$ is said 
to be the star of $\mathcal{V}_\mu$, and similarly  $\str \mathcal{T}_{\mu\nu}$ 
is said to be the star of  $\mathcal{T}_{\mu\nu}$.

Note that $\str V_{\mu}$ is an $SO(2)$ pseudovector for any vector $V_{\mu}$ since the $\str$ operation projects onto the subspace orthogonal to 
both $u^{\mu}$ and $\zeta^{\mu}$ where $\zeta^{\mu}$ was defined in \eqref{E:zetadef},
$$
	\zeta^{\mu} = P^{\mu\nu} \xi_{\nu}\,.
$$
In Table \ref{tb:fdpv} we list the independent one derivative 
pseudo-scalars, vectors and tensors which we will use in this section. It 
proves convenient to choose our on-shell independent scalars $S_i^c$ 
to differ from both the choices made  in section \ref{S:ParityInvariant}. 
The linear independence of this choice of scalars is demonstrated in 
Appendix \ref{calc}.
Our choice of on-shell independent vectors is also  
different from the one is made in Table \ref{tb:fdsv} of section \ref{S:ParityInvariant}. We have used the superscript $c$ for the pseudo vectors 
and vectors to flag this difference. We have demonstrated the linear 
independence of this choice of vectors in Appendix \ref{calc}. 

{\tiny
\TABLE{
\renewcommand\arraystretch{2.2}
\caption{Parity odd independent one derivative data which we use in this section. The $\str$ operation that takes us from vectors $\mathcal{V}$ and tensors $\mathcal{T}$ to pseudovectors $\tilde{\mathcal{V}}$ and pseudotensors, $\tilde{\mathcal{T}}$ is given in \eqref{E:starop}. We list a basis for pseudovectors, and a new basis for scalars and vectors which we will work with in this section. Note that $\tilde{\mathcal{V}}^c_i = * \mathcal{V}^c_i$.}
\label{tb:fdpv}
\begin{tabular}{| c | c | c | c | c | c | c | c |}
\hline
	$i$&1&2&3&4&5&6&7\\
\hline
	$\tilde{{\cal V}}_i^{c\,\mu}$&
	$*V_{1\,\mu}$&
%	$*\left(\frac{E_{\sigma}}{T} - \partial_{\sigma} \frac{\mu}{T}\right)$&
	$*\left(\zeta^{\alpha}\sigma_{\alpha\beta}\right)$&
	$\str \nabla_\sigma T$&
	$\str \nabla_\sigma \left(\frac{\mu}{T}\right)$&
	$\str \nabla_\sigma\left(\frac{ \zeta^2}{T^2}\right)$&
	$\tilde P^{\mu \nu} \omega_{\nu}$&
	$\tilde P^\mu_\sigma B^\sigma$\\	
\hline
	${{\cal V}}_i^{c\,\mu}$&
	$\tilde{P}^{\mu\sigma}V_{1\,\sigma}$&
%	$\tilde{P}^{\mu\sigma} \left(\frac{E_{\sigma}}{T} - \partial_{\sigma} \frac{\mu}{T}\right)$&
	$\tilde{P}^{\mu\beta}\left(\zeta^{\alpha}\sigma_{\alpha\beta}\right)$&
	$\tilde{P}^{\mu\sigma} \nabla_\sigma T$&
	$\tilde{P}^{\mu\sigma} \nabla_\sigma \left(\frac{\mu}{T}\right)$&
	$\tilde{P}^{\mu\sigma} \nabla_\sigma\left(\frac{ \zeta^2}{T^2}\right)$&
	$ \frac{\mathcal{V}_2^{c\,\mu} - \tilde{P}^{\mu\alpha}\zeta^{\nu}\partial_{\alpha} u_{\nu}}{ \zeta^{2}}$&
	$-\frac{P^{\mu\nu} F_{\nu\alpha}\zeta^{\alpha}} {\zeta^{2}}$\\	
\hline
	$\mathcal{S}^c$&
	$\zeta_{\mu}V_1^{\mu}$&
%	$\zeta_{\mu}P^{\mu\sigma} \left(\frac{E_{\sigma}}{T} - \partial_{\sigma} \frac{\mu}{T}\right)$ &
	$u \cd \partial T$&
	$u \cd \partial \frac{\mu}{T}$&
	$u \cd \partial \frac{\zeta^2}{T^2} $&
	$\zeta^{\mu} P^{\mu\alpha} \partial_{\alpha}T$& 
	$\zeta^{\mu}E_{\mu}$ &
	$\zeta^{\mu} P^{\mu\alpha}\partial_{\alpha} \frac{\zeta^2}{T^2}$ \\
\hline
	$\tilde{\mathcal{T}}_{i\,\mu\nu} $&
	$\str \sigma^u_{\mu\nu}$ &
	$\str \sigma^{\xi}_{\mu\nu}$ &
	-- & -- & -- & -- & -- \\
\hline
	$\tilde{\mathcal{S}}_i$ &
	$\omega \cd \xi$ &
	$B \cd \xi$ &
	-- & -- & -- & -- & -- \\
\hline	
\end{tabular}
}
}

In table \ref{tb:potd} we list all two derivative pseudo scalars. The format we use is identical to that of previous 
sections. We list all possible two derivative pseudo scalars that 
can be constructed out of fluid and field data in the left column. 
The equations of motion appear in the second column. 
For technical reasons we find it convenient to formally treat $E_\mu$ and $B_\mu$ as independent fields in the full list of two derivative data (the left column in table \ref{tb:potd}) and to impose the Bianchi identity (which relates some 
of the derivatives of these fields to others) at the same stage that we impose the equations of motion. In other words, in our listings below 
we will pretend that $E_\mu$ and $B_\mu$ are independent fields constrained by 
``equations of motion'' which are given by the Bianchi identities. In the last column we list a basis of independent two derivative pseudo scalars.
In Appendix \ref{calc} we have demonstrated that the two derivative pseudo 
scalars presented in Table \ref{tb:potd} are independent. 

\TABLE{
\renewcommand\arraystretch{1.1}
\caption{Parity odd two derivative scalar data. The first column lists all possible independent two derivative scalars with $B^{\mu}$ and $E^{\mu}$ treated as independent, the second column lists the equations of motion (first three rows) and Bianchi identities (last two rows) and the last column the independent data. We have used the shorthand notation $e^{\lambda\sigma} = \epsilon^{\mu\nu\lambda\sigma}u_{\mu}\zeta_{\nu}$ and $C_{\mu\nu} \equiv \nabla_\mu \xi_\nu - \nabla_\nu \xi_\mu - F_{\mu\nu}$.}
\label{tb:potd}
\begin{tabular}{|c|r|c|}
\hline
All data & equations of motion & Two derivative data \\
\hline
	$\zeta^\mu u^\nu \nabla_\nu B_\mu$ &
	& 
	\\
	$\zeta^\mu \zeta^\nu \nabla_\nu B_\mu$ &
	$e^{\lambda\sigma}(\zeta.\nabla)C_{\lambda\sigma} = 0$&
	\\
	$\tilde P^{\mu\nu} \nabla_\nu B_\mu$ &
	$e^{\lambda\sigma}(\nabla_\theta T^\theta _\sigma - F_{\sigma\theta}j^\theta)=0$ &
	$e^{\lambda\sigma}\nabla_{\lambda}E_{\sigma}$\\
	$e^{\lambda\sigma}\nabla_\lambda E_\sigma $ &
	$e^{\lambda\sigma} u^\theta\nabla_\lambda C_{\theta\sigma}=0$&
	$e^{\lambda\sigma} (\zeta \cd \nabla) \nabla_{\lambda}u_{\sigma}$ \\
	$e^{\lambda\sigma}(u\cd\nabla)\nabla_\lambda u_\sigma $ &
	$\epsilon^{\mu\nu\lambda\sigma}\zeta_\mu \nabla_\nu F_{\lambda\sigma} = 0$&
	$\zeta^{\mu}\zeta^{\nu} \nabla_{\mu} B_{\nu}$\\
	$e^{\lambda\sigma}(\zeta\cd\nabla)\nabla_\lambda u_\sigma$ &
	$\epsilon^{\mu\nu\lambda\sigma}u_\mu \nabla_\nu F_{\lambda\sigma}=0 $&
	\\
	$e^{\lambda\sigma}(u\cd\nabla)\nabla_\lambda \zeta_\sigma$ &
	&
	\\
	$e^{\lambda\sigma} (\zeta\cd\nabla)\nabla_\lambda \zeta_\sigma $ &
	& 
	\\
\hline
\end{tabular}
}
%

%=================================================================================
\subsection{Constraints from vanishing of two derivative and curvature terms}
%=================================================================================
%
We denote the total entropy current by
\begin{equation}
	J_S^{\mu} = \Jsc + \Jsn + \partial_{\nu} \left(c_0 \mathcal{Q}^{\mu\nu}\right) \,,
\end{equation}
where $\Jsn$ is a parity odd contribution to the entropy current and $ \partial_{\nu} \left(c_0 \mathcal{Q}^{\mu\nu}\right)$ is a residual ambiguity in the definition of the entropy current in the parity preserving sector described in section \ref{S:ParityInvariant}.
The most general contribution to $\Jsn$ is given by 
\begin{equation}\label{parent}
 \Jsn = 
\sum_{i = 1}^7 \tilde v_i \tilde{{\cal V}}^{c\,\mu}_i+ u^{\mu} \sum_{i=1}^2 \tilde{s}_i^a \tilde{\mathcal{S}}_i + \zeta^{\mu} \sum_{i=1}^2 \tilde{s}_i^b \tilde{\mathcal{S}}_i\,.
\end{equation}
This entropy current has eleven new coefficients, each of which is an arbitrary
function of $T$, $\mu$ and $\xi$. In what follows we will swap the variables $\mu$, and $\xi$ with
\begin{equation}
	\nu = \frac{\mu}{T} 
	\qquad \hbox{and} \qquad
	\chi = \frac{\zeta^2}{T^2}\,.
\end{equation}
As in previous sections, the eleven parameter 
set of currents is significantly constrained by the requirement that two 
derivative and curvature terms in its divergence vanish. 
It will prove convenient to rewrite the entropy current in the form
\begin{multline}\label{gencuri}
\Jsn = \epsilon^{\mu\nu\rho\sigma} \partial_{\nu} \left( \sigma_1 T
u_{\rho} \zeta_{\sigma} \right)
+ \sthr \tilde{\mathcal{V}}^{c\,\mu}_3
+ T \sfou \tilde{\mathcal{V}}_4^{c\,\mu}
+ T \sfiv \tilde{\mathcal{V}}_5^{c\,\mu}\\
+ \frac{\sigma_8}{2} \epsilon^{\mu\nu\rho\sigma} \xi_{\nu} F_{\rho\sigma}
+T^2 \snin \omega^{\mu}
+T \sten B^{\mu} \\
+\alpha_1 \tilde{\mathcal{V}}^{c\,\mu}_1 + \alpha_2 \tilde{\mathcal{V}}^{c\,\mu}_2
+\zeta^\mu\left[\alpha_3 (\omega\cd\zeta) + \alpha_4 (B\cd\zeta)\right]
\end{multline}
The coefficient functions $\sigma_i$ in \eqref{gencur} are linearly related 
to the coefficient functions in \eqref{parent}. Although we will never need 
these relations, for the sake of completeness we present them in 
Appendix \ref{calc}. 

The first term in \eqref{gencur} is the divergent-less combination 
\begin{equation}
\label{E:Totald}
	\frac{1}{2}\epsilon^{\mu\nu\lambda\sigma}\nabla_{\nu} \left(\sone T u_{\lambda} \zeta_{\sigma}\right)
\end{equation}
with $\sone$ an undetermined function of $T, \nu$ and $\chi$. 
This expression contributes to the entropy current but does not contribute 
to its divergence. We've observed a similar expression in the parity even 
sector of the entropy current in equation \eqref{fecf}. As was the case in the previous 
section, this term is physically irrelevant (it does not contribute in any 
way to the constraints on dissipative terms) and can be ignored in what 
follows.

We now proceed to compute the correction to the entropy current 
\eqref{gencuri}. It is not difficult to verify that the two derivative 
and curvature contribution to the divergence of the entropy current is 
given by 
\begin{equation}\label{tdcoef}
\begin{split}
\nabla_{\mu} J_{new}^\mu &= \alpha_1
\left[\epsilon^{\mu\nu\lambda\sigma}u_\mu\zeta_\nu
\right]\nabla_\lambda E_\sigma +
\frac{(\alpha_2 + \zeta \alpha_3)}{2} \left[\epsilon^{\mu\nu\lambda\sigma}u_\mu\zeta_\nu
\right](\zeta.\nabla)\nabla_\lambda u_\sigma
\\&+\alpha_4 \zeta^\mu\zeta^\nu \nabla_\mu B_\nu
+ \frac{\alpha_2}{2} ~\epsilon^{\mu\nu\lambda\sigma}\zeta_\mu u_\nu u^\rho
\zeta^\theta\left[ R_{\rho\sigma\lambda\theta} - \frac{R_{\rho\theta\sigma\lambda}}{2}\right]
+\dots
\end{split}
\end{equation}
where the dots denote the derivative squared terms. 
The expression on the right hand side of \eqref{tdcoef} is 
a sum over a set of linearly independent two derivative data and curvature 
terms (see table \ref{tb:potd}). Positivity of the divergence 
of the entropy current requires that the  coefficients of the two derivative data and 
curvature terms vanish. Thus, 
\begin{equation}
\label{E:PoddResults}
	\alpha_1=\alpha_2=\alpha_3=\alpha_4 =0.
\end{equation}
Had we not demanded the vanishing of curvature terms above, 
we would have erroneously concluded that $\alpha_2$ is non zero. 
Inserting \eqref{E:PoddResults} into \eqref{gencuri} we conclude that the most 
general parity odd addition to the entropy current whose divergence has no 
two derivative data or curvature terms is given by 
\begin{multline}\label{gencur}
	\Jsn = \epsilon^{\mu\nu\rho\sigma} \partial_{\nu} \left( \sigma_1 T u_{\rho} \zeta_{\sigma} \right)
		+ \sthr \tilde{\mathcal{V}}^{c\,\mu}_3 
		+ T \sfou \tilde{\mathcal{V}}_4^{c\,\mu}
		+ T \sfiv \tilde{\mathcal{V}}_5^{c\,\mu} 
		\\
		+ \frac{\sigma_8}{2} \epsilon^{\mu\nu\rho\sigma} \xi_{\nu} F_{\rho\sigma}
		+T^2 \snin \omega^{\mu} 
		+T \sten B^{\mu} \,.
\end{multline}
Expression \eqref{gencur} is the final result of this subsection.

Note that positivity of the divergence of the entropy current implies a correction to the entropy density of the form
\begin{equation}
\label{E:correction}
	-u_{\mu}J_S^{\mu} = s + \seig \tilde{\mathcal{S}}_2 + \sone T \tilde{\mathcal{S}}_1 +\partial_{\mu} \left(c_0 T \zeta^{\mu}\right) - c_0 T \zeta^{\mu} u\cd \partial u_{\mu}\,.
\end{equation}
Equation \eqref{E:correction} implies that in a parity violating system the entropy density, as defined via the Gibbs Duhem relation (i.e., the log of the partition function in equilibrium), would receive gradient corrections if $c_0$, $\sigma_1$ or $\sigma_8$ are non zero.
%
%======================================================================
\subsection{Constraints from positivity of the quadratic form}
%======================================================================
%
The divergence of the sum of the canonical entropy current and 
the the new piece \eqref{gencur} has no two derivative 
or curvature terms, and so is a quadratic form in one derivative data. 
In this subsection we will determine the corresponding quadratic form. The requirement that this quadratic form is positive will enable us to deduce the restrictions on the $\sigma_i$'s in \eqref{gencur} and on the parameters describing the equations of motion of the theory.
 
From \eqref{diventcur} and taking into account the anomaly, we find that
\begin{equation}
	\partial_{\mu} J_S^{\mu} = -\partial_{\mu} \left(\frac{u_{\nu}}{T}\right)T_{diss}^{\mu\nu} + \left(\frac{E_\mu}{T} -\partial_\mu \left( \frac{\mu}{T} \right)\right) J_{diss}^{\mu} + \frac{\mu_{diss}}{T}\partial_{\mu}\left(f \xi^{\mu}\right) + \partial_{\mu} \Jsn - c \nu E \cd B\,.
\end{equation}
On general grounds the divergence of the non canonical part of the entropy 
current, together with the contribution from the anomaly, takes the form:\footnote{This follows from the observation that the final result is a pseudo 
scalar. The absence of a tensor times pseudo tensor term in this expression 
is not completely obvious but may be checked.}
\begin{equation}
\label{E:dJsn}
	\partial_{\mu} \Jsn- c \nu E \cd B = \sum_{i=1}^7 \sum_{j=1}^7\tilde{\mathcal{V}}^c_i\cd B_{ij}^v {\mathcal{V}}^c_j + \sum_{i=1}^{7}\sum_{j=1}^2 \tilde{\mathcal{S}}^c_i B^s_{ij}{\mathcal{S}}_j\,.
\end{equation}
Recall that $\tilde{\mathcal{V}}^{c\,\mu} = *\mathcal{V}^c_{\mu}$ so that $B_{ij}^v$ is an antisymmetric $ 7 \times 7$ matrix. 

The expressions for $B_{ij}^v$ and $B^s_{ij}$ are straightforward but 
rather tedious to compute. We provide some details relevant to the 
computation in Appendix \ref{calc}. We will not completely list 
these matrices here, but will unveil relevant aspects of their 
 explicit form as we go along.

To complete our analysis, we need to re-express the divergence of the 
canonical entropy current \eqref{diventcur} in terms of on-shell 
independent data. We carry out this computation in a frame invariant manner, 
similar to that of section \ref{S:NormalPEven}. 

We find it convenient to decompose the 
explicit fluid first derivative data that appears in \eqref{diventcur}, namely 
$\nabla_{\mu} \left(\frac{u_{\nu}}{T}\right)$  and $E_{\mu}/T - \partial_{\mu} \nu$ 
into $SO(2)$ invariant scalars, vectors, and tensors as follows
\begin{align}
\begin{split}
\label{E:dudecomposition}
	\nabla_{\mu} \left(\frac{u_{\nu}}{T}\right) =& 
	\left(\frac{1}{3T} \nabla \cd  u - \frac{1}{2\zeta^2 T} \zeta \cd \sigma \cd \zeta \right) \tilde{P}_{\mu\nu}
	+\frac{1}{\zeta^2}\left(\frac{1}{3 T} \nabla \cd u + \frac{1}{\zeta^2 T} \zeta \cd \sigma \cd \zeta\right)
 \zeta_{\mu}\zeta_{\nu} \\
	&+ \frac{u \cd \partial T}{T^2} u_{\mu}u_{\nu} 
	-\frac{1}{\zeta^2 T} \left(\zeta^{\alpha} u\cd\nabla u_{\alpha} 
+ \frac{\zeta^{\alpha}\partial_{\alpha}T}{T}\right) u_{(\mu}\zeta_{\nu)} \\
	&-\frac{1}{T} \left( u \cd \nabla u_{\alpha} + \frac{\partial_{\alpha}T}{T}\right)
 u_{(\mu}\tilde{P}_{\nu)}^{\phantom{\nu}\alpha} 
	+\frac{2}{\zeta^2 T} \zeta_{(\mu}\mathcal{V}_{2\,\nu)} 
	+\frac{1}{T}\mathcal{T}_{1\,\mu\nu}\\
	\frac{E_{\mu}}{T} - \partial_{\mu}\nu & = \mathcal{V}_{1\,\mu} + u_{\mu} u \cd \partial \nu  + 
\zeta_{\mu} \frac{\zeta \cd V_1}{\zeta^2}\,.
\end{split}
\end{align}
We will find it convenient to organize the 7 scalars and three vectors 
above into two columns $\mathbf{s}_d$ and $\mathbf{v}_d$,
\begin{equation} \label{cd}
\mathbf{s}_d=\begin{pmatrix}
		\frac{1}{3T} \nabla \cd  u - \frac{1}{2\zeta^2 T} \zeta \cd \sigma \cd \zeta \\
		\frac{1}{3 T} \nabla \cd u + \frac{1}{\zeta^2 T} \zeta \cd \sigma \cd \zeta \\
		\frac{u \cd \partial T}{T^2} \\
		-\frac{1}{\zeta^2 T} \left(\zeta^{\alpha} u\cd\partial u_{\alpha} 
+ \frac{\zeta^{\alpha}\partial_{\alpha}T}{T}\right) \\
		- u \cd \nabla \nu \\
		-\frac{\zeta \cd V_1}{\zeta^2} \\
		\frac{\partial_\mu (f \xi^\mu) }{T}
	\end{pmatrix}
~~~
\mathbf{v}_d=	\begin{pmatrix}
		-\frac{1}{T} \left( u \cd \nabla u_{\alpha} + \frac{\partial_{\alpha}T}{T}\right)
 \tilde{P}^{\alpha}_{\phantom{\alpha}\nu} \\
		\frac{2}{\zeta^2 T} \mathcal{V}^c_{2\,\nu} \\
		-\mathcal{V}^c_{1\,\nu}
	\end{pmatrix}\,.
\end{equation}

The pieces of scalar and vector data in \eqref{cd} multiply particular 
scalar and vector components of $T^{\mu\nu}_{diss}$ and $J^\mu_{diss}$. 
It is convenient to classify the scalar, vector and tensor combinations 
as 
\begin{align}\label{pio}
	\mathbf{s}_1 &= T_{diss}^{\mu\nu} \tilde{P}_{\mu\nu} & 
	\zeta^2 \mathbf{s}_2 &= \zeta \cd T_{diss} \cd \zeta \\ \notag
	\mathbf{s}_3 &= u \cd T_{diss} \cd u & 
	\mathbf{s}_4 &= u \cd T_{diss} \cd \zeta \\ \notag
	\mathbf{s}_5 &= u \cd J_{diss} & 
	\mathbf{s}_6 & = \zeta \cd J_{diss} \\ \notag
	\mathbf{s}_7 & = -\mu_{diss} && \\ \notag
	\mathbf{v}_1^{\nu} & = u_{\mu} T_{diss}^{\mu\alpha} \tilde{P}^{\nu}_{\phantom{\nu} \alpha} & 
	\mathbf{v}_2^{\nu} & = \zeta_{\mu} T_{diss}^{\mu\alpha} \tilde{P}^{\nu}_{\phantom{\nu} \alpha} \\ \notag
	\mathbf{v}_3^{\nu} & = \tilde{P}^{\nu}_{\phantom{\nu}\alpha}J^{\alpha} & & \\
\notag
	\mathbf{t}  &= \tilde{P}_{\mu}^{\phantom{\mu}{\alpha}}\tilde{P}_{\nu}^{\phantom{\mu}{\beta}}T_{diss\,\alpha\beta} - \frac{1}{2}\tilde{P}^{\mu\nu} \tilde{P}^{\alpha\beta}T_{diss,\alpha\beta}\,,&&
\end{align}
and to group these into row vectors 
\begin{equation} \label{sd} \begin{split}
\mathbf{s}& = \begin{pmatrix}
		\mathbf{s}_1 & \mathbf{s}_2 & \mathbf{s}_3 & 
\mathbf{s}_4 & \mathbf{s}_5 & \mathbf{s}_6 & \mathbf{s}_7 \\
              \end{pmatrix} \\
\mathbf{v}&=\begin{pmatrix} \mathbf{v}_1 & \mathbf{v}_2 & \mathbf{v}_3 
\end{pmatrix}\,.
\end{split}
\end{equation}
These definitions permit a very simple formal 
expression for the divergence of the canonical entropy current:
\begin{equation}
\label{Jhh}
	\partial_{\mu}\Jsc = -\mathbf{s} ~\mathbf{s_d} - \mathbf{v}_\mu~
\mathbf{v_d}^\mu - \frac{1}{T}\mathbf{t}_{\mu\nu} \mathcal{T}^{\mu\nu}_1 \,.
\end{equation}
In presenting \eqref{Jhh} we have used matrix notation: the expressions on the right hand side of 
\eqref{Jhh} are each the product of a row and a column.

Equation  \eqref{Jhh} is not our final expression for the divergence of 
the canonical entropy current for an important reason; the 
entries in the columns $\mathbf{s}_d$ 
and $\mathbf{v}_d$ are not independent on-shell. In fact, on-shell,
each of the 7 $\mathbf{s}_d$ is a linear combination 
of only four independent scalar terms. Similarly, on-shell, each of the three entries in $\mathbf{v}_d$
is a linear sum over two on-shell vectors. Specifically
\begin{equation}
\mathbf{s}_d= 
		A^s \mathcal{S} ,
	\qquad
	\mathbf{v}_d
	=
	A^v \mathcal{V} 
\end{equation}		
where
\begin{equation} \label{svdef} 
\mathcal{S}=\begin{pmatrix}
		\mathcal{S}_1^c \\
		\mathcal{S}_2^c \\
		\mathcal{S}_3^c \\
		\mathcal{S}_4^c
	\end{pmatrix} \,
~~~
\mathcal{V}=	\begin{pmatrix}
		\mathcal{V}^c_{1\,\nu} \\
		\mathcal{V}^c_{2\,\nu}
	\end{pmatrix}
\end{equation}
and 
\begin{equation}\label{adef}
	A^{s}=	\begin{pmatrix}
			\frac{R s}{2 q_n T \chi}\quad  & \quad\frac{B_3}{3 T}-\frac{A_3}{2 T \chi}\quad  &
 \quad\frac{B_2}{3 T} - \frac{A_2}{2 T \chi}\quad  & \quad\frac{B_1}{3 T} - \frac{A_1}{2 T \chi} \\
			-\frac{R s }{q_n \chi T}\quad  & 	\quad\frac{B_3}{3 T} + \frac{A_3}{T \chi}\quad  &
 \quad\frac{B_2}{3 T} + \frac{A_2}{T \chi}\quad  & \quad \frac{B_1}{3 T} + \frac{A_1}{T \chi}  \\
			0 & \frac{1}{T^2} & 0 & 0 \\
			-\frac{R}{T^2 \chi} & \frac{K_3}{T} & \frac{K_2}{T} & \frac{K_1}{T} \\
			0 & 0 & -1 & 0 \\
			-\frac{1}{T^2 \chi} & 0 & 0 & 0  \\
			0 & \frac{(\rho +P)K_3}{T} & \frac{(\rho +P)K_2}{T} & \frac{(\rho+P)K_1}{T} \\
		\end{pmatrix}\,,
		\quad
	A^v = \begin{pmatrix}
		-R & 0 \\
		0 & \frac{2}{T^3 \chi} \\
		-1 & 0
		\end{pmatrix}\,
\end{equation}
with
\begin{equation}
\label{E:Rdef}
	R = \frac{q}{\rho+P}\,.
\end{equation}
Note that $\mathcal{S}$ is a column composed of 4 of the 7 on-shell independent
scalars listed in table \ref{tb:fdpv}, while $\mathcal{V}$ is a 
column composed of 2 of the 6 on-shell independent vectors listed in the 
same table. Note also that $A^s$ is a $7 \times 4$ matrix while $A^v$ is a $3 \times 2$ matrix.
The computation of the entries of the matrix $A^s$ and $A^v$ is outlined 
in Appendix \ref{calc} where we also provide expressions for $A_i$, $B_i$ and $K_i$.
It follows that when re-expressed in terms of on-shell independent data, the 
divergence of the canonical part of the entropy current takes the form
\begin{equation}
\label{E:dJsc}
	\partial_{\mu}\Jsc = - \mathbf{s} A^s \mathcal{S} - 
\mathbf{v}_\mu A^v \mathcal{V}^\mu - \frac{1}{T}\mathbf{t}_{\mu\nu} \mathcal{T}^{\mu\nu}_1 \,.
\end{equation}

Equation \eqref{E:dJsc} depends only on those 4 linear 
combinations of the 7 scalars $\mathbf{s}_i$  
that appear in the 4 columns of $\mathbf{s} A^s $. 
Similarly \eqref{E:dJsc} depends on only 
the 2 linear combinations of the three vectors $\mathbf{v}_i$
 that appear in the 2 columns of 
$\mathbf{v} A^v $. These numbers could have been anticipated on general grounds. 
As in the 
case of ordinary fluid dynamics (see section \ref{cfm}) 
the equations of superfluid dynamics suffer from a field redefinition 
ambiguity. The field redefinitions in question are $u^\mu$, $T$ 
and $\mu$, and may be decomposed into 3 $SO(2)$ scalars and one $SO(2)$ 
vector. All of the seven scalars $\mathbf{s}_i$ 
described above transform under the scalar field redefinitions. As these redefinitions 
depend on three parameters, $7-3=4$ independent linear combinations of the 
scalars are field redefinition invariant. Similarly we expect to find 
$3-1=2$ field redefinition independent linear combinations of the three 
vectors  $\mathbf{v}_i$. 
Since the canonical entropy current 
is field redefinition invariant, the expression for its divergence 
is necessarily also frame invariant. It follows that 
$\mathbf{s}_i A_{ij}^s$ and $\mathbf{v}_i A_{ij}^v$ must be  
linear combinations of the four frame invariant scalars and the two frame invariant vectors 
respectively. We have explicitly checked 
(see Appendix \ref{calc}) that the the four linear combinations 
$\mathbf{s}A^s $ and the 2 linear combinations $\mathbf{v}A^v $ 
are indeed field redefinition invariant.

Putting together \eqref{E:dJsn} and \eqref{E:dJsc}, positivity of the entropy current implies that
\begin{subequations}
\label{E:allsectors}
\begin{align}
\label{E:tsector}
	- \frac{1}{T}\mathbf{t}_{\mu\nu} \mathcal{T}^{\mu\nu}_1 &\geq 0 \\
\label{E:vsector}
	- \mathbf{v}^{\mu}_i A^v_{ij} \mathcal{V}^c_j + \tilde{\mathcal{V}}^c_i \cd B_{ij}^v {\mathcal{V}}^c_j &\geq 0 \\
	-\mathbf{s}_i A^s_{ij} \mathcal{S}^c_j +	\tilde{\mathcal{S}}^c_i B_{ij}^s {\mathcal{S}}^c_j &\geq 0\,.
\label{E:ssector}
\end{align}
\end{subequations}

In the rest of this section we will deduce the constraints imposed by 
\eqref{E:allsectors} on the one derivative constitutive relations for the tensor 
$\mathbf{t}_{\mu\nu}$, the frame invariant vectors $\mathbf{v}_\mu A^v$ 
and the frame invariant scalars $\mathbf{s} A^s$. 
%

%----------------------------------------------------------
\subsubsection{Positivity in the Tensor Sector}
%----------------------------------------------------------
%
The most general one derivative constitutive 
relation for the frame invariant tensor $\mathbf{t}$ (on symmetry 
grounds) is given by 
\begin{equation} \label{tres}
	\mathbf{t}_{\mu\nu}  
=  \tilde{P}_{\mu}^{\phantom{\mu}{\alpha}}\tilde{P}_{\nu}^{\phantom{\mu}{\beta}}T_{diss\,\alpha\beta} 
- \frac{1}{2}\tilde{P}^{\mu\nu} \tilde{P}^{\alpha\beta}T_{diss,\alpha\beta} = -\eta_i \mathcal{T}_{i\,\mu\nu} - \tilde{\eta}_i \tilde{\mathcal{T}}_{i\,\mu\nu}
\end{equation}
where $\mathcal{T}_1 = \sigma^u$, $\mathcal{T}_2 = \sigma^\xi$ and $\tilde{\mathcal{T}}_i = \str \mathcal{T}_i$ and the $\eta$'s and $\tilde\eta$'s are four possible coefficients.
Positivity of the entropy current requires that
\begin{equation}\label{tresb}
	\eta_1 > 0 \qquad \tilde{\eta}_1 \in \mathbf{R} \qquad \eta_2 = \tilde{\eta}_2 =  0\,.
\end{equation}
In other words, the tensor part of $T^{\mu\nu}_{diss}$ has no term
proportional to $\mathcal{T}_2$ and to $\str\mathcal{T}_2$. 
The latter result could have been anticipated.
The divergence of the entropy current does not have a piece 
proportional to $\mathcal{T}_2$. Therefore terms in the divergence of the entropy current which are linear 
in  $\mathcal{T}_2$ must vanish. On the other hand $\mathbf{t}_{\mu\nu}$ does, in general, have a 
term proportional to  $\mathcal{T}_1$; the coefficient of this term is 
the shear viscosity of the fluid and is constrained to be positive.  
Positivity also allows a term proportional to
$\str\mathcal{T}_1$. The coefficient of this term, ${\tilde \eta}_1$, 
is a new parity odd coefficient (it is of course an arbitrary 
function of the thermodynamical scalars).  Note that ${\tilde \eta}_1$ 
drops out of the formula for entropy production. This is because the 
contraction of any tensor with its own star vanishes. 
Consequently, ${\tilde \eta}_1$ is non dissipative; its coefficient is 
unconstrained by the requirement of positivity of entropy production. 
%

%----------------------------------------------------------
\subsubsection{Positivity in the Vector Sector}
%----------------------------------------------------------
%
The most general constitutive relations allowed by symmetries for 
the two frame invariant vectors takes the form 
\begin{equation}
\label{E:decompositionV}
	\mathbf{v}^{\mu}_i A^v_{ij}  = -\mathcal{V}_i \kappa_{ij} - \tilde{\mathcal{V}_i} \tilde{\kappa}_{ij}\,.
\end{equation}
The matrix of possible 
transport coefficients $\kappa$ is a $7 \times 2$ matrix. The same is true 
for $ \tilde{\kappa}$.
The first index runs over a basis of on-shell inequivalent vectors, while the 
second index runs over a basis of field redefinition invariant combinations 
of vectors in $T^{\mu \nu}_{diss}$ and $J^\mu_{diss}$. 

A useful observation is that the quadratic form in the vector sector contains 
no terms proportional to the square of $\mathcal{V}^c_i$ for $i =3 \ldots 7$. The first expression in \eqref{E:vsector} has an explicit factor of either 
$\mathcal{V}^c_1$ or $\mathcal{V}^c_2$. Since $B_{ij}^v$ is antisymmetric, the second expression in \eqref{E:vsector} has no terms proportional to the square of any of the on-shell independent vectors.
Positivity thus demands that the quadratic form in the vector sector 
be completely independent of  $\mathcal{V}^c_i$ for $i =3 \ldots 7$.
This implies that 
\begin{subequations}
\label{E:cases}
\begin{align}
\label{Case1}
	B_{ij}^v & = 0 				& i &=3,\ldots,7, \quad j=3,\ldots,7\\
\label{Case2}
	\tilde{\kappa}_{ij} &= -B^v_{ij}	& i &=3,\ldots,7, \quad j=1,2\\
\label{Case3}
	\kappa_{ij} &= 0 			& i &=3,\ldots,7, \quad j=1,2
\end{align}
\end{subequations}
We now proceed to investigate the implications of \eqref{E:cases}.

Let us first focus on \eqref{Case1}. 
A priori, this equality gives us a set of $5 \times 4/2=10$ 
partial differential equations that constrain the six independent $\sigma_i$ coefficients
in \eqref{gencur}. One of these equations is a tautology. 
The remaining nine equations are as follows 
\begin{alignat}{2}
\notag
	B^v_{34}=0 \Rightarrow\quad &\frac{\partial( \sigma_4-\sigma_8)}{\partial T}  
-\frac{1}{T}\frac{\partial \sigma_3}{\partial \nu}=0 \quad&
	B^v_{35}=0 \Rightarrow\quad &\frac{\partial \sigma_5}{\partial T}  
-\frac{1}{T}\frac{\partial \sigma_3}{\partial \chi}=0\\
\notag
	B^v_{45}=0 \Rightarrow\quad &\frac{\partial( \sigma_4-\sigma_8)}{\partial \chi}  
-\frac{\partial \sigma_5}{\partial \nu}=0 &
	B^v_{37}=0 \Rightarrow\quad &\frac{\partial (\sigma_{10}-\nu \sigma_8)}{\partial T}+ \frac{\sigma_3}{T} = 0\\
\notag
	B^v_{57}=0 \Rightarrow\quad &\frac{\partial (\sigma_{10}-\nu \sigma_8)}{\partial \chi}+ \sigma_5 = 0 &
	B^v_{46}=0 \Rightarrow \quad &\frac{\partial \sigma_9}{\partial \nu}-2\sigma_{10}+ 2\nu\sigma_4 =0\\
\notag
B^v_{36}=0 \Rightarrow\quad &\frac{\partial \sigma_9}{\partial T}+ 2\nu\frac{\sigma_3}{T} = 0 &
B^v_{56}=0 \Rightarrow\quad &\frac{\partial \sigma_9}{\partial \chi}+ 2\nu\sigma_5 = 0\\
B^v_{47}=0 \Rightarrow\quad &\frac{\partial (\sigma_{10}-\nu \sigma_8)}{\partial \nu}+ (\sigma_4-\sigma_8) =c \nu\,. &&
\label{setcond1a}
\end{alignat}
Equations \eqref{setcond1a} are far from independent.
They admit the following two parameter set of solutions 
\begin{equation}\label{arbsol}
\begin{split}
\sigma_3&= -T \frac{\partial (\sigma_{10}-\nu \sigma_8)}{\partial T}\\
\sigma_4&= \sigma_8 + c\nu -\frac{\partial (\sigma_{10}-\nu \sigma_8)}{\partial \nu}\\
\sigma_5&=-\frac{\partial (\sigma_{10}-\nu \sigma_8)}{\partial \chi}\\
\sigma_9&=s_9 - \frac{2}{3} c\nu^3 + 2\nu (\sigma_{10}-\nu \sigma_8)
\end{split}
\end{equation}
In \eqref{arbsol} we have chosen $\sigma_8$ and $\sigma_{10}$ as our free parameters. The term $s_9$ in the last line of \eqref{arbsol} is an integration
constant. It turns out that the requirement of CPT invariance forces 
$$s_9=0.$$ 
The argument for this conclusion is very similar to that given below 
\eqref{res}. While we present all formulae below at nonzero $s_9$, the 
reader should keep in mind that $s_9$ actually vanishes in any superfluid 
that enjoys invariance under CPT invariance, i.e. any superfluid that arises 
from a quantum field theory (and so presumably for any superfluid in the 
real world, or obtained via the AdS/CFT correspondence).

 Equation \eqref{arbsol} is the most general solution to \eqref{setcond1a}.
As an aside we note that 
\eqref{arbsol} also implies that $B_{ij}^v = 0$ for $i,j = 1,2$. We will use 
this information shortly. 

We now turn to the implications of \eqref{Case2}. In this case, the coefficients $\tilde{\kappa}$ whose second index lies between 
$3$ and $7$ are completely determined in terms of $B_{ij}$, which is, in 
turn, a function of the two free parameters $\sigma_8$ and $\sigma_{10}$. 
We find that $B_{i2}=0$ for $i = 3 \ldots 7$. Thus, 
$$
	\tilde{\kappa}_{i2}=0 \quad i=3,\ldots,7\,.
$$ 
However
$\tilde\kappa_{i1}$ are in general nonzero. They are given by 
\begin{align}
\begin{split}
\label{E:solv}
%	\kappa_{i2}& = \kappa_{i1} = 0\\
%	\tilde{\kappa}_{i2}& =-B^v_{i2}= 0 \quad i>2\\
	\tilde{\kappa}_{31} & = -B^v_{31}= -R T \sigma_3 - T \partial_T \sigma_8 \\
	\tilde{\kappa}_{41} & = -B^v_{41}=- R T^2 \sigma_4 - T \partial_\nu \sigma_8 \\
	\tilde{\kappa}_{51} &= -B^v_{51} = -R T^2 \sigma_5 - T \partial_{\chi} \sigma_8 \\
	\tilde{\kappa}_{61} & = -B^v_{61}= -2 R T^3 \sigma_9 +2 T^2 \sigma_{10} \\
	\tilde{\kappa}_{71} & = -B^v_{71}= -R T^2 \sigma_{10} + 2 T \sigma_8 + c T \nu\,.
\end{split}
\end{align}
To be clear we reiterate \eqref{Case3}
\begin{equation}
	\kappa_{ij} = 0 \quad i =3,\ldots,7, \quad j=1,2\,.
\end{equation}

Once we have implemented all these conditions, the quadratic form 
in the vector sector takes the form 
$\mathcal{V}^c_i \cd \kappa_{ij} \mathcal{V}^c_j+\tilde{\mathcal{V}}^c_i \cd \tilde\kappa_{ij} \mathcal{V}^c_j$ where $i,j=1,2$. Positivity
of the divergence of entropy current implies
\begin{equation} \label{diventcvec}
	\sum_{i=1,2}\sum_{j=1,2}\left( \mathcal{V}^c_i \cd \kappa_{ij} \mathcal{V}^c_j 
+  \tilde{\mathcal{V}}^c_i \cd \tilde\kappa_{ij} \mathcal{V}^c_j\right) \geq 0.
\end{equation}
The Onsager relations (see \cite{LLvol6}) imply that 
$\kappa_{12} = \kappa_{21}$ and $\tilde{\kappa}_{12} = \tilde{\kappa}_{21}$
\footnote{In an earlier version of this paper we had claimed the relation 
$\tilde{\kappa}_{12} =- \tilde{\kappa}_{21}$. Our sign error was pointed out in 
\cite{Neiman:2011mj}. Together with N. Banerjee, S. Jain and T. Sharma we have verified that
the sign claimed in \cite{Neiman:2011mj} is indeed correct by checking that the stress tensor and 
current two point functions obey the relations required by time reversal invariance
only when the Onsager relations are given by $\tilde{\kappa}_{12} = \tilde{\kappa}_{21}$.
We thank Y. Oz, N. Banerjee, S. Jain and T. Sharma for very useful discussions on this point.}. 
We are left with 6 independent transport coefficients
$\kappa_{11}$, $\kappa_{22}$, $\kappa_{12}=\kappa_{21}$, 
${\tilde \kappa}_{11}$, $ {\tilde \kappa}_{22}$ and ${\tilde \kappa}_{12}=
{\tilde \kappa}_{21}$, each of which is an arbitrary function of $\mu$, 
$T$ and $\xi$. Curiously, on imposing the condition 
${\tilde \kappa}_{12}={\tilde \kappa}_{21}$, the second term on the right hand side of \eqref{diventcvec}, 
vanishes. This removes any dependence on $\tilde \kappa_{ij}, \ j=1,2$ from 
the condition of positivity of the divergence of entropy current. Thus given the 
Onsager relations, only the coefficients 
$\kappa_{11}$, $\kappa_{22}$, $\kappa_{12}=\kappa_{21}$
are constrained by the inequality that the matrix 
\begin{equation}
	\mathcal{K} = \begin{pmatrix}
		\kappa_{11} & \kappa_{12} \\
		\kappa_{12}  & \kappa_{22}
		\end{pmatrix}
\end{equation}
is a positive semi-definite matrix. This condition simply reads
\begin{equation} \label{ineq}
	\kappa_{11}\kappa_{22} \geq \kappa_{12}^2.
\end{equation}
Note that the 3 $\kappa$ transport coefficients occur even in parity 
preserving superfluids, while the 3 ${\tilde \kappa}$ coefficients violate 
parity and are new. Note also that all of the three new parity violating 
transport coefficients, ${\tilde \kappa}_{11}$, ${\tilde \kappa}_{22}$, and ${\tilde \kappa}_{12}=
{\tilde \kappa}_{21}$ do not appear in the expression for entropy production, and so are non 
dissipative \footnote{In an earlier version of this paper we had reported  ${\tilde \kappa}_{12}$ to be dissipative
and our erroneous conclusion was based on the incorrect additional sign, that we had in the relation between ${\tilde \kappa}_{12}$ and 
${\tilde \kappa}_{21}$. We again thank the authors of \cite{Neiman:2011mj} for pointing this to us.}. 
%

%-----------------------------------------------------------------
\subsubsection{Positivity in the Scalar Sector}
%-----------------------------------------------------------------
%
The most general
expansion of the constitutive relations in the scalar 
sector takes the form 
\begin{equation} \label{sdis}
	\mathbf{s}_i A^s_{ij}  = -\sum_{i=1}^{7} \mathcal{S}^c_i \beta_{ij} -  \sum_{i=1}^2 \tilde{\mathcal{S}}_{i}\tilde{\beta}_{ij}\,,
\end{equation}
implying the inequality
\begin{equation}
\label{E:scalarinequality}
	\sum_{i=1}^{7}\sum_{j=1}^4 \mathcal{S}^c_i \beta_{ij}\mathcal{S}^c_j + \sum_{i=1}^2 \sum_{j=1}^4\tilde{\mathcal{S}}_{i}\tilde{\beta}_{ij}\mathcal{S}_j^c + \sum_{i=1}^2\sum_{j=1}^7 \tilde{\mathcal{S}}^c_i B_{ij}^s \mathcal{S}_j \geq 0\,.
\end{equation}
Terms in \eqref{E:scalarinequality} involving products of scalars and pseudo-scalars must vanish since they can not be positive definite for an arbitrary flow. 
Thus, focusing on the parity odd sector we must impose
\begin{equation}\label{scv}
\sum_{i=1}^2 \sum_{j=1}^4\tilde{\mathcal{S}}_{i}\tilde{\beta}_{ij}\mathcal{S}_j^c 
+ \sum_{i=1}^2\sum_{j=1}^7 \tilde{\mathcal{S}}^c_i B_{ij}^s \mathcal{S}_j^c = 0\,.
\end{equation}
Only the second piece on the right hand side of \eqref{scv} involves $\mathcal{S}_j^c$ 
with $j = 4 \ldots 7$. We immediately deduce the equation 
\begin{equation}
\label{E:Bs0}
	B^s_{ij}=0 \qquad  = 1 , 2\,,  \quad j=4, \ldots, 7.
\end{equation}
Equations \eqref{E:Bs0} are automatically satisfied 
once we impose the solution \eqref{arbsol}; we have no further restrictions 
on the free parameters $\sigma_8$ and $\sigma_{10}$. For $j =1, \ldots, 4$,
equation \eqref{scv} implies  
\begin{equation}\label{sdi}
	{\tilde \beta}_{ij}  = -B_{ij}^s\,.
\end{equation}
It follows that there are no undetermined 
 parity odd transport coefficients in the 
scalar sector; all parity odd contributions to frame invariant scalar
 combinations of $T^{\mu\nu}_{diss}$, $J^{\mu}_{diss}$ and $\mu_{diss}$ are 
completely determined in terms of $\sigma_{8}$ and $\sigma_{10}$ 
by \eqref{sdi} and the explicit listing 
{\small
\begin{equation}
	B^s_{ij} = \begin{pmatrix}
		\frac{2 R T \snin}{\chi} - \frac{2 \sten}{\chi} \quad&
		\quad -2\sthr - 2 T^2 K_3  \snin \quad&
		\quad -2 T \sfou -2 T^2  K_2 \snin \quad&
		\quad -2 T \sfiv - 2 K_1 T^2  \snin \\
		-\frac{c \nu}{T \chi} - \frac{2 \seig}{T \chi} + \frac{R \sten}{\chi}  \quad&
		\quad \partial_T \seig - K_3 T \sten \quad &
		\quad \partial_{\nu} \seig - K_2 T \sten \quad &
		\quad \partial_{\chi} \sten - K_1 T \sten
	\end{pmatrix}\,.
\end{equation}
}
Note that $B^s_{11} = -\tilde{\kappa}_{61}/\zeta^2$ and $B^s_{21}=-\tilde{\kappa}_{71}/\zeta^2$ with $\tilde{\kappa}_{i1}$ defined in \eqref{E:solv}.

The remaining undetermined constitutive relations in the scalar sector are associated with the parity
even coefficients. These are parameterized by $\beta_{ij}$ where $i$ and $j$ 
both range between $1$ and $4$. The Onsager relations imply that the 
antisymmetric part of $\beta_{ij}$ is zero. So we have a total of 10 parity 
even dissipative coefficients in the scalar sector. These 10 coefficients 
parameterize a symmetric matrix that is constrained to be positive
\begin{align}
	\sum_{i=1}^{4}\sum_{j=1}^4 \mathcal{S}^c_i \beta_{ij}\mathcal{S}^c_j  &\geq 0 \,.
\end{align}
%

%**************************************************************
\section{Summary of results and special limits}
\label{S:Summary}
%**************************************************************
% 
In sections \ref{S:ParityInvariant} and \ref{S:SuperfluidPodd} we have 
built up a complete theory of relativistic superfluid hydrodynamics that 
may or may not enjoy a parity invariance. Our work 
generalizes the recent work of \cite{Bhattacharya:2011ee} (see also 
\cite{Herzog:2011ec}) for parity conserving superfluids, and the work of 
\cite{Son:2009tf} for charged gauge-theory fluids with triangle anomalies 
(see also \cite{Erdmenger:2008rm,Banerjee:2008th}). 

The computations presented in sections \ref{S:ParityInvariant} and 
\ref{S:SuperfluidPodd} are lengthy and algebraically rather intensive. 
In this section we present the final results of our calculations in 
a manner that is self contained and makes no reference to the method 
of derivation.
%

%==========================================================
\subsection{The problem addressed} 
%==========================================================
%
As discussed in section \ref{S:ParityInvariant}, the energy momentum tensor $T_{\mu\nu}$ and charged current $J_{\mu}$ of an $s$ wave superfluid are given by
\begin{equation} \label{ltconstb}
\begin{split}
	T^{\mu\nu}&=(\rho+P) u^\mu u^\nu + P \eta^{\mu\nu} + 
	f \xi^\mu \xi^\nu + T_{diss}^{\mu\nu} \\
	J^\mu &=q_n u^\mu - f{\xi}^{\mu} + J^\mu_{diss}\\
	\mu \cd \xi & = \mu + \mu_{diss}\,
\end{split}
\end{equation}
where $\xi_{\mu}$ is related to the background gauge potential $A_{\mu}$ and 
Goldstone Boson $\psi$ through
\begin{equation}
	\xi_{\mu} = -\partial_{\mu} \psi +A_{\mu}\,.
\end{equation}
($\psi$ is the phase of the charged scalar condensate.)
We often use the projected variable
\begin{equation}
	\zeta^{\mu} = (\eta^{\mu\nu} + u^{\mu} u^{\nu})\xi_{\nu} \equiv P^{\mu\nu}\xi_{\nu}\,.
\end{equation}
in what follows. 
The equations of motion of superfluidity are given by
\begin{equation} \label{backeq}\begin{split}
	\partial_\mu T^{\mu\nu}&=F^{\nu\mu} J_\mu\\
	\partial_\mu J^\mu &= c E_\mu B^\mu\\
	\partial_\mu \xi_\nu-\partial_\nu \xi_\mu &= F_{\mu\nu}
\end{split}
\end{equation}
where we have allowed for the presence of triangle anomalies via the 
non conservation of the charged current. In writing \eqref{ltconstb} we have 
not specified a frame---a particular definition of the temperature $T$, chemical potential $\mu$ and four-velocity $u^{\mu}$ once we deviate from thermodynamic equilibrium. The thermodynamic functions $\rho$, $P$ and $f$ are functions 
of the variables
\begin{equation}
	T ~,\qquad \nu = \frac{\mu}{T} ~, \qquad \chi = \frac{\mu^2 -\xi^2}{T^2}\,.
\end{equation}
The equations \eqref{backeq} and \eqref{ltconstb} together specify a complete
set of equations for the 9 fluid dynamical fields $u^\mu(x)$, $T(x)$, 
$\mu(x)$, $\xi^\mu(x)$ once $T_{diss}^{\mu\nu}$, $ J^\mu_{diss}$ and 
$\mu_{diss}$ are specified as functions of the fluid dynamical fields and their 
derivatives. The equations that determine  $T_{diss}^{\mu\nu}$, $ J^\mu_{diss}$ and 
$\mu_{diss}$ in terms of derivatives of fluid dynamical fields are called 
constitutive relations. The main result of this paper is the 
determination of  the most 
general form of the constitutive relations consistent with Lorentz invariance, time-reversal invariance,
positivity of the divergence of the entropy current and the Onsager relations.

The most general allowed constitutive relations turn out to depend on 
20 unspecified functions of $T$, $\nu$ and $\chi$ which we will refer to as free parameters. In this section we will 
present our final results for the constitutive relations in terms of 
these 20 parameters. 

Superfluid flows are also accompanied by an entropy current. The requirement
that this entropy current be of positive divergence for every fluid flow 
played a key role in our derivation of the allowed form of the constitutive
relations. At first order in the derivative expansion the entropy current 
takes the form
\begin{equation}
	J_S = \Jsc + \Jsn\,.
\end{equation}
where $\Jsc$ is the so called canonical entropy current (in a fluid frame)
\begin{equation}\label{entcurs} 
\Jsc= s u^\mu - \frac{\mu}{T} J^\mu_{diss} - \frac{u_\nu T_{diss}^{\mu\nu}}{T}
\end{equation}
and $\Jsn$ is a correction proportional to a sum of single derivatives of fluid 
fields. In this section we also present our results for the correction 
$\Jsn$ to the entropy current.

After summarizing our results we proceed to explain 
the specialization of these results to the case of conformal or Weyl invariant
superfluids, and  certain simplified but often used limit where the 
superfluid velocity is relatively small---in liquid Helium once the 
superfluid velocity is too large superfluidity breaks down.
%

%==============================================================
\subsection{Listing our main result}\label{S:FrameInvariant}
%==============================================================
%
The most general correction to the entropy current assuming time-reversal invariance is given by 
\begin{multline}\label{gencurs}
	\Jsn = \partial_{\nu} \left( c_0  {\cal Q}^{\nu \mu} \right)
	+\epsilon^{\mu\nu\rho\sigma} \partial_{\nu} \left( \sigma_1 T u_{\rho} \zeta_{\sigma} \right) 
		+ \sthr \tilde{\mathcal{V}}^{c\,\mu}_3 
		+ T \sfou \tilde{\mathcal{V}}_4^{c\,\mu}
		+ T \sfiv \tilde{\mathcal{V}}_5^{c\,\mu} 
		\\
		+ \frac{\sigma_8}{2} \epsilon^{\mu\nu\rho\sigma} \xi_{\nu} F_{\rho\sigma}
		+T^2 \snin \omega^{\mu} 
		+T \sten B^{\mu} \,.
\end{multline}
where \begin{equation}\label{qdefs}
{\cal Q}_{\mu \nu} = T(\xi_{\mu} u_{\nu}-\xi_{\nu}u_{\mu})\,,
\end{equation} 
and $\sigma_3$, $\sigma_4$, $\sigma_5$ and $\sigma_9$ 
are determined in terms of $\sigma_8$ and $\sigma_{10}$ by 
\begin{equation}\label{arbsols}
\begin{split}
\sigma_3&= -T \frac{\partial}{\partial T} (\sigma_{10}-\nu \sigma_8)\\
\sigma_4&= \sigma_8 + c\nu -\frac{\partial}{\partial \nu}  (\sigma_{10}-\nu \sigma_8)\\
\sigma_5&=-\frac{\partial}{\partial \chi} (\sigma_{10}-\nu \sigma_8) \\
\sigma_9&=s_9 - \frac{2}{3} c\nu^3 + 2\nu (\sigma_{10}-\nu \sigma_8)
\end{split}
\end{equation}
In the above relations $\omega$ and $B$ were defined in \eqref{E:omegaB}, 
$c$ is the anomaly coefficient and $s_9$ is an integration constant. 
In any CPT invariant theory 
$$s_9=0.$$
The terms in the entropy current proportional 
to $c_0$ and $\sigma_1$ are divergence free and so are physically irrelevant 
(they have no effect on the equations of motion). Therefore, we ignore these terms 
and do not consider them in to be a part of the 20 parameters, required to describe time-reversal 
invariant superfluid dynamics (as mentioned in \S \ref{S:intro}).
The final result for the entropy current is expressed in terms of $\sigma_8$ and $\sigma_{10}$ which are 
undetermined functions of $T$, $\nu$ and $\chi$. As we will see below, 
$\sigma_8$ and $\sigma_{10}$ enter the constitutive relations, and so are 
free parameters \footnote{ Assuming parity invariance in addition to time reversal 
invariance forces the entropy current to take its canonical form.}.

We now turn to our results for the constitutive relations. We express our results in a field 
redefinition invariant manner. Since the expressions
$T^{\mu\nu}_{diss}$, $J^\mu_{diss}$ and $\mu_{diss}$ are not separately 
field redefinition invariant, we first list
%. It is only physically meaningful to specify constitutive relations for 
field redefinition invariant linear combinations of these quantities. 
%Before stating our results for the constitutive equations, we must specify  the nature of these linear combinations. 
Let 
\begin{align}\label{pios}
	\mathbf{s}_1 &= T_{diss}^{\mu\nu} \tilde{P}_{\mu\nu} & 
	\zeta^2 \mathbf{s}_2 &= \zeta \cd T_{diss} \cd \zeta \\ \notag
	\mathbf{s}_3 &= u \cd T_{diss} \cd u & 
	\mathbf{s}_4 &= u \cd T_{diss} \cd \zeta \\ \notag
	\mathbf{s}_5 &= u \cd J_{diss} & 
	\mathbf{s}_6 & = \zeta \cd J_{diss} \\ \notag
	\mathbf{s}_7 & = -\mu_{diss} && \\ \notag
	\mathbf{v}_1^{\nu} & = u_{\mu} T_{diss}^{\mu\alpha} \tilde{P}^{\nu}_{\phantom{\nu} \alpha} & 
	\mathbf{v}_2^{\nu} & = \zeta_{\mu} T_{diss}^{\mu\alpha} \tilde{P}^{\nu}_{\phantom{\nu} \alpha} \\ \notag
	\mathbf{v}_3^{\nu} & = \tilde{P}^{\nu}_{\phantom{\nu}\alpha}J^{\alpha} & & \\
\notag
	\mathbf{t}  
&= \tilde{P}_{\mu}^{\phantom{\mu}{\alpha}}\tilde{P}_{\nu}^{\phantom{\mu}{\beta}}T_{diss\,\alpha\beta} - \frac{1}{2}\tilde{P}^{\mu\nu} \tilde{P}^{\alpha\beta}T_{diss,\alpha\beta}\,,&&
\end{align}
where
\begin{equation}
	P^{\mu\nu} = \eta^{\mu\nu} + u^{\mu}u^{\nu} 
	\qquad
	\tilde{P}^{\mu\nu} = P^{\mu\nu} + \frac{\zeta^{\mu} \zeta^{\nu}}{\zeta^2}\,.
\end{equation}
We define the row vectors 
\begin{equation} \label{sds} \begin{split}
\mathbf{s}& = \begin{pmatrix}
		\mathbf{s}_1 & \mathbf{s}_2 & \mathbf{s}_3 & 
\mathbf{s}_4 & \mathbf{s}_5 & \mathbf{s}_6 & \mathbf{s}_7 \\
              \end{pmatrix} \\
\mathbf{v}&=\begin{pmatrix} \mathbf{v}_1 & \mathbf{v}_2 & \mathbf{v}_3 
\end{pmatrix}\,.
\end{split}
\end{equation}
We also define the matrices
\begin{equation}\label{adefs}
	A^{s}=	\begin{pmatrix}
			\frac{R s}{2 q_n T \chi}\quad  & \quad\frac{B_3}{3 T}-\frac{A_3}{2 T \chi}\quad  &
 \quad\frac{B_2}{3 T} - \frac{A_2}{2 T \chi}\quad  & \quad\frac{B_1}{3 T} - \frac{A_1}{2 T \chi} \\
			-\frac{R s }{q_n \chi T}\quad  & 	\quad\frac{B_3}{3 T} + \frac{A_3}{T \chi}\quad  &
 \quad\frac{B_2}{3 T} + \frac{A_2}{T \chi}\quad  & \quad \frac{B_1}{3 T} + \frac{A_1}{T \chi}  \\
			0 & \frac{1}{T^2} & 0 & 0 \\
			-\frac{R}{T^2 \chi} & \frac{K_3}{T} & \frac{K_2}{T} & \frac{K_1}{T} \\
			0 & 0 & -1 & 0 \\
			-\frac{1}{T^2 \chi} & 0 & 0 & 0  \\
			0 & \frac{(\rho +P)K_3}{T} & \frac{(\rho +P)K_2}{T} & \frac{(\rho+P)K_1}{T} \\
		\end{pmatrix}\,,
		\quad
	A^v = \begin{pmatrix}
		-R & 0 \\
		0 & \frac{2}{T^3 \chi} \\
		-1 & 0
		\end{pmatrix}\,
\end{equation}
where
$$
	R=\frac{q}{\rho+P} \qquad 
	V_1^{\mu}=\frac{E^{\mu}}{T} - P^{\mu\nu}\partial_{\nu}\nu
$$ 
and the $A_i$'s $B_i$'s, $C_i$'s and $K_i$'s defined in appendix \ref{S:SuperfluidPodd}.

In terms of \eqref{pios}-\eqref{adefs}, the frame invariant scalar, vector and 
tensor combinations of $T^{\mu\nu}_{diss}$, $J^\mu_{diss}$ and $\mu_{diss}$
are given by the row vectors
\begin{equation}\label{fic}
\mathbf{s} A^s, ~~~~\mathbf{v}_\mu A^v, ~~~~~\mathbf{t}_{\mu\nu}\,.
\end{equation}
By scalars, vectors and tensors we mean expressions which transform 
as spin 0, $\pm 1$ and $\pm 2$ representations of the $SO(2)$ symmetry that 
is left invariant by the two vectors $u^\mu$ and $\xi^\mu$ at each point in spacetime.

We have 4 frame invariant scalars, 2 frame invariant vectors and one 
frame invariant tensor. The constitutive relations determine these 
quantities as functions of first derivative fluid expressions through 
the equations  
\begin{align} \label{hios}
\begin{split}
	\mathbf{t}^{\mu\nu}  &= -\eta \mathcal{T}^{\mu\nu}_1 - \tilde{\eta} \tilde{\mathcal{T}}^{\mu\nu}_1 \\
	\mathbf{v}^{\mu}_i A^v_{ij}  &= -\sum_{i=1}^2  \mathcal{V}_i \kappa_{ij} 
- \sum_{i=1}^2 \tilde{\mathcal{V}_i} \tilde{\kappa}_{ij}
- \delta_{j1} \left( \sum_{i=3}^7 \tilde{\mathcal{V}_i} \tilde{\kappa}_{i1} 
\right) \\
	\mathbf{s}_i A^s_{ij} & = -\sum_{i=1}^{4} \sum_{j=1}^4\mathcal{S}_i 
\beta_{ij} -  \left( 
\sum_{i=1}^2\sum_{j=1}^4 \tilde{\mathcal{S}}_{i}\tilde{\beta}_{ij} \right)
\end{split}
\end{align}
with $\mathcal{T}$, $\tilde{\mathcal{T}}$, $\mathcal{V}$, $\tilde{\mathcal{V}}$, $\mathcal{S}$ and $\tilde{\mathcal{S}}$ a basis of onshell independent 
 $SO(2)$ invariant tensors, vectors and scalars given in table 
\ref{tb:mainresultbasis}.
{\small
\TABLE{
\renewcommand\arraystretch{1.4}
\caption{
The basis in which we present our main results. The pseudo vectors and pseudo tensors $\tilde{\mathcal{V}}$ and $\tilde{\mathcal{T}}$ are given through $\tilde{\mathcal{V}}^{\mu}  = \epsilon^{\mu\nu\alpha\beta}u_{\nu} \xi_{\alpha} \mathcal{V}_{\beta}$ and $\tilde{\mathcal{T}}^{\mu\nu}  =  \epsilon^{\mu\rho\alpha\beta}u_{\rho}\xi_{\alpha} \mathcal{T}_{\beta}^{\nu} +  \epsilon^{\nu\rho\alpha\beta}u_{\rho}\xi_{\alpha} \mathcal{T}_{\beta}^{\mu}$ Here $\tilde{P}^{\mu\nu} = \eta^{\mu\nu} + u^{\mu} u^{\nu} - \zeta^{\mu}\zeta^{\nu} / \zeta^2$.
}
\label{tb:mainresultbasis}
\begin{tabular}{| c | c | c | c | c | c | c | c |}
\hline
	$i$&1&2&3&4&5&6&7\\
\hline
	${{\cal V}}_i^{\,\mu}$&
	$\tilde{P}^{\mu\sigma}V_{1\,\sigma}$&
	$\tilde{P}^{\mu\beta}\left(\zeta^{\alpha}\sigma_{\alpha\beta}\right)$&
	$\tilde{P}^{\mu\sigma} \nabla_\sigma T$&
	$\tilde{P}^{\mu\sigma} \nabla_\sigma \left(\frac{\mu}{T}\right)$&
	$\tilde{P}^{\mu\sigma} \nabla_\sigma\left(\frac{ \zeta^2}{T^2}\right)$&
	$ \frac{\mathcal{V}_2^{c\,\mu} - \tilde{P}^{\mu\alpha}\zeta^{\nu}\partial_{\alpha} u_{\nu}}{ \zeta^{2}}$&
	$-\frac{P^{\mu\nu} F_{\nu\alpha}\zeta^{\alpha}} {\zeta^{2}}$\\
\hline
	$\mathcal{S}$&
	$\zeta_{\mu}V_1^{\mu}$&
	$u \cd \partial T$&
	$u \cd \partial \frac{\mu}{T}$&
	$u \cd \partial \frac{\zeta^2}{T^2} $&
	$\zeta^{\mu} P^{\mu\alpha} \partial_{\alpha}T$& 
	$\zeta^{\mu}E_{\mu}$ &
	$\zeta^{\mu} P^{\mu\alpha}\partial_{\alpha} \frac{\zeta^2}{T^2}$ \\
\hline
	${\mathcal{T}}_{i\,\mu\nu} $&
	$\str \sigma^u_{\mu\nu}$ &
	$\str \sigma^{\xi}_{\mu\nu}$ &
	-- & -- & -- & -- & -- \\
\hline
	$\tilde{\mathcal{S}}_i$ &
	$\omega \cd \xi$ &
	$B \cd \xi$ &
	-- & -- & -- & -- & -- \\
\hline	
\end{tabular}
}
}
In equation \eqref{hios} $\kappa$ and $\beta$ are symmetric matrices and $\tilde{\kappa}_{12} = \tilde{\kappa}_{21}$. Those coefficients that occur
in the big brackets on the right hand side of that equation, namely
$\tilde{\kappa}_{1i}$ for $i=3 \ldots 7$ and 
${\tilde \beta}_{ij}$ for $i=1, 2$ and $j=1 \ldots 4$, 
are not free but are determined in terms of $\sigma_8$ and $\sigma_{10}$ through 
the equations 
\begin{align}
\begin{split}
\label{E:solvs}
%	\kappa_{i2}& = \kappa_{i1} = 0\\
%	\tilde{\kappa}_{i2}& =-B^v_{i2}= 0 \quad i>2\\
	\tilde{\kappa}_{31} & = -R T \sigma_3 - T \partial_T \sigma_8 \\
	\tilde{\kappa}_{41} & =- R T^2 \sigma_4 - T \partial_\nu \sigma_8 \\
	\tilde{\kappa}_{51} & = -R T^2 \sigma_5 - T \partial_{\chi} \sigma_8 \\
	\tilde{\kappa}_{61} & = -2 R T^3 \sigma_9 +2 T^2 \sigma_{10} \\
	\tilde{\kappa}_{71} & = -R T^2 \sigma_{10} + 2 T \sigma_8 + c T \nu
\end{split}
\end{align}
{\small
\begin{equation}
	-\tilde{\beta}_{ij} = \begin{pmatrix}
		\frac{2 R T \snin}{\chi} - \frac{2 \sten}{\chi} \quad&
		\quad -2\sthr - 2 T^2 K_3  \snin \quad&
		\quad -2 T \sfou -2 T^2  K_2 \snin \quad&
		\quad -2 T \sfiv - 2 K_1 T^2  \snin \\
		-\frac{c \nu}{T \chi} - \frac{2 \seig}{T \chi} + \frac{R \sten}{\chi}  \quad&
		\quad \partial_T \seig - K_3 T \sten \quad &
		\quad \partial_{\nu} \seig - K_2 T \sten \quad &
		\quad \partial_{\chi} \sten - K_1 T \sten
	\end{pmatrix}\,.
\end{equation}
}
where $\sigma_3$, $\sigma_4$, $\sigma_5$ and $\sigma_9$ are related to $\sigma_8$ and $\sigma_{10}$ through the relations
\eqref{arbsols}.

The remaining coefficients in \eqref{hios}, those that occur outside 
the big brackets on the right hand side of \eqref{hios}, are free  
parameters. The symmetric $2\times 2$ matrix $\kappa$, $4\times 4$ matrix $\beta$, and $\eta$ are constrained to obey
\begin{align}
\begin{split}
\label{E:constraints}
	\eta &>0 \\
%	\tilde{\eta} &\in \mathbf{R} \\
	\kappa_{11}\kappa_{22} &\geq \kappa_{12}^2  \\
%	\tilde{\kappa}_{11} &\in \mathbf{R} \\
%	\tilde{\kappa}_{22} &\in \mathbf{R} \\
%	\tilde{\beta}_{ij} &= -B^s_{ij} \\
	\beta_{ij} &\quad \hbox{is symmetric positive semi-definite}\,.
%	\tilde{\kappa}_{i3} & = R T \sthr + T \partial_T \seig \\
%	\tilde{\kappa}_{i4} & = R T^2 \sfou + T \partial_\nu \seig \\
%	\tilde{\kappa}_{i5} & = R T^2 \sfiv + T \partial_{\chi} \seig \\
%	\tilde{\kappa}_{i6} & = 2 R T^3 \snin - 2 T^2 \sten \\
%	\tilde{\kappa}_{i7} & = R T^2 \sten - 2 T \seig + c T \nu\,.
\end{split}
\end{align}
The remaining parameters
\begin{equation}
	\tilde{\kappa}_{11}\,\quad
	\tilde{\kappa}_{22}\,\quad
	\tilde{\kappa}_{12} = \tilde{\kappa}_{21} \, \quad
	\tilde{\eta}\,\quad
	\sigma_8\, \quad
	\sigma_{10}
\end{equation}
are unconstrained.

In \S \ref{S:intro} we counted twenty free parameters. Let us enumerate them explicitly.
The parity even parameters consist of a $4\times 4$ symmetric matrix $\beta_{ij}$ in 
the scalar sector (10 parameters),  a $2\times 2$ symmetric matrix 
$\kappa_{ij}$ in the vector sector (3 parameters) and the shear viscosity 
$\eta$ in the tensor sector (1 parameter) yielding a total of 14 parity even 
free parameters. These are the only  parameters in a 
theory that conserves parity. The parity odd  parameters consist of the three parameters 
$\tilde{\kappa}_{11}$, $\tilde{\kappa}_{22}$, $\tilde{\kappa}_{21} =\tilde{\kappa}_{12}$, in the vector sector and the one parameter $\tilde{\eta}$
in the tensor sector, amounting to a total of 4 free parameters.
The two undetermined functions $\sigma_8$ and $\sigma_{10}$ are 
two additional free parameters. 

None of the six parity free
parameters, namely $\tilde{\kappa}_{11}$, $\tilde{\kappa}_{22}$, 
$\tilde{\eta}$, $\sigma_8$ and $\sigma_{10}$, and $\tilde{\kappa}_{21} =\tilde{\kappa}_{12}$ result in entropy production. 
On the other hand  $\sigma_8$ and $\sigma_{10}$ multiply expressions that do not vanish in equilibrium and are referred to 
non-dissipative in \cite{Bhattacharyya:2012xi}.
The remaining four parity odd parameters multiply expressions that vanish in equilibrium. These terms 
are unconstrained by the analysis of \cite{Bhattacharyya:2012xi}and are referred to as dissipative in that paper. It appears slightly 
non-intuitive to refer to a parameter that does not appear in entropy production as dissipative and this suggests 
a need for modified nomenclature.

The remaining 14  parity even constitutive parameters are non-dissipative in every sense. They multiply expressions 
that vanish in equilibrium. Moreover these parameters also appear in the formula for entropy production and so are 
forced to obey appropriate inequalities listed  in \eqref{E:constraints}.

% After completing the work reported in this paper we received the preprint 
% \cite{Lin:2011mr} which addresses the same question dealt with in this 
% section. While we have not yet performed a detailed comparison between 
% our results and those of \cite{Lin:2011mr}, a superficial comparison
% of final results reveals several differences. We list some of these 
% differences here. 
% \begin{enumerate} 
% \item The author of \cite{Lin:2011mr} finds a unique entropy current, 
% while the entropy current in this paper has a two parameter (function) 
% freedom.
% \item The author of  \cite{Lin:2011mr} finds a total of 2 new parity odd dissipative 
% parameters, while we find 6 such parameters in this paper. 
% The parameter we refer to as $\tilde{\eta}$ seems unaccounted for in \cite{Lin:2011mr}.
% \item The author of  \cite{Lin:2011mr} reports that his new parameters are all functions 
% of a single thermodynamical variable, while the six new dissipative parameters
% in this paper are all unconstrained functions of all three thermodynamic 
% scalars.
% \item The author of \cite{Lin:2011mr} finds no mixing between parity even and parity 
% odd sectors, while in this paper we find nontrivial mixing between parity 
% even and parity odd vector contributions.
% \end{enumerate}
% We do not yet understand the reason for the discrepancy between our results
% and those of \cite{Lin:2011mr}. We leave an investigation of this issue to 
% future work.
%

%==========================================================
\subsection{Weyl invariant superfluid dynamics}
%==========================================================
%
The equations of superfluidity simplify somewhat in conformal field theories, 
i.e. theories that enjoy invariance under Weyl transformations. The source 
of the simplification is twofold: first the stress tensor of a CFT is 
always traceless. Second, the dependence of all thermodynamic functions on temperature can be deduced by dimensional analysis if we take the remaining variables to be the dimensionless quantities $\nu$ and $\chi$.
In the rest of this subsection we briefly
outline the special simplifications in Weyl invariant superfluid hydrodynamics.

Let us first consider the entropy current. 
The coefficient $\sigma_3$ in the expression \eqref{gencurs} for the entropy current 
of a Weyl invariant theory must vanish. This follows because $\tilde{\mathcal{V}}_3$ does
not transform homogeneously under Weyl transformations. Since $\sigma_8$ and 
$\sigma_{10}$ are dimensionless and therefore independent of $T$, equation \eqref{arbsols} does not pose an extra restriction on these two parameters.

Let us now turn to the consequences of Weyl invariance on constitutive 
relations. It turns out that (beside implying that all constitutive 
parameters are independent of $T$) Weyl invariance implies no special 
simplifications for constitutive relations in the tensor sector.
In the vector sector we expect that $\tilde{\kappa}_{13}=0$ since, like $\sigma_3$, it multiplies  $\tilde{\mathcal{V}}_3$ which does not transform homogeneously under Weyl transformations. Since $\sigma_3$ vanishes and $\sigma_8$ is independent of the temperature, this condition is automatic from 
equation \eqref{E:solvs}, i.e. it  does not pose an extra restriction on $\sigma_8$. The remaining terms in the vector sector remain unchanged.

In the scalar sector dimensional analysis implies that
\begin{equation}
\label{E:Ccondition3}
	A_3 = K_3 = 0 
	\qquad
	B_3 = -\frac{3}{T}\,.
\end{equation}
The fact that the scalar $\mathcal{S}_2$ does not transform
homogeneously under Weyl transformations implies that both indices of 
$\beta$ in  the third equation in \eqref{hios} run only over the indices 
$1, 3, 4$. Consequently $\beta$ is effectively a $3 \times 3$ symmetric 
matrix (with 6 parameters) in the Weyl invariant case. 
More importantly, the tracelessness of the stress tensor implies
that the number of scalar terms in $T_{diss}^{\mu\nu}$, $J^\mu_{diss}$ and 
$\mu_{diss}$ is six rather than seven. In other words, the seven terms $\mathbf{s}_i$ 
listed in \eqref{pios} are restricted by the linear relation
\begin{equation}
\label{E:Ccondition1}
	\mathbf{s}_1 + \mathbf{s}_2 - \mathbf{s}_3 = 0\,.
\end{equation}
Since there are still three field redefinitions in the scalar sector, this implies 
that we have three rather than four field redefinition invariant scalar terms which can be constructed out of  $T_{diss}^{\mu\nu}$, $J^\mu_{diss}$ and $\mu_{diss}$. Accordingly we find that 
\begin{align}
\begin{split}
\label{E:Ccondition4}
	\mathbf{s}_{i}A^s_{i2} &= 0  \\
	\tilde{\beta}_{i2} & = 0
\end{split}
\end{align}
so that the row vector $\mathbf{s}_{i}A^s_{ij}$ has three rather than four independent
entries. Equation \eqref{E:Ccondition4} follows from \eqref{E:Ccondition3}, \eqref{E:Ccondition1} and $\sigma_3=0$.

It follows that Weyl invariant superfluid hydrodynamics is governed by 
10 constitutive parameters (when it preserves parity) and 16 constitutive 
parameters (when it violates parity).
%

%==============================================================
\subsection{The $\zeta \to 0$ limit}
\label{S:collinear}
%==============================================================
%
In liquid Helium, superfluidity breaks down once the velocity of the superfluid component is too large relative to the velocity of the normal component. The projected superfluid velocity $\zeta^{\mu} = P^{\mu\nu}\xi_{\nu}$ can be thought of as a control parameter for that limit. Once $\zeta = 0$ one can locally boost a fluid element to a frame where neither its normal component nor its superfluid component are in motion. 

When $\zeta$ is non zero, the velocity of the normal component $u^{\mu}$ together with $\zeta^{\mu}$ break the $SO(3,1)$ symmetry to $SO(2)$. In the $\zeta \to 0$ limit (the collinear limit) this symmetry is enhanced to $SO(3)$. The number of possible expressions for $SO(3)$ invariant scalars, tensors, vectors and their parity odd counterparts is significantly smaller than that of the $SO(2)$ symmetric case. As discussed in \cite{Herzog:2011ec,Bhattacharya:2011ee} there are two possible scalars, $\partial \cd u$ and $\partial_{\mu} \left(f \xi^{\mu}\right)$ (which we can swap with $u\cd\partial T$ and $u\cd \partial \nu$ via the equations of motion), one vector $V_1^{\mu} = -P^{\mu\nu}\partial_{\nu} \nu + E^{\mu}/T$ and two tensors in the parity even sector, $\sigma_{\mu\nu}$ and the symmetric traceless projection of $\partial_{\mu} \zeta_{\nu}$ onto the space orthogonal to $u_{\mu}$. In the parity odd sector we have two pseudo vectors 
\begin{equation}
\label{E:omegaB}
	\omega^{\mu} = \frac{1}{2}\epsilon^{\mu\nu\rho\sigma}u_{\nu}\partial_{\rho} u_{\sigma}
	\quad \hbox{and} \quad
	B^{\mu} = \frac{1}{2} \epsilon^{\mu\nu\rho\sigma}u_{\nu}F_{\rho\sigma}\,.
\end{equation}
Thus, in the collinear limit the transport coefficients of the theory reduce to a $2\times 2$ symmetric matrix $\beta_{ij}$ (in the parity even scalar sector), a diffusion coefficient $\kappa$ (in the parity even vector sector), 
a shear viscosity $\eta$ (in the parity even tensor sector), and two 
parity odd terms $\tilde{\kappa}_{\omega}$ and $\tilde{\kappa}_B$. 
These enter the constitutive relations through
\begin{align}
\begin{split}
\label{E:zeta0Sol}
	-\frac{\partial \cd u}{3 T} T_{diss}^{\mu\nu}P_{\mu\nu} - \frac{u \cd \partial T}{T^2} 
u\cd T_{diss} \cd u+u \cd\partial\nu\, u \cd J_{diss} + 
\frac{\mu_{diss}}{T} \partial_{\mu} \left(f \xi^{\mu}\right) &=  S_i \beta_{ij} S_j \\ 	  \left( R  P^{\mu}_{\phantom{\mu}\alpha} T_{diss}^{\alpha\nu} u_{\nu}  + P^{\mu}_{\phantom{\mu}\alpha} J^{\alpha}\right) &= \kappa V_{1}^{\mu} - \tilde{\kappa}_B B^{\mu} - \tilde{\kappa}_{\omega} \omega^{\mu} \\
	P^{\mu}_{\phantom{\mu}\alpha}P^{\nu}_{\phantom{\nu}\beta}T_{diss}^{\alpha\beta} 
- \frac{1}{3} P^{\mu\nu} P_{\alpha\beta} T_{diss}^{\alpha\beta} & = -\eta \sigma^{\mu\nu}\,.
\end{split}
\end{align}
It must also be true that 
\begin{equation}
\label{E:JSlimit}
	J_{S~new}^{\mu} = \sigma_{\omega} \omega^{\mu} + \sigma_B B^{\mu}+ ... 
\end{equation}
where we have written only the odd part of the entropy current, omitting 
the total derivative term in the parity even sector (the second term on the right hand side of \eqref{gencurs}).
This omission is represented by the $\ldots$ in \eqref{E:JSlimit}.

As we have explained, we expect \eqref{E:zeta0Sol} and \eqref{E:JSlimit} to 
hold on general grounds. We should, therefore, be able to verify 
these equations---and read off the values of 
$\tilde{\kappa_B}$, $\tilde{\kappa_\omega}$---by studying the $\zeta \to 0$ 
limit of our general results. We now describe how to take this limit. 

Let us first consider the entropy current. The third, fourth and fifth
terms in \eqref{gencurs} each have an explicit factor of $\zeta$ and so simply
vanish in the $\zeta \to 0$ limit. (We assume that all physical quantities like
the entropy current---and therefore all all coefficient functions that 
appear in the entropy current---are analytic functions of $\zeta$.) 
The last two terms in \eqref{gencurs} are already proportional to 
$\omega^\mu$ and $B^\mu$ respectively. The sixth term in 
\eqref{gencurs} is proportional to $B^\mu$ as $\zeta \to 0$, while the 
first term has a piece proportional to $\omega^\mu$ and a piece proportional 
to $B^\mu$. Putting it all together we recover the form \eqref{E:JSlimit}
with 
\begin{align}
\begin{split}
\label{E:zeta0sigma}
	\sigma_{\omega} &= T^2 \left(s_9 - \frac{2}{3} c\nu^3 
+ 2 \nu \left(\sten-\nu \seig-\frac{1}{2}  \sone\right)\right) \\
	\sigma_B &= T \left(\sten-\nu \seig-\frac{1}{2}  \sone\right)\,.
\end{split}
\end{align}
where all functions are evaluated, of course, at $\zeta^2=0$.

Let us now turn to the constitutive relations. Notice that 
$-(\mathbf{v}A^v)_1$ is simply the $SO(2)$ vector part of the $SO(3)$ frame 
invariant vector presented on the left hand side of the second equation in  \eqref{E:zeta0Sol}. 
Note also that at leading order in small $\zeta$  
$$\frac{ 2 T (\mathbf{s} A^s)_2}{A_3} \sim 
\frac{ 2 T (\mathbf{s} A^s)_2}{A_2} \sim \frac{ 2 T (\mathbf{s} A^s)_1}{A_1}
\sim \frac{2 \mathbf{s}_2- \mathbf{s}_1}{\chi}\,.$$
Since the constitutive relations \eqref{hios} equate this combination to 
an analytic function of $\zeta$, it follows that in the $\zeta \to 0$ limit 
$2 \mathbf{s}_2-\mathbf{s}_1$ is frame invariant and, in fact, vanishes at 
$\zeta=0$.\footnote{This 
could have been anticipated on general grounds, as this combination is 
proportional to the $SO(2)$ scalar in the $SO(3)$ stress tensor, i.e. it is 
proportional to $\zeta\cd \sigma \cd \zeta$. }  Using this fact it follows that 
$-(\mathbf{s}A^s)_1$ is proportional to the scalar part of the vector 
on the left hand side of the second equation in \eqref{E:zeta0Sol}.  Thus,
\begin{align}
\begin{split}
\label{E:zeta0Result}
	\kappa & \geq 0 \\
	\eta & \geq 0 \\
	\beta_{ij} & \quad \hbox{is positive semi-definite} \\
	\tilde{\kappa}_{\omega} & =-\tilde{\kappa}_{61}= -\zeta^2 \beta_{11}=
 2 R T^3 s_9 - \frac{4}{3} c R T^3 \nu^3 - 4 R T^3 \nu^2 \seig + 2 T^2(2 R T \nu -1) \sten \\
	\tilde{\kappa}_B & = -\tilde{\kappa}_{62}= 
-\zeta^2 \beta_{21}=R T^2 \sten - 2 T \seig - c T \nu\,.
\end{split}
\end{align}
It was necessary that $-\tilde{\kappa}_{61}= -\zeta^2 \beta_{11}$ and that 
$\tilde{\kappa}_{62}  = -\zeta^2 \beta_{21}$ in order that our results
group into an $SO(3)$ vector in the $\zeta \to 0$ limit. This gives us a mild consistency check on our results. 

Let us summarize. The parity odd contributions to the entropy current and 
constitutive relations are much simpler in the collinear limit than in 
the general case. The only constitutive relation that receives corrections 
in this limit is the expression for the frame invariant vector listed 
in the left hand side of the second equation of  \eqref{E:zeta0Sol} (equal to the 
dissipative part of the charge current in a Landau like frame). The corrections 
to both this vector and the entropy current are linear combinations of 
$B^\mu$ and $\omega^\mu$. These linear combinations are specified in terms of 
3 free functions, $\sigma_8$, $\sigma_{10}$ and $\sigma_1$ and one 
integration constant (which vanishes on assuming CPT symmetry).  As 4 
coefficient parameters are determined in terms of 3 parameters, the coefficients
obey a single linear identity. This identity is given by 
\begin{equation}
\label{E:relationf}
	\frac{1}{2}\sigma_{\omega} - \mu \sigma_B 
		= - \frac{\mu^3}{3 T} c +\frac{1}{2} s_9 T^2
\end{equation}
Under the eminently reasonably assumption of CPT invariance $s_9=0$ and 
\begin{equation}
\label{E:relation}
	\frac{1}{2}\sigma_{\omega} - \mu \sigma_B 
		= - \frac{\mu^3}{3 T} c
\end{equation}
The relation \eqref{E:relation} 
is the only real prediction in the $SO(3)$ 
invariant sector. Note that it constrains coefficients in the entropy current
alone. In the next section we will use the AdS/CFT correspondence to 
test this relationship (of course $s_9=0$ in the holographic context). 

%-------------------------------------------------------------------------------
\cl{The $\zeta \to 0$ for Weyl invariant superfluids}
%-------------------------------------------------------------------------------
%
As we have seen above, Weyl invariance removes one of scalar field redefinition
invariant combinations of one derivative corrections to the constitutive relations 
(because the trace of the stress tensor is set to zero) and also removes one 
of the scalars in terms of which the parity even part of these constitutive
relations is expressed (because $u \cd \partial T$  and $\partial \cd u$ 
do not transform homogeneously under Weyl rescaling). In more detail, since 
in a conformal field theory, $q_n \left(\frac{\partial}{\partial \nu} \ln \frac{q_n}{s}  \right) \,u \cd \partial \nu = \partial_{\mu}\left(f \xi^{\mu}\right)$, there is only one permissible on-shell scalar.
In the $\zeta \to 0$ limit 
the $2 \times 2$ matrix $\beta$ described above collapses to a $1 \times 
1$ matrix. Thus Weyl invariant superfluid dynamics is characterized by 
3 constitutive parameters in the $\zeta \to 0$ limit, one in the 
tensor sector (the shear viscosity), one in the vector sector (conductivity) 
and one in the scalar sector (a new coefficient).

Apart from determining the dependence of $\tilde{\kappa}_{\omega}$, $\tilde{\kappa}_B$, $\sigma_{\omega}$ and $\sigma_B$ on the temperature (using 
dimensional analysis), Weyl invariance does not impact the discussion presented above 
for parity odd corrections to constitutive relations and the entropy 
current. In particular 
\eqref{E:zeta0Sol}-\eqref{E:relation} continue to hold, and do not 
significantly simplify in a theory that enjoys Weyl invariance.
%

%*******************************************************
\section{A holographic computation}
\label{S:Holographic}
%*******************************************************
%
  
The AdS/CFT correspondence \cite{Gubser:1998bc,Witten:1998qj,Maldacena:1997re} can be used to study the superfluid phase of appropriate large $N$ gauge theories in the limit of infinite t' Hooft coupling \cite{Gubser:2008px,Hartnoll:2008vx,Herzog:2008he,Basu:2008st,Amado:2009ts}. We will refer to such phases of gauge theories as holographic superfluids. Various features of holographic superfluids have been studied in the literature. These include sound modes \cite{Herzog:2009ci,Yarom:2009uq,Amado:2009ts,Herzog:2009md,Herzog:2011ec}, critical superfluid velocities \cite{Herzog:2008he,Gubser:2009qf} and vortex structure \cite{Keranen:2009ss,Keranen:2009re,Keranen:2010sx}. Holographic superfluids with a non uniform temperature, chemical potential and velocity fields have been recently constructed in \cite{Herzog:2011ec,Bhattacharya:2011ee} (see also \cite{Sonner:2010yx}). These dynamical superfluids were constructed for a certain specialized bulk action which, following the work of \cite{Herzog:2010vz}, allows for an analytic treatment.

In this subsection we will demonstrate that the prediction 
\eqref{E:relation}, that follows from our general analysis, is indeed true 
for superfluids that admit a dual description via the AdS/CFT correspondence.
Our demonstration uses the technology of the so called fluid gravity 
correspondence \cite{Bhattacharyya:2008jc,VanRaamsdonk:2008fp,Bhattacharyya:2008xc, Bhattacharyya:2008ji,Haack:2008cp,Erdmenger:2008rm,Banerjee:2008th,Bhattacharyya:2008mz} but allows for the implementation of this map in an 
abstract manner that does not require us to know the explicit form of the 
solutions dual even to stationary fluid flows. In this respect our analysis 
is reminiscent of the thermodynamical analysis of Sonner and 
Withers \cite{Sonner:2010yx} (see \cite{Kalaydzhyan:2011vx} for related work).

Our starting point for the analysis is the bulk action 
\begin{equation}
\label{E:BulkAction}
	S = S_{EH} + S_{matter} + S_{CS}
\end{equation}
where 
\begin{align}
\begin{split}
	S_{EH} &= \frac{1}{2\kappa^2} \int \sqrt{-g} \left(R + 12\right) \\
	S_{matter} &= \frac{1}{2\kappa^2} \int \sqrt{-g} \left( - \frac{1}{4} V_F(|\psi|) F_{mn}F^{mn} 
- V_{\psi}(|\psi|) |\partial_{m}\psi - i q A_{m}\psi |^2  - V(|\psi|) \right) \\
	S_{CS}&= \frac{c}{24} \int \sqrt{-g}  \epsilon^{mnpqr}A_m F_{np}F_{qr} 
\end{split}
\end{align}
with $\epsilon^{01234} = 1/\sqrt{-g}$, $F=dA$, $V_F(0)=1$, $V_{\psi}(0)=1$ and $V(0)=0$. We will often use $D_m \psi = \partial_m \psi - i q A_m \psi$. Roman indices run from $0$ to $3$ and $5$. Later we will use Greek indices to denote the boundary coordinates $\mu = 0,\ldots,3$ and $i,j = 1,\ldots,3$ will denote the spatial coordinates along the boundary. We have chosen the action \eqref{E:BulkAction} for several reasons. As has been discussed in detail in \cite{Gubser:2008px,Hartnoll:2008vx,Herzog:2008he,Basu:2008st}, to construct a holographic superfluid one needs to spontaneously break a gauge symmetry in an asymptotically AdS geometry. The minimal required fields for such a construction involve a gauge field $A_{\mu}$, a charged scalar $\psi$ and a metric $g_{\mu\nu}$. The parity conserving part of the action \eqref{E:BulkAction}, i.e., $S_{EH}+S_{matter}$, is the most general action one can write down which involves the fields described above. The parity violating sector consists of a Chern-Simons term, $S_{CS}$, which naturally appears in consistent truncations of type IIB supergravity. For example, the actions of the consistent supergravity truncations described in \cite{Gubser:2009qm,Bhattacharyya:2010yg} can be described by the action \eqref{E:BulkAction}.

The equations of motion which follow from \eqref{E:BulkAction} are 
\begin{align}
\begin{split}
\label{E:BulkEOM}
	\frac{1}{\sqrt{-g}} \left( \left( \partial_m - i q A_m \right) V_{\psi} \sqrt{-g} D^m\psi \right) 
&= \frac{1}{4} \frac{\partial V_F}{\partial \psi^*} F^2 +  \frac{\partial V_{\psi}}{\partial \psi^*} | D \psi |^2 + \frac{\partial V}{\partial \psi^*} \\ 
	\frac{1}{\sqrt{-g}} \partial_m \left(\sqrt{-g} V_F F^{mn} \right)& =iq V_{\psi}\left( \psi^* (D^n \psi ) 
- \psi (D^n \psi)^* \right) - \frac{c \kappa^2}{4} \epsilon^{nmpqs}F_{mp}F_{qs} \\
	R_{mn} - \frac{1}{2} R g_{mn} - 6 g_{mn} = \mathcal{T}_{mn}
\end{split}
\end{align}
with
\begin{multline}
	\mathcal{T}_{mn} = \frac{1}{2} V_F F_{mp} F^{np} - \frac{1}{8}g_{mn} F_{pq}F^{pq} 
- \frac{1}{2}g_{mn} V - \frac{1}{2} g_{mn} V_{\psi} |D\psi|^2 \\
	+ \frac{1}{2} V_{\psi}\left( (D_m\psi) (D_n \psi)^* + (D_n\psi) (D_m \psi)^* \right) \,.
\end{multline}

Our strategy for computing the parity odd corrections to the entropy current and the parity odd transport coefficients closely follows that of  \cite{Bhattacharyya:2008jc,Bhattacharyya:2008xc,Erdmenger:2008rm,Banerjee:2008th,Herzog:2011ec,Bhattacharya:2011ee}. We first construct a stationary solution to the equations of motion \eqref{E:BulkEOM} in which neither the superfluid not the normal component are in motion. Being interested in the limit of small relative superfluid velocity, we then allow for the gauge field to have a small non vanishing spatial component. By boosting and rotating the latter solution, we find a generic bulk configuration involving six free parameters which are associated with the temperature, chemical potential, velocity of the normal component and the expectation value of the goldstone boson of the dual superfluid. We then promote these six parameters to be spacetime dependent fields. Once we do so the equations of motion \eqref{E:BulkEOM} will no longer be satisfied. Therefore, we add corrections to the our previous solution so that the equations of motion \eqref{E:BulkEOM} will be satisfied at least up to second order gradients of the six parameters. We can then compute the one point functions of the energy momentum tensor, charged current and gradient of the Goldstone boson in the boundary theory and from them read off the transport coefficients associated with the first order terms in a gradient expansion.\footnote{The authors of \cite{Amado:2011zx} recently argued that the transport coefficients associated with the parity odd sector can also be computed using a Kubo formula.}
%

%======================================================
\subsection{A stationary solution}
%======================================================
%
The ansatz for a stationary solution to the equations of motion \eqref{E:BulkEOM}---corresponding to a configuration in which the normal fluid and 
superfluid are both at rest---is
\begin{align} \label{ansatz}
	ds^2 &= -r^2 f(r) dt^2 + r^2 dx^2 + \sigma(r) dt dr \\
	\psi &= \rho(r) e^{i q \varphi(r)} \\
	A_m &= (A_0(r),0,0,0,A_5(r))\,.
\end{align}
In this ansatz we have not completely fixed a gauge. In what follows we will present our analysis in the gauge invariant variables
\begin{equation}
	G_m = A_m - \partial_m\varphi\,.
\end{equation}

In this section we will not attempt to find an explicit solution for the 
unknown functions in \eqref{ansatz}, such a solution will not be needed in 
what follows below. We will instead note certain abstract aspects of this 
solution that can be obtained without explicitly solving the equations.
\begin{itemize}
\item
	The equation for $G_5$ is algebraic and is given by
	\begin{equation}
	\label{E:G5}
		G_5 = -\frac{G_0 \sigma}{r^2 f}\,.
	\end{equation}
\item
The remaining equations of motion can be chosen to be a set of 4 linear 
differential equations in the variables $f$, $\sigma$,  $\rho$ and $G_0$. 
The equations have the property that they involve only first derivatives 
(in the variable $r$) of $f$ and $\sigma$ and second order derivatives (in the 
variable $r$) of $\rho$ and $G_0$. 
\item
	Assuming that the solution is a black hole, the fields $\sigma$ and $\rho$ are 
non vanishing at $r_h$, the black hole horizon, while $f$ and $G_0$ have a simple zero at $r_h$. 
The fact that $G_0$ vanishes at $r_h$ follows from regularity of $G_5$ (defined in \eqref{E:G5}) 
near the horizon. The temperature, $T$, and entropy density $s$ of the black hole are given by
	\begin{equation}
	\label{E:sandT}
		T = \frac{r_h^2 f'(r_h)}{4 \pi \sigma(r_h)}
		\qquad
		s = \frac{\pi r_h^3}{\kappa^2}\,.
	\end{equation}
\item
	One linear combination of the 4 equations can be shown to 
express the constancy (in the radial direction) of the `Noether' charge $Q_1$ \cite{Gubser:2009cg},
	\begin{equation}
	\label{E:Q1def}
		2 \kappa^2 Q_1 = \frac{r^5 f'}{\sigma} - \frac{r^3 V_F G G'}{\sigma}\,.
	\end{equation}
The equation in question asserts that 
$$Q_1'=0$$ 
where a prime denotes a derivative in the $r$ direction. Using \eqref{E:sandT} we find that
	\begin{equation}
	\label{E:Q1Val}
		Q_1 = s T \,.
	\end{equation}
\item
	It is not difficult to work out a Graham Fefferman style solution 
of the equations of motion at large $r$. Putting in the physical constraint 
that the non normalizable mode dual to the scalar is turned off (i.e. 
that the operator dual to the scalar field is not sourced in the dual 
boundary theory), we can use \eqref{E:Q1def} and the equations of motion for $f$, $\sigma$, $\rho$ and $G_0$ to show that near the asymptotically AdS boundary  ($r \to \infty$)
	\begin{align}
	\label{E:Nearb}
	\begin{split}
		f & = 1 + \frac{1}{b^4 r^4} + \mathcal{O}(r^{-5}) \\
		\sigma & = 1 -\frac{1}{6} C_{\Delta} |\langle {\rm{O}}_{\psi} \rangle|^2 \Delta r^{-2\Delta} 
+  {O}(r^{-2\Delta-2}) \\
		\rho & = r^{-\Delta} \left(C_{\Delta} |\langle {\rm{O}}_{\psi} \rangle| + {O}(r^{-2}) \right)\\
		G_0 & = \mu - \kappa^2 q_t r^{-2} + {O}(r^{-3})
	\end{split}
	\end{align}
	where $C_{\Delta}$ is a real number, ${\rm{O}}_{\psi}$ is the operator dual to $\psi$, and $q_t$ 
represents the total charge density as seen in the rest frame of the normal component; using the standard 
AdS/CFT prescription \cite{Witten:1998qj,Gubser:1998bc} to convert bulk to boundary quantities we find that 
$$
	q_t = - u^{\mu}J_{\mu}
$$ 
with $J_{\mu}$ the boundary theory charged current. Using \eqref{E:Q1def} together with \eqref{E:Q1Val} we find
	\begin{equation}
	\label{E:GDRelation}
		s T + \mu q_t = \frac{2}{b^4 \kappa^4} \equiv 4 P
	\end{equation}
which is the Gibbs Duhem relation for a conformal superfluid (see \eqref{E:gdrelation}).
\end{itemize}
%

%=====================================================================
\subsection{Adding in a small uniform superfluid velocity}
%=====================================================================
%
As we have explained above, in this section we focus on the collinear limit, 
i.e. the limit in which the normal and superfluid velocities are equal. 
Following a long tradition we do, however, wish to allow the derivatives 
of the normal and superfluid velocities to be independent variables. Thus while 
we can set the normal and superfluid velocities equal at a given point, 
it would be inconsistent to do the same in the neighborhood of that point. 
In order to implement the fluid gravity map, even in our limited context, 
we need more general stationary background solutions than the ones 
described in the previous paragraph. In particular, since we are interested only in first 
derivative corrections, we need control over background solutions 
in which the fluid is at rest and the superfluid in motion to linear order 
in the superfluid velocity. In this section we describe the relevant solutions.

A constant superfluid velocity is an $SO(3)$ vector. The solution that 
describes infinitesimal superfluid motion (at first order) is a regular, normalizable, translationally invariant fluctuation about 
the static background described in the previous subsection. A vector 
fluctuation involves linear fluctuations of the vector modes $g_{ti}$ and $G_i$. 
We find it more convenient to work in terms of the variables $g$ and $\gamma$ 
defined by 
$$G_i = -g \partial_i \phi ~~~~g_{ti} = -r^2 \gamma \partial_i \phi$$ 
where $\partial_i\phi$ is the expectation value of the spatial component of 
the Goldstone boson in the dual field theory (i.e. $\partial_{\mu} \phi$ is a 
function of the four boundary rather than the 5 bulk dimensions).

From the Einstein equations we obtain the following equation of motion for 
$\gamma$ at linear order 
\begin{equation}
\label{E:gammaEOM}
	\left(\frac{ \gamma' r^5 + G_0' g r^3 V_F}{\sigma} \right)' = 0\,.
\end{equation}
Integrating \eqref{E:gammaEOM} we obtain
\begin{equation}
\label{E:DefQ2}
	2 \kappa^2 Q_2 = \frac{ \gamma' r^5 + G_0' g r^3 V_F}{\sigma} 
\end{equation}
with $Q_2$ an integration constant. The asymptotic behavior of $\gamma$ can be related to thermodynamic 
quantities in the solution via 
\begin{equation}
\label{E:gammaNearb}
	\gamma = \frac{1}{2} (q_t-q) r^{-4} + {O}(r^{-5})\,.
\end{equation}
Integrating \eqref{E:DefQ2} near the boundary, it is not difficult to 
verify the form \eqref{E:gammaNearb} and to check that 
\begin{equation}
	Q_2 = q\,.
\end{equation}

The equation of motion for $g$ is obtained from the spatial components of the Maxwell equation and can also be written as a total derivative,
\begin{equation}
\label{E:gEOM}
	\left(\frac{ f r^3 (g G_0' - g' G_0) V_F}{\sigma}  + 2 \kappa^2 \gamma Q_1  \right)' =  2 \kappa^2 f'  Q_2\,.
\end{equation}
Integrating it once we get
\begin{equation}
\label{E:igEOM}
	\frac{ f r^3 (g G_0' - g' G_0) V_F}{\sigma}  + 2 \kappa^2 \gamma Q_1  = Q_3 + 2 \kappa^2 f  Q_2
\end{equation}
with $Q_3$ an integration constant. We will now argue that $Q_3$ vanishes. 
The argument follows by demanding regularity at the event horizon.
The fact that $r_h$ is an event horizon (and so that $dr$ is a null one-form 
at $r=r_h$) implies that 
\begin{equation}
	\gamma(r_h) = 0\,.
\end{equation}
Every term in \eqref{E:igEOM} vanishes at $r=r_h$ 
except $Q_3$, so we conclude that $Q_3 = 0$ for $T>0$. 
The near boundary expansion of $g$ then takes the form
\begin{equation}
\label{E:gNearb}
	g = 1-\frac{(q_t-q) \kappa^2}{\mu}r^{-2} + {O}(r^{-3})\,.
\end{equation}

The second order linearized equations \eqref{E:gammaEOM} and \eqref{E:gEOM} have four linearly independent solutions. 
One of these is
\begin{equation}
\label{E:soldef}
	g = 0 \qquad \gamma = \hbox{constant}
\end{equation}
which corresponds to a deformation of the dual field theory metric. Another solution is
\begin{equation}
	g = G_0 \qquad \gamma = \frac{Q_2}{Q_1} f
\end{equation}
which corresponds to a boost of the stationary solution. Of the remaining two linearly independent solutions only 
one can be chosen to be regular at the horizon. In terms of the two first order integrated equations 
\eqref{E:DefQ2} and \eqref{E:igEOM}, imposing $Q_2 \neq 0$ precludes \eqref{E:soldef} and imposing $Q_3 =0$ 
removes the solution which diverges at the horizon.

\subsection{Boosting}
So far we have worked in a frame where $u^{\mu} = (1,0,0,0)$. As explained in \cite{Herzog:2011ec} we can always boost to a frame where the metric and gauge field take the form 
\begin{align}
\begin{split}
\label{E:BoostedSol}
	ds^2 &= -r^2 f(r) u_{\mu}u_{\nu} dx^{\mu} dx^{\nu} + r^2 P_{\mu\nu}dx^{\mu}dx^{\nu} 
- 2\sigma u_{\mu} dx^{\mu} dr + 2 r^2 \gamma u_{(\mu}\bar{\zeta}_{\nu)} dx^{\mu}dx^{\nu} \\
	A_{\mu} & = G u_{\mu} dx^{\mu} + g \bar{\zeta}_{\mu}dx^{\mu} + G_5 dr\,,
\end{split}
\end{align}
with
\begin{align}
\begin{split}
\label{E:Ndef}
	G &= G_0 \\
	\lim_{r \to \infty} A_\beta &= a_{\beta} \\
	\bar{\zeta}^{\alpha} &= P^{\alpha\beta} (\partial_{\beta} \phi - a_{\beta})\,,
\end{split}
\end{align}
$a_{\beta}$ the external gauge field and $\phi$ the Goldstone boson. Our conventions here slightly differ from the main text where we've defined $\bar{\zeta} = - \zeta$, see equations \eqref{E:xidef}, \eqref{E:zetadef}.

In our current holographic formulation, it is convenient to use a frame where the energy momentum tensor, charged current and 
Josephson condition are given by 
\begin{align}
\begin{split}
\label{E:AdSframe}
	T^{\mu\nu} &= (\rho + P+f \mu^2) u^{\mu}u^{\nu} + \eta^{\mu\nu} P + 2 f \mu \bar{\zeta}_{(\mu}u_{\nu)} 
+ f \bar{\zeta}_{\mu}\bar{\zeta}_{\nu} + \bar{T}_{diss}^{\mu\nu}\\
	J^{\mu} &= q_t u^{\mu} + f \bar{\zeta}^{\mu} + \bar{J}_{diss}^{\mu} \\
	u^{\mu} \left(\partial_{\mu}\phi - a_{\mu}\right) &= -\mu + \bar{\mu}_{diss} \equiv -\mu_T
\end{split}
\end{align}
where
\begin{equation}
	q_t = q + f \mu\,,
\end{equation}
and our frame choice is $u_{\mu}\bar{T}_{diss}^{\mu\nu} = 0$ and $u_{\mu}\bar{J}_{diss}^{\mu} = 0$.\footnote{Actually, in our computation we are using a frame where
\begin{align*}
%\begin{split}
%\label{E:AdSframe}
	T^{\mu\nu} &= (\rho + P+f \mu^2) u^{\mu}u^{\nu} + \eta^{\mu\nu} P + 2 f \mu \bar{\zeta}_{(\mu}u_{\nu)} 
+ \frac{f \mu}{\mu_T} \bar{\zeta}_{\mu}\bar{\zeta}_{\nu} + \bar{T}_{diss}^{\mu\nu}\\
	J^{\mu} &= q_t u^{\mu} + \frac{f \mu}{\mu_T} \bar{\zeta}^{\mu} + \bar{J}_{diss}^{\mu} \\
	u^{\mu} \left(\partial_{\mu}\phi - a_{\mu}\right) &= -\mu + \bar{\mu}_{diss} \equiv -\mu_T\,.
%\end{split}
\end{align*}
The dissipative corrections $\bar{J}_{diss}$, $\bar{T}_{diss}$ and $\bar{\mu}_{diss}$ in this frame are different from $J_{diss}$, $T_{diss}$ and $\mu_{diss}$ used throughout this work. Since in this section we will only be computing the transport coefficients in the vector sector, the difference between the two frames is immaterial. The reader is referred to \cite{Herzog:2011ec,Bhattacharya:2011ee} for an extensive discussion.} One can check that expanding \eqref{E:AdSframe} to linear order in $\bar{\zeta}^{\mu}$ and neglecting $\bar{T}_{diss}^{\mu\nu}$, $\bar{J}_{diss}^{\mu}$ and $\bar{\mu}_{diss}$, the boundary theory energy momentum tensor, charged current and chemical potential associated with the solution \eqref{E:BoostedSol} are obtained.
%

%=============================================================================
\subsection{The gradient expansion}
%=============================================================================
%
We now have in hand all stationary solutions that will be needed for 
our analysis. 
As explained at the beginning of this section, the next step in our analysis is to allow the thermodynamic variables to depend on the spacetime coordinates and look for corrections to the metric, $\delta g_{mn}$, gauge field, $\delta A_m$ and scalar $\delta \psi$, so that the equations of motion \eqref{E:BulkEOM} are satisfied. Since we will be interested in the collinear limit, we will set $\bar{\zeta}=0$ but keep its derivatives non zero.  In this limit, the equations of motion for the corrections to the metric, gauge field and scalar naturally decompose themselves into scalar, vector and tensor equations under the $SO(3) \subset SO(3,1)$ symmetry retained by the background. We write the corrections to the metric in the form
\begin{multline}
	\delta g_{mn}\ dx^m dx^n = -2 u_{\mu} dx^{\mu} r \left(u^{\alpha}\partial_{\alpha} u_{\nu} 
+ \frac{1}{3} \partial_{\alpha} u^{\alpha} u_{\nu} \right) dx^{\nu} - 2 \delta \sigma\  u_{\mu} dx^{\mu} dr \\
	-r^2 \delta f\  u_{\mu} dx^{\mu} u_{\nu} dx^{\nu} + 2 r^2 \delta V_{\mu}\ u_{\nu} dx^{\mu}dx^{\nu} 
+ r^2 \delta \pi_{\mu\nu} \ dx^{\mu}dx^{\nu}\,,
\end{multline}
the corrections to the gauge invariant combination $G_m$ in the form
\begin{equation}
	\delta A_m dx^m = -\delta G\, u_{\mu} dx^{\mu} + \delta g_{\mu} dx^{\mu} + \delta G_5 dr \,,
\end{equation}
and corrections to the magnitude of the scalar as $\delta \rho$.

The equation of motion for $\delta G_5$ turns out to be an algebraic equation which can be solved. The remaining kinematic equations for $\delta G$, $\delta \rho$, $\delta f$ and $\delta \sigma$ are coupled. The vector sector contains two coupled equations involving $\delta V_{\mu}$ and $\delta g_{\mu}$ and the equation of motion for the tensor mode $\delta \pi_{\mu\nu}$ can be solved for since it is decoupled from the rest of the equations. We have verified that the five constraint equations imply energy momentum conservation and current conservation in the dual field theory. We have also checked that the ratio of the shear viscosity to entropy density retains its universal value as has been alluded to in \cite{Buchel:2010wf,Natsuume:2010ky,Herzog:2011ec}. 

Since we are interested only in the parity odd sector, and since in the collinear limit this sector receives contributions only from the vector modes, we can focus entirely on the equations of motion for $\delta V_{\mu}$ and $\delta g_{\mu}$ which read
\begin{subequations}
\label{E:fodelta}
\begin{align}
	\left(\frac{r^5 \delta V_{\mu} ' }{\sigma} - \frac{G' \delta g_{\mu} r^3 V_F}{\sigma} \right)' 
&= \mathbf{S}^{\delta V}_{\mu} \\
\label{E:idgeqn}
	\left(\frac{V_F f r^3 (\delta g_{\mu} G' - \delta g_\mu' G)}{\sigma} 
-  2 \kappa^2 Q_1 \delta V_{\mu} \right) ' & = \mathbf{S}^{\delta g}_\mu \,.
\end{align}
\end{subequations}
(Recall that $G=G_0$ from \eqref{E:Ndef}.)
Note that the homogeneous parts of these equations agree exactly 
with \eqref{E:gammaEOM} as expected.

After integrating these once we can eliminate $\delta V$ in place of $\delta g$ and write the resulting differential equation in Sturm-Liouville form,
\begin{equation}
\label{E:deltagO2}
	\left(\frac{r^3 f V_F}{\sigma} \delta g_{\mu}' \right)' - \left( \frac{r V_F^2 G^{\prime\,2}}{\sigma} 
+ 2 q^2 r \sigma V_{\psi} \rho^2 \right)\delta g_{\mu} 
	= 
	-\frac{\mathbf{S}^{\delta g}_{\mu}}{G} + \frac{2 \kappa^2 Q_1 \sigma}{r^5 g} \int_{r}^{\infty} 
\mathbf{S}^{\delta V}_{\mu}(x) dx
	\equiv
	\mathbf{S}^{\rm total}\,.
\end{equation}
In order to simplify the term linear in $\delta g_{\mu}$ we used the equation of motion for $G$.
The limits of integration on the right hand side of \eqref{E:deltagO2} have been determined by the boundary values of $\delta g_{\mu}$ and $\delta V_{\mu} $ and the requirement that we are in the frame described in \eqref{E:AdSframe}.

Since \eqref{E:deltagO2} is a linear second order differential equation we can use the method of Green's functions to solve it. Consider first the homogeneous version of \eqref{E:deltagO2}. We denote the two linearly independent solutions of the homogeneous equation by $a$ and $\hat{a}$ such that
\begin{align}
	a & = 1 + \mathcal{O}(r^{-2}) & a(r_h) & = a_h \\
	\hat{a} & = r^{-2} + \ldots & \hat{a}(r_h) & = \infty \,.
\end{align}
We find that
\begin{equation}
\label{E:hom}
	a = \frac{Q_1 g + Q_2 G}{Q_1 + Q_2 \mu}\,.
\end{equation}
The solution \eqref{E:hom} corresponds to a shift in the superfluid velocity
described in the previous subsection.
(One can use the method of variation of parameters to find $\hat{a}$ in integral form though we will not be needing it in our analysis.) The solution to 
\begin{equation}
\label{E:Greens}
	\frac{d}{dr} \left(\frac{r^3 f V_F}{\sigma} \frac{d \mathcal{G}(r,\tilde{r})}{dr}  \right) - \left( \frac{r V_F^2 G^{\prime\,2}}{\sigma} + 2 q^2 r \sigma V_{\psi} \rho^2 \right)\mathcal{G}(r,\tilde{r}) = \delta(r-\tilde{r})
\end{equation}
which vanishes at the asymptotically AdS boundary and also at the horizon is given by
\begin{equation}
	\mathcal{G}(r,\tilde{r}) = \begin{cases}
		-\frac{1}{2} a(\tilde{r}) \hat{a}(r) & r>\tilde{r} \\
		-\frac{1}{2} a({r}) \hat{a}(\tilde{r}) & r<\tilde{r} \,.
	\end{cases}
\end{equation}
Thus, if $S^{\rm total}(\infty) = 0$, a near boundary series expansion of $\delta g_{\mu}$ is given by
\begin{equation}
	\delta g_{\mu} = -\frac{1}{2 r^2} \int_{r_h}^{\infty} a \mathbf{S}^{\rm total}_{\mu} dx 
+ \frac{1}{2 r^2} \lim_{x \to \infty}  \left(  x^2 \frac{d}{dx} \mathbf{S}^{\rm total}_{\mu}(x)  \right) + \ldots \,.
\end{equation}
Then, according to the standard AdS/CFT prescription, the dissipative corrections to the charged current in the boundary theory take the form
\begin{equation}
\label{E:Jdiss}
	\bar{J}_{diss\,\mu} = -\frac{1}{2 \kappa^2} \left( \int_{r_h}^{\infty} a 
\mathbf{S}^{\rm total}_{\mu} dx - \lim_{x \to \infty}  \left(  x^2 \frac{d}{dx} \mathbf{S}^{\rm total}_{\mu}(x)  \right)  \right) \,. 
\end{equation}

Let us focus on parity odd contributions to $\mathbf{S}^{\rm total}$. we find that
\begin{equation}
	\mathbf{S}^{\rm total}_{\mu} =4 c \kappa^2 (G -\mu g) G' \omega_{\mu} -2 c \kappa^2 g G' B_{\mu}  
+ \left( \substack{\hbox{\small parity even} \\ \hbox{\small terms}} \right)
\end{equation}
where $\omega^{\mu}$ and $B^{\mu}$ were defined in Table \ref{tb:PEoneD}. We reproduce their expression here for convenience
\begin{equation}
	\omega^{\mu} \equiv \frac{1}{2}\epsilon^{\mu\nu\rho\sigma} u_{\nu}\partial_{\rho}u_{\sigma} \qquad
	B^{\mu} \equiv \frac{1}{2} \epsilon^{\mu\nu\rho\sigma} u_{\nu} F_{\rho \sigma} \,.
\end{equation}
Hence
\begin{multline}
\label{E:JdissResult}
	\bar{J}_{diss}^{\mu}  = \frac{c}{Q_1 + Q_2 \mu} \left(  \int_{r_h}^{\infty} (Q_1 g + Q_2 G) g G' dr B^{\mu} 
	+ \int_{r_h}^{\infty} 2 (\mu g - G) (Q_1 g+Q_2 G) G' dr \omega^{\mu}  \right) 
	\\
	+ \left( \substack{\hbox{\small parity even} \\ \hbox{\small terms}} \right)\,.
\end{multline}
Following the work of \cite{Bhattacharyya:2008xc}, the contribution of the parity odd terms to the entropy current come from the horizon value of $\delta V_{\mu}$, i.e., if we denote the entropy current by $J_S^{\mu}$ then
\begin{equation}
	J_S^{\mu} =  s \delta V^{\mu}(r_h)  +  \left( \substack{\hbox{\small parity even} 
\\ \hbox{\small terms}} \right)\,.
\end{equation}
It remains to evaluate the parity odd contributions to $\delta V^{\mu}$ at the horizon. Integrating \eqref{E:idgeqn} once we find that
\begin{equation}
\label{E:idgeqnV2}
		\frac{V_F f r^3 (\delta g_{\mu} G' - \delta g_\mu' G)}{\sigma} -  2 \kappa^2 Q_1 \delta V_{\mu}  
= \int_{r_h}^r \mathbf{S}^{\delta g}_\mu + \delta Q
\end{equation}
with $\delta Q$ an integration constant. Evaluating \eqref{E:idgeqnV2} at the horizon and at the boundary, we find that
\begin{align}
\begin{split}
\label{E:JSResult}
	\delta V^{\mu}(r_h) &= -\frac{\mu \bar{J}^{\mu}_{diss}}{Q_1} + \frac{\int_{r_h}^{\infty} 
\mathbf{S}^{\delta g} dr}{2 Q_1 \kappa^2} + \left( \substack{\hbox{\small parity even} \\ \hbox{\small terms}} \right)\\
	&=-\frac{\mu}{T} \bar{J}_{diss}^{\mu} + \frac{c}{T } \left( \int_{r_h}^{\infty} 2 (\mu g-G) G G' dr  
\omega^{\mu} + \int_{r_h}^{\infty} g G G' dr B^{\mu} \right) + \left( \substack{\hbox{\small parity even} 
\\ \hbox{\small terms}} \right)\,.
\end{split}
\end{align}
Restricting ourselves to parity odd contributions, the first term on the right hand side of the first line of \eqref{E:JSResult}  represents the contribution of the canonical part of the entropy current $\Jsc$. The second term on the right hand side of the first line of \eqref{E:JSResult} will give us the corrections to $\Jsc$.

Following the notation in \eqref{E:Rdef}, we denote
\begin{equation}
	R = \frac{q}{\rho+P}\,
\end{equation}
and write the parity odd transport coefficients and the corrections to the entropy current as in \eqref{E:zeta0Sol}
\begin{align}
\begin{split}
	\bar{J}^{\mu}_{diss} & =- \tilde{\kappa}_{\omega} \omega^{\mu} - \tilde{\kappa}_B B^{\mu} 
+ \left( \substack{\hbox{\small parity even} \\ \hbox{\small terms}} \right)\\
	J_S^{\mu} & = -\frac{\mu}{T} \bar{J}_{diss}^{\mu} + \sigma_B \omega^{\mu} + \sigma_{\omega} \omega^{\mu} 
+ \left( \substack{\hbox{\small parity even} \\ \hbox{\small terms}} \right)\,.
\end{split}
\end{align}
Then, \eqref{E:JSResult} and \eqref{E:JdissResult} imply
\begin{subequations}
\label{E:oddRules}
\begin{align}
	\tilde{\kappa}_{B} &= - c \int_{r_h}^{\infty}g^2 G' +R (G-g\mu)g G' dr \\
	\tilde{\kappa}_B &= 2 c \int_{r_h}^{\infty} (G-\mu g) g G' + R (G-\mu g)^2 G' dr \\
	\sigma_{\omega} & = \frac{2 c}{T} \int_{r_h}^{\infty} (G-\mu g) G G' dr \\
	\sigma_{B} & =  \frac{c}{T } \int_{rh}^{\infty} g G G' dr\,.
\end{align}
\end{subequations}

While we can not solve the integrals in \eqref{E:oddRules} explicitly, we find that the following relations are satisfied: 
\begin{subequations}
\label{E:relations}
\begin{align}
\label{E:relation1}
	\frac{1}{2}\sigma_\omega - \mu \sigma_{B} & = -\frac{c \mu^3}{3 T} \\
	\frac{1}{2} \tilde{\kappa}_{\omega} - \mu \tilde{\kappa}_B + T \left(1-\mu R\right) \sigma_B & 
= -\frac{c R \mu^3}{3}\,.
\end{align}
\end{subequations}
(To obtain \eqref{E:relations} from \eqref{E:oddRules} we used $G(\infty) = \mu$ and $G(r_h)=0$.) 
The first of these relations is precisely the universal prediction 
 \eqref{E:zeta0sigma} between entropy current coefficients predicted in the 
last section. 
Using  \eqref{E:zeta0Result} and \eqref{E:zeta0sigma} we can also 
obtain holographic expressions  for $s_9$, $\sigma_1$, $\sigma_8$ and $\sigma_{10}$ 
\begin{subequations}
\begin{align}
	s_9 & = 0 \\
	\sigma_{1} & = \frac{c }{ T^2} \left(\mu^2 - \int_{r_h}^{\infty} g(4 G - \mu g) G' dr \right) \\
	\sigma_8 & = \frac{c}{2 T}\left(-\mu + \int_{r_h}^{\infty} g^2 G' dr \right) \\
	\sigma_{10} & = -\frac{c}{T^2} \int g (G-\mu g ) G' dr\,.
\end{align}
\end{subequations}

To study the values of $\tilde{\kappa}_{\omega}$, $\tilde{\kappa}_B$, $\sigma_{\omega}$ and $\sigma_B$ at the phase transition we set $g=1$. Then, all the integrals in \eqref{E:oddRules} can be carried out explicitly. We find that the resulting expressions exactly match the zero superfluid density values of  $\tilde{\kappa}_{\omega}$, $\tilde{\kappa}_B$, $\sigma_{\omega}$ and $\sigma_B$ studied in \cite{Erdmenger:2008rm,Banerjee:2008th,Son:2009tf}, implying that these coefficients are continuous across the phase transition. 
%
%
%**************************************************
 \section{Discussion}
 \label{S:discussion}
%**************************************************
%
In this paper we have described a framework for describing superfluid hydrodynamics at 
first order in the derivative expansion. We have determined the most general
form of the hydrodynamical equations that are consistent with Lorentz 
invariance, the Onsager principle, and the requirement that the second 
law of thermodynamics apply in every conceivable situation. We begin this 
section by summarizing our results, and then turn to a discussion of possible
applications and extensions.

We have found that the most general equations of Lorentz and time-reversal invariant but 
non-parity conserving superfluids requires the specification of twenty 
parameters. Fourteen of these are associated with the parity even sector and 
were described already in \cite{Bhattacharya:2011ee}, generalizing 
an earlier 13 parameter framework spelt out in \cite{Clark,Putterman}). 
In this work we have found that  six more parameters need to be specified 
in the parity odd sector of the theory. All six of these parameters are 
non dissipative;  they drop off from the expression of the 
divergence of the entropy current, and are unconstrained by inequalities. 
Four of these six parameters ($\tilde{\eta}$, $\tilde{\kappa}_{11}$,
$\tilde{\kappa}_{22}$ and $\tilde{\kappa}_{12} =\tilde{\kappa}_{21}$ in the language of the main text) are relatively 
simple. $\tilde{\eta}$ is the transport coefficients which is associated with the 
pseudo-tensor mode
\begin{subequations}
\begin{equation}
\label{E:teta}
	\tilde{\mathcal{T}}^{\mu\alpha} 
= \epsilon^{\mu\nu\rho\sigma}u_{\nu}\zeta_{\nu} \sigma^{u\phantom{\sigma}\alpha}_{\sigma}
	+
	\epsilon^{\alpha\nu\rho\sigma}u_{\nu}\zeta_{\nu} \sigma^{u\phantom{\sigma}\mu}_{\sigma}
\end{equation}
while $\tilde{\kappa}_{11}$,
$\tilde{\kappa}_{22}$ and $\tilde{\kappa}_{12} =\tilde{\kappa}_{21}$ are the coefficient of 
 the pseudo vector modes
\begin{align}
\label{E:tk11}
	\tilde{\mathcal{V}}_1 &= \epsilon^{\mu\nu\rho\sigma}u_{\nu}\zeta_{\rho}
\left(\partial_{\sigma}\frac{\mu}{T} - \frac{E_{\sigma}}{T}\right) \\
\label{E:tk22}
	\tilde{\mathcal{V}}_2 & = \epsilon^{\mu\nu\rho\sigma}u_{\nu}\zeta_{\rho}\zeta^{\alpha}\sigma_{\alpha\sigma}\,.
\end{align}
\end{subequations}
that appear in the two field redefinition invariant combinations of vectors, in the corrections 
to the stress tensor and charge current.

Note that $\tilde{\eta}$ closely resembles the hall viscosity in $2+1$ dimensional theories \cite{PhysRevLett.75.697, Saremi:2011ab}. There, when parity is violated it is possible to have a transport coefficient $\eta_H$ associated with the tensor mode
\begin{equation}
	\epsilon^{\mu\nu\rho}u_{\nu}\sigma_{\rho}^{\phantom{\rho}\alpha} 
+ \epsilon^{\alpha\nu\rho}u_{\nu}\sigma_{\rho}^{\phantom{\rho}\mu}\,.
\end{equation}
The hall viscosity is known to be associated with the ground state's intrinsic angular momentum. It would be interesting to see whether $\tilde{\eta}$
 is also associated with similar quantities.

The remaining two non dissipative constitutive parameters, $\sigma_8$ and 
$\sigma_{10}$, multiply relatively complicated expressions in the 
constitutive relations.  These two parameters determine the eight transport coefficients associated with the pseudo scalars $\omega \cd \xi$ and 
$B \cd \xi$ which appear in the constitutive relations, and the transport 
coefficients of the pseudo vectors $\str \partial_{\mu} T$, $\str \partial_{\mu} \frac{\mu}{T}$,  $\star \partial_{\mu} \frac{\zeta^2}{T^2}$, $\omega^{\mu}$ and $B^{\mu}$.
It is possible that these expressions admit 
significant simplifications when  expressed in terms of variables that 
might turn out to be more natural than those adopted in this paper. 
$\sigma_8$ and $\sigma_{10}$ also parameterize deviations of the entropy 
current away from its canonical form.

% The sixth parity odd parameter is dissipative since it enters into 
% the formula for entropy production. In the main text we have denoted this parameter by $\tilde{\kappa}_{21}$ (and also $\tilde{\kappa}_{12} = - \tilde{\kappa}_{21}$). It is associated with the pseudo vector modes \eqref{E:tk11} and \eqref{E:tk22}. While undetermined, it, and the transport coefficients $\kappa_{ij}$ $i,j=1,2$ associated with the parity even sector, need to satisfy the inequality 
% \begin{equation*}
% \kappa_{11}\kappa_{22} \geq \kappa_{12}^2 + \zeta^2 \tilde{\kappa}_{12}^2\,.
% \end{equation*} 
% Note that this equation mixes parity odd and parity even coefficients.

The triangle anomaly, if it exists, enters the expression for the 
pseudo scalar and pseudo vector coefficients whose value depends on $\sigma_8$ and 
$\sigma_{10}$. It is interesting to compare this result with that of 
\cite{Son:2009tf} which studied (non-superfluid) hydrodynamics in the 
presence of triangle anomalies. There, the only addition to the parity-odd sector are the two pseudo vectors $\omega^{\mu}$ and $B^{\mu}$ whose transport coefficients are completely fixed in terms of the anomaly and a single 
integration constants.  When considering superfluids the functional dependence of the transport coefficients of $\omega^{\mu}$ and $B^{\mu}$ on $T$, $\mu$ and $\zeta$ is completely arbitrary. This novel feature opens the possibility for interesting physical phenomenon in a parity violating theory, even in the absence of an anomaly which is an intrinsically relativistic effect. 

The most immediate application of our work would be to the modeling of the 
long distance behaviors of real world superfluids (or superconductors) 
whose hydrodynamics violates parity. We do not yet know of any candidate 
experimental systems of this sort. We do note, however, that non centro symmetric
superconductors live on parity violating lattices. It would be exciting to 
investigate whether parity violation in these (or analogous systems) could
lead to experimentally observable nonzero values for any of the 6 parameters
described above. 

In this paper we have worked out the general theory of superfluidity in 4 
dimensions. Several aspects of our analysis depended on the existence of 
an $\epsilon$ tensor with 4 indices. It seems likely that the theory of 
superfluidity in 5 and higher dimensions will differ in qualitative aspects
from the theory we have worked out. It would be interesting to flesh this out. 

We have already verified certain aspects of our general construction of the 
equations of superfluid hydrodynamics using explicit computations within the 
AdS/CFT framework. However all the computations reported in this paper 
work in the collinear limit of vanishing superfluid velocity. It would be 
instructive to demonstrate that all the numerous other relations, implied by 
our work, between transport coefficients that do not preserve $SO(3)$ 
invariance, are also borne out by AdS/CFT computations. In this context 
we point out that the generality of the expected results suggests that their 
derivation may also be carried out for a generic holographic superfluid (perhaps along the 
lines of Section \ref{S:Holographic}) and should not  involve 
the details of the corresponding dynamical systems.

An interesting feature of our AdS/CFT analysis is that in the collinear limit all the transport coefficients in the parity odd sector of the theory are continuous at the phase transition (i.e., they match their value in the uncondensed phase where $f=0$). This is similar to the behavior of the shear viscosity $\eta$ and diffusion coefficient $\kappa$ of the parity even sector, but differs from the divergent behavior of the transport coefficient associated with the  scalar $\partial_{\mu} (f \xi^{\mu})$ \cite{Herzog:2011ec}. It would be interesting to understand this result in terms of the theory of dynamical critical exponents \cite{RevModPhys.49.435}.

It would also be interesting to work out the properties of stationary superfluid 
flows (like rotating superfluids on an $S^3$) and to investigate how the 
properties of such configurations are affected by the parity odd non dissipative
terms that can be `turned on' in such configurations. Such configurations 
would be dual, under the AdS/CFT correspondence, to rotating hairy black 
holes in global $AdS$ space.

To end this paper let us highlight a structural 
aspect of our analysis that we find quite remarkable. 
For this purpose it is sufficient to focus on the case of parity and time-reversal invariant 
superfluids. The total number of constitutive coefficients allowed, merely
on symmetry grounds, is 50. In this paper we have shown that the 
requirement that the equations that follow from these constitutive relations
are consistent with positivity of divergence of {\it any} entropy current 
(plus time-reversal invariance and the Onsager relations) cuts down the number of constitutive 
coefficients from 50 to 14. In other words, the second law of thermodynamics
gives very powerful and precise constraints on dynamical equations. Given 
the duality between fluid dynamics and gravity, it is natural to wonder 
whether similar results may also be true for extensions of the theory of 
gravity. We leave detailed investigation of this exciting possibility to 
future work.
%
%
%
%

%
%
%
%
%&&&&&&&&&&&&&&&&&&&&&&&&&&&&&&&&&&&&
\acknowledgments
%&&&&&&&&&&&&&&&&&&&&&&&&&&&&&&&&&&&&
%
We would like especially to thank D. Son for very useful discussions 
and correspondence. We would like to acknowledge useful discussions and 
correspondences with
K.~Damle, D.~Dhar,  R.~Gopakumar, C.~Herzog, R.~Loganayagam,  D.~Son, V.~Tripathy and  S.~Wadia.
The work of S.~M. was supported in part by a Swarnajayanti Fellowship.
J.~B., S.~B., and S.~M. would also like to acknowledge their debt to the
steady and generous support of the people of India for research in basic science. S. M. Would
like to thank the organizers of the Great Lakes Strings Meeting, Iranian
school for  applications of AdS/CFT and the Solvay conference for hospitality
during the completion of this work. A.~Y. is supported in part by the
Department of Energy under Grant No. DE-FG02-91ER40671.
%

%
%&&&&&&&&&&&&&&&&&&&&&&&&&&&&&
\appendix
%&&&&&&&&&&&&&&&&&&&&&&&&&&&&&
%
%******************************************************************************************
\section{The linear independence of data for the parity even superfluid}\label{linind}
%******************************************************************************************
%
%==========================================================================================
\subsection{The linear independence of first order terms}
%==========================================================================================
%
%----------------------------------------------------------------
\subsubsection{The vector sector}
%----------------------------------------------------------------
%
The equations of motion in the vector sector are 
\begin{equation}\label{vse}
 \begin{split}
  \tilde P_{\mu \beta} \partial_{\nu} T^{\nu \beta} &= \tilde P_{\mu \beta}F^{\beta \nu} J_{\nu}\\
  \tilde P^{\mu \beta} u^{\nu} \left( \partial_{\beta} \xi_{\nu} - \partial_{\nu} \xi_{\beta}\right)
 &= \tilde P^{\mu \beta} E_{\beta}\\
   \tilde P^{\mu \beta} \xi^{\nu} \left( \partial_{\beta} \xi_{\nu} - \partial_{\nu} \xi_{\beta}\right)
 &= \tilde P^{\mu \beta} F_{\beta \nu} \xi^{\nu}\,.
 \end{split}
\end{equation}
A basis of ten one derivative vectors (before using the equations of motion) 
was listed in table \ref{tb:PESoneD}. It is given by 
\begin{equation}\label{aq}
\begin{split}
&\tilde P^{\mu \beta} (u\cd\partial) u_{\beta},~ \tilde P^{\mu \beta} (u\cd\partial)\xi_{\beta},~
\tilde P^{\mu \beta} (\xi\cd\partial) u_{\beta},~ \tilde P^{\mu \beta} (\xi\cd\partial)\xi_{\beta},~
\tilde P^{\mu \beta} \partial_{\beta} \left( \frac{\mu}{T}\right), \\
&\tilde P^{\mu \beta} \partial_{\beta} T,~\tilde P^{\mu \beta} \partial_{\beta} \left( \frac{\xi}{T}\right),~
\tilde P^{\mu \beta} \xi^{\nu} \partial_{\beta} u_{\nu},~
\tilde P^{\mu \beta} E_{\beta},~ \tilde P^{\mu \beta} F_{\beta \nu} \xi^{\nu}\,.
\end{split}
\end{equation}
The quantities in \eqref{aq} are not all on-shell inequivalent as they are constrained by the relations \eqref{vse}. In this subsection we will argue that it is 
consistent to choose the seven vectors listed in the third column of 
Table \ref{tb:PESoneD} as independent vector data. That is, we will show that it
is possible to use the equations \eqref{vse} to solve for
$\tilde P^{\mu \beta} \partial_{\beta} T,
~\tilde P^{\mu \beta} \partial_{\beta} \left( \frac{\xi}{T}\right),~
\tilde P^{\mu \beta} \xi^{\nu} \partial_{\beta} u_{\nu}$ 
in terms of
\begin{equation}\label{indvec}
\begin{split}
&\tilde P^{\mu \beta} (u\cd\partial) u_{\beta},~ \tilde P^{\mu \beta} (u\cd\partial)\xi_{\beta},~
\tilde P^{\mu \beta} (\xi\cd\partial) u_{\beta},~ \tilde P^{\mu \beta} (\xi\cd\partial)\xi_{\beta},~
\tilde P^{\mu \beta} \partial_{\beta} \left( \frac{\mu}{T}\right), \\
&
\tilde P^{\mu \beta} E_{\beta},~ \tilde P^{\mu \beta} F_{\beta \nu} \xi^{\nu}
\end{split}
\end{equation}

If we rewrite the equations of motion in \eqref{vse} in terms of the quantities in \eqref{aq} we find
\begin{equation}\label{eeom}
 \begin{split}
  \tilde P^{\mu \beta} \left((P + \rho) (u\cd\partial) u_{\beta} 
+ \partial_T P \partial_{\beta} T + \partial_{\frac{\mu}{T}}\partial_{\beta}\frac{\mu}{T} + 
\partial_{\frac{\xi}{T}} \partial_{\beta} \frac{\xi}{T}  + f (\xi. \partial) \xi_{\beta} \right) = & 
  \tilde P^{\mu \beta} \left( q E_{\mu} - f F_{\mu \nu} \xi^{\nu}\right) \\
\tilde P^{\mu \beta} \left( (u\cd\partial) \xi_{\beta} - T \partial_{\beta} \frac{\mu}{T} 
- \frac{\mu}{T} \partial_{\beta} T + \xi^{\nu} \partial_{\beta} u_{\nu} \right) =&
  \tilde P^{\mu \beta} E_{\beta} \\
\tilde P^{\mu \beta} \left( (\xi. \partial) \xi_{\beta} + T \xi \partial_{\beta} \frac{\xi}{T} 
+ \frac{\xi}{T}  \partial_{\beta} T \right) =& \tilde P^{\mu \beta} F_{\beta \nu} \xi^{\nu}.
 \end{split}
\end{equation}
It is possible to use \eqref{eeom} to solve for the scalars listed in 
\eqref{indvec} if and only if the $3\times 3$ matrix of the three 
vectors $\tilde P^{\mu \beta} \partial_{\beta} T,
~\tilde P^{\mu \beta} \partial_{\beta} \left( \frac{\xi}{T}\right),~
\tilde P^{\mu \beta} \xi^{\nu} \partial_{\beta} u_{\nu}$ in the three 
equations \eqref{eeom} has nonzero determinant. This $3 \times 3 $ matrix 
is given by 
\begin{equation}
M_{(v)} = \left( \begin{array}{ccc}
  \partial_T P & ~~\partial_{\xi/T} P~~  & 0 \\
 -\mu/T & 0 & 1 \\
\xi^2/T & T \xi & 0 \\
 \end{array} \right)
\end{equation}
and its determinant is given by 
\begin{equation}
 Det\left(M_{(v)}\right) = \xi \left( \frac{\xi}{T} \partial_{\frac{\xi}{T}} P - T \partial_T P \right)\,.
\end{equation}
It is nonzero for a generic functional form for $P(T, \mu, \xi)$. We conclude
that the vectors \eqref{indvec} form a basis for onshell independent one 
derivative vectors. 
%
%-------------------------------------------------
\subsubsection{The scalar sector}
%-------------------------------------------------
%
The equations of motion in the scalar sector are given by
\begin{equation}\label{sseom}
 \begin{split}
  \xi_{\nu} \partial_{\mu} T^{\mu \nu} &= q E\cd \xi \\
  u_{\nu} \partial_{\mu} T^{\mu \nu} &= f E\cd \xi \\
  \partial_{\mu} J^{\mu} &= c E\cd B \\
 u^{\nu} \xi^{\mu} \left( \partial_{\mu} \xi_{\nu}- \partial_{\nu} \xi_{\mu}\right) &= E \cd \xi\,.
 \end{split}
\end{equation}
A basis of 11 one derivative vectors (before using the equations of motion) 
was listed in Table \ref{tb:PESoneD}. We denote them by 
$\{ {\cal L}^{(a)}_j, S^{(a)}_i \} $ for the first set of on shell independent scalars 
and $\{ {\cal L}^{(b)}_j, S^{(b)}_i \} $ for the second set. Here $j$ runs from 1 to 4 and 
$i$ runs from 1 to 7. We have used the notation in Table \ref{tb:fdsv}. The new quantities $ {\cal L}^{(b)}_j, {\cal L}^{(b)}_j$ are defined as follows
\begin{align}
\label{depdata}
 {\cal L}^{(a)}_1 &= u\cd\partial \Sigma_1 , &{\cal L}^{(b)}_i=\xi\cd\partial \Sigma_1, \\ \notag
 {\cal L}^{(a)}_2 &= u\cd\partial \Sigma_2, &{\cal L}^{(b)}_i=\xi\cd\partial \Sigma_2, \\ \notag
 {\cal L}^{(a)}_3 &= u\cd\partial \Sigma_3, &{\cal L}^{(b)}_i=\xi\cd\partial \Sigma_3, \\ \notag
 {\cal L}^{(a)}_4 &= \xi^{\mu} u\cd\partial u_{\mu}, &{\cal L}^{(b)}_i= \xi^{\mu} \xi\cd\partial u_{\mu}\,.
\end{align}
The quantities defined in \eqref{depdata} are the dependent data for the two choices of
bases among the on-shell inequivalent quantities. These quantities are to be determined 
by the equation of motion \eqref{sseom} in terms of the dependent quantities $S_i$. Note that the sets 
$\{ {\cal L}^{(a)}_i, S^{(a)}_i \} $ and $\{ {\cal L}^{(b)}_i, S^{(b)}_i \} $
are different partitioning of the same set of quantities.

The equation of motion in \eqref{sseom} expressed in terms of the quantities in \eqref{depdata} has the 
form
\begin{equation}\label{eomsch}
\begin{split} 
&\sum_{i=1}^7 ({\bf e}^{(a)}_i)_p ~S^{(a)}_i + \sum_{j=1}^{4} (\ell^{(a)}_j)_{p} ~{\cal L}^{a}_{j} = 0. \\ 
&\sum_{i=1}^7 ({\bf e}^{(b)}_i)_p ~S^{(b)}_i + \sum_{j=1}^{4} (\ell^{(b)}_j)_{p} ~{\cal L}^{b}_{j} = 0. 
\end{split}
\end{equation}
In the equations above the index $p$ runs from 1 to 4 denoting the 4 equations in \eqref{sseom}. Again note 
that both the equations in \eqref{eomsch} refer to the same set of equations.
We find it convenient to define the new set of quantities
\begin{equation}
 A = - \frac{\chi^2}{T^2(\nu^2 - \chi^2)}; 
 \quad B = -\frac{1}{T^2(\mu^2 - \xi^2)}; 
 \quad C= -\frac{\nu}{T(\nu^2 - \xi^2)}. 
\end{equation}
so that the projector 
\begin{equation}
 \tilde P^{\mu \nu} = \eta^{\mu \nu} + A u^{\mu} u^{\nu} + B \xi^{\mu} \xi^{\nu} 
+ C \left( \xi^{\mu} u^{\nu} + u^{\mu} \xi^{\nu} \right).
\end{equation}
The coefficients in \eqref{eomsch} are given by
\begin{align} \label{indcoef}
 ({\bf e}^{(a)}_1)_1 &= -(P+\rho), & ({\bf e}^{(a)}_2)_1 &= f \nu , \\ \notag
 ({\bf e}^{(a)}_3)_1 &= B (P+\rho) + C f \mu - f, &
 ({\bf e}^{(a)}_4)_1 &= B \xi T f \mu + \mu \partial_\chi f, \\ \notag
({\bf e}^{(a)}_5)_1 &= \mu \partial_\nu f -CTf\mu +fT,
&({\bf e}^{(a)}_6)_1 &= \mu \partial_T f + 2 f \nu, \\ \notag
({\bf e}^{(a)}_7)_1 &= -f \\ \notag
({\bf e}^{(a)}_1)_2 &= (P+\rho) \mu, & ({\bf e}^{(a)}_2)_2 &= - f \chi^2 T, \\ \notag
({\bf e}^{(a)}_3)_1 &
= - (\mu B (\rho +P) + C T f \xi^2), &
({\bf e}^{(a)}_4)_2 &= \partial_\chi P - \xi^2 \partial_\chi f - \xi f T - B \xi^3 T f, \\ \notag
({\bf e}^{(a)}_5)_2 &= \partial_\nu P - \xi^2 \partial_\nu f +C T f \xi^2,
& ({\bf e}^{(a)}_6)_2 &=\partial_T P - \xi^2 \partial_T f - 2 f \chi^2 T, \\ \notag
({\bf e}^{(a)}_7)_2 &= - q, \\ \notag
({\bf e}^{(a)}_1)_3 &= q,& ({\bf e}^{(a)}_2)_3 &= \frac{f}{T} \\ \notag
({\bf e}^{(a)}_3)_3 &= -(B q +C f), & ({\bf e}^{(a)}_4)_3 &= - \partial_\chi f - B \xi f T \\ \notag
({\bf e}^{(a)}_5)_3 &= \partial_\nu f + C T f, & ({\bf e}^{(a)}_6)_3 &= -\partial_T f - \frac{f}{T} \\ \notag
({\bf e}^{(a)}_7)_3 &= 0, \\ \notag
({\bf e}^{(a)}_1)_4 &= 0, & ({\bf e}^{(a)}_2)_4 &=0, \\ \notag
({\bf e}^{(a)}_3)_4 &= -1 , & ({\bf e}^{(a)}_4)_4 &= 0 \\ \notag
({\bf e}^{(a)}_5)_4 &= T, & ({\bf e}^{(a)}_5)_6 &= \nu, \\ \notag
({\bf e}^{(a)}_7)_4 &= -1.
\end{align}
and
\begin{align}\label{depcoef}
 (\ell^{(a)}_1)_{1} &= C f \mu \xi T - \partial_\chi \rho, & 
(\ell^{(a)}_2)_{1} &= -(A f \mu T + \partial_\nu \rho), \\ \notag
(\ell^{(a)}_3)_{1} &= -\partial_T \rho, & (\ell^{(a)}_4)_{1} &= A f \mu + C(P+\rho), \\ \notag
(\ell^{(a)}_1)_{2} &= \mu \partial_\chi (P+\rho) - C f \xi^3 T, &
(\ell^{(a)}_2)_{2} &= \mu \partial_\nu (P+\rho) + T A \xi^2 f, \\ \notag
(\ell^{(a)}_3)_{2} &= \mu \partial_T (P+\rho), &
(\ell^{(a)}_1)_{2} &= (P+\rho) - C\mu (P+\rho)- A f \xi^2 \\ \notag
(\ell^{(a)}_1)_{3} &= \partial_\chi q -C \xi f T, &
(\ell^{(a)}_2)_{3} &= \partial_\nu q + ATf, \\ \notag
(\ell^{(a)}_3)_{3} &= \partial_T q, & (\ell^{(a)}_4)_{3} &= -(C q + A f) \\ \notag
(\ell^{(a)}_1)_{4} &= \xi T,&(\ell^{(a)}_2)_{4} &= 0 , \\ \notag
(\ell^{(a)}_3)_{4} &= \chi^2 T,& (\ell^{(a)}_4)_{4} &= 0 . \\ \notag
\end{align}
The other set of coefficients, with index ($b$), can be read from \eqref{indcoef} and \eqref{depcoef} using
\begin{equation}
\begin{split}
 (\ell^{(b)}_1)_{i} =& ({\bf e}^{(a)}_4)_i ;
~(\ell^{(b)}_2)_{i}= ({\bf e}^{(a)}_5)_i;
~(\ell^{(b)}_3)_{i}=({\bf e}^{(a)}_6)_i;
~(\ell^{(b)}_4)_{i}=({\bf e}^{(a)}_3)_i;\\ 
 ({\bf e}^{(b)}_4)_i =& (\ell^{(a)}_1)_{i}  ;
~({\bf e}^{(b)}_5)_i = (\ell^{(a)}_2)_{i};
~({\bf e}^{(b)}_6)_i=(\ell^{(a)}_3)_{i};
~({\bf e}^{(b)}_3)_i=(\ell^{(a)}_4)_{i};\\
\end{split}
\end{equation}
We can express all the derivatives in \eqref{indcoef} and \eqref{depcoef} 
as derivatives of a single function, say, the pressure.
Thermodynamic relations that enable us to do so are
\begin{equation}\label{threl}
 q = \frac{1}{T} \partial_{\mu/T} P~;~~ f = \frac{1}{T \xi} \partial_{\xi/T} P~; ~~~
\rho=-P+T \partial_T P -\frac{\xi}{T} \partial_{\frac{\xi}{T}} P.
\end{equation}

We make the following observations:
\begin{itemize} 
\item[a)]
We can use the equations of motion \eqref{eomsch}
to solve for the 4 scalars $\xi^{\mu} u\cd\nabla u_{\mu}$, $u\cd\partial \Sigma_i$ ($i=1,\ldots,3$)
in terms of the 7 independent scalars in the 3rd column of the first row of 
Table \ref{tb:PESoneD}. This is possible if and only if the $ 4 \times 4$ matrix of coefficients 
of the four quantities in the first equation in \eqref{eomsch} has nonzero determinant. 
This matrix is given by  
\begin{equation}\label{mat1}
 M^{(a)}_{ij} = (\ell^{(a)}_i)_j;
\end{equation}
\item[b)] We can use the equations of motion 
\eqref{eomsch} to solve for the quantities
$\xi^{\mu} \xi\cd\partial u_{\mu}$, $\xi\cd\partial \Sigma_i$
in terms of the 7 independent scalars in the 3rd column of the second row of 
Table \ref{tb:PESoneD}. This is possible if and only if the $ 4 \times 4$ matrix of coefficients 
of the 4 quantities in the second equation in \eqref{eomsch} has nonzero determinant. 
This matrix is given 
\begin{equation}\label{mat2}
 M^{(b)}_{ij} = (\ell^{(b)}_i)_j;
\end{equation}
 \end{itemize}
The relations \eqref{threl} allow us to express the matrices \eqref{mat1} and \eqref{mat2} in terms of the pressure. Using several reasonable equations of state we have used \verb+Mathematica+ to  verify  that the determinant of the matrices in \eqref{mat1} and \eqref{mat2} is generically non-zero.
%

%==========================================================================================
\subsection{The linear independence of the second order terms}
%==========================================================================================
%
A list of second order scalar data, the second order equations of motion and a  choice of second order independent scalar data can be found in Table \ref{tb:std}.
The second order scalar equations that follows from the first order vector equations are 
\begin{equation}
 \begin{split}
  \nabla_{\alpha}\left(\tilde P^{\alpha \mu}u^{\nu}\left( \partial_{\mu} \xi_{\nu}-\partial_{\mu} \xi_{\nu}\right) 
\right)&= \nabla_{\alpha}\left(\tilde P^{\alpha \mu}E_{\mu} \right),\\
\nabla_{\mu}\left(\tilde P^{\mu \nu} \nabla_{\beta} T^{\beta}_{~\nu} \right)
&= \nabla_{\mu}\left(\tilde P^{\mu \nu} F_{\nu \beta} J^{\beta} \right)\\
\nabla_{\alpha}\left(\tilde P^{\alpha \mu}\xi^{\nu}
\left( \partial_{\mu} \xi_{\nu}-\partial_{\mu} \xi_{\nu}\right) \right)
&= \nabla_{\alpha}\left( \tilde P^{\alpha \mu}F_{\mu \nu} \xi^{\nu}\right)\\
 \end{split}
\end{equation}
The two derivative terms in these equations take the form
\begin{align}\label{2dseomv1}
  \tilde P^{\mu \nu} \left( (P+\rho) u^{\beta} \nabla_{\mu} \nabla_{\beta} u_{\nu} 
+ \partial_T \rho \nabla_{\mu} \partial_{\nu} \frac{\mu}{T} 
+ \partial_{\xi/T} \nabla_{\mu} \partial_{\nu} \frac{\xi}{T} 
+ f \xi^{\beta} \nabla_{\mu} \nabla_{\beta} \xi_{\nu} 
\right) &= \dots \\
\label{2dseomv2}
  \tilde P^{\mu \nu} \left(
u^{\beta} \nabla_{\mu} \nabla_{\beta} \xi_{\nu} -T \nabla_{\mu} \partial_{\nu}\frac{\mu}{T}
- \frac{\mu}{T} \nabla_{\mu} \partial_{\nu}T 
+ \xi^{\beta} \nabla_{\mu} \nabla_{\nu} u_{\beta}
\right) &= \dots \\
\label{2dseomv3}
  \tilde P^{\mu \nu} \left(
\xi^{\beta} \nabla_{\mu} \nabla_{\beta} \xi_{\nu} + T \xi \nabla_{\mu} \partial_{\nu}\frac{\xi}{T}
+ \frac{\xi^2}{T} \nabla_{\mu} \partial_{\nu} T 
\right) &= \dots\,.
\end{align}
The quantities $\tilde P^{\mu \nu} \nabla_{\mu} \partial_{\nu} T , 
\tilde P^{\mu \nu} \nabla_{\mu} \partial_{\nu} \frac{\xi}{T}, 
\tilde P^{\mu \nu} \xi^{\beta} \nabla_{\mu} \nabla_{\nu} u_{\beta}$
can be solved for using equations 
\eqref{2dseomv1}, \eqref{2dseomv2}, and \eqref{2dseomv3}.
Note that these two derivative scalar quantities do not appear in any other equations of motion. 
We can then use the remaining 8 equations of motion to solve for the other
8 dependent data,  
\begin{equation}\label{2ddepterm}
u^{\mu} u^{\nu} \nabla_{\mu} \partial_{\nu} \Sigma_i, \quad
u^{\mu} u^{\nu} \xi^{\beta} \nabla_{\mu} \nabla_{\nu} u_{\beta}, \quad
\xi^{\mu} \xi^{\nu} \xi^{\beta} \nabla_{\mu} \nabla_{\nu} u_{\beta}, \quad
\xi^{\mu} \xi^{\nu} \nabla_{\mu} \partial_{\nu} \Sigma_i
\end{equation} 
where $i$ runs from 1 to 3.
The reaming 8 two derivative scalar equation of motion are 
\begin{subequations}\label{8eom2d}
\begin{eqnarray}
u^{\beta}\nabla_{\beta}\left( u_{\mu} \nabla_{\nu} T^{\mu \nu} \right)
&=& u^{\beta}\nabla_{\beta}\left(-E_{\mu} J^{\mu}\right), \label{8eom2d1} \\
u^{\beta}\nabla_{\beta}\left(\xi_{\mu} \nabla_{\nu} T^{\mu \nu} \right)
&=& u^{\beta}\nabla_{\beta}\left(\xi^{\mu} F_{\mu \nu} J^{\nu}\right),\\ \label{8eom2d2}
  u^{\beta}\nabla_{\beta}\left( \nabla_{\mu} J^{\mu} \right)
&=&u^{\beta}\nabla_{\beta}\left( c E^{\mu} B_{\mu}\right),\\ \label{8eom2d3}
u^{\beta}\nabla_{\beta}\left(\xi^{\mu}u^{\nu} \left( \partial_{\mu} 
\xi_{\nu}-\partial_{\nu} \xi_{\mu}\right) \right)
&=&u^{\beta}\nabla_{\beta}\left(\xi^{\mu} E_{\mu} \right)\\ \label{8eom2d4}
\xi^{\beta}\nabla_{\beta}\left( u_{\mu} \nabla_{\nu} T^{\mu \nu} \right)
&=& \xi^{\beta}\nabla_{\beta}\left(-E_{\mu} J^{\mu}\right),\\ \label{8eom2d5}
\xi^{\beta}\nabla_{\beta}\left(\xi_{\mu} \nabla_{\nu} T^{\mu \nu} \right)
&=& \xi^{\beta}\nabla_{\beta}\left(\xi^{\mu} F_{\mu \nu} J^{\nu}\right),\\ \label{8eom2d6}
\xi^{\beta}\nabla_{\beta}\left( \nabla_{\mu} J^{\mu} \right)
&=&\xi^{\beta}\nabla_{\beta} \left( c E^{\mu} B_{\mu}\right),\\ \label{8eom2d7}
\xi^{\beta}\nabla_{\beta}\left(\xi^{\mu}u^{\nu} \left( \partial_{\mu} 
\xi_{\nu}-\partial_{\nu} \xi_{\mu}\right) \right)
&=&\xi^{\beta}\nabla_{\beta}\left(\xi^{\mu} E_{\mu} \right) \label{8eom2d8}
\end{eqnarray}
\end{subequations}
The matrix of coefficients of the terms in \eqref{2ddepterm} as they appear in the equation 
of motion \eqref{8eom2d} may be expressed as 
\begin{equation}\label{coefmat}
N_{ij} = \begin{pmatrix}
  M^{(a)}_{ij} & 0\\
  0            & M^{(b)}_{ij}\\
 \end{pmatrix}\,
\end{equation}
where the rows represent the 
ordered equations in \eqref{8eom2d} and the columns represent the ordered quantities in 
\eqref{2ddepterm}.
It follows that
\begin{equation}  
\text{Det}[N_{ij}] =  \text{Det} [M^{(a)}_{ij}] \text{Det}[M^{(b)}_{ij}]. 
\end{equation}
In the previous section we concluded that both $\text{Det} [M^{(b)}_{ij}]$
and $\text{Det} [M^{(b)}_{ij}]$ are generically non-zero. Therefore we can infer that 
$\text{Det}[N_{ij}]$ is also generically non-zero.

In order to understand the structure of the matrix $N$ we note that the first four equations in \eqref{8eom2d} are generated by the action of $u\cd\partial$ on the first equation in \eqref{eomsch}. We then find that $u\cd\partial$ acting on $S^{(a)}_i$ generates all the independent second order data as presented in Table \ref{tb:std}. Likewise, the action of $u\cd\partial$ on the ${\cal L}^{(a)}_i$ generates the  four terms $u^{\mu} u^{\nu} \xi^{\beta} \nabla_{\mu} \nabla_{\nu} u_{\beta}$, $ u^{\mu} u^{\nu} \nabla_{\mu} \partial_{\nu} \Sigma_i$ ($i=1,\ldots,3$). In fact these dependent two derivative terms appear only in equations \eqref{8eom2d1}, \eqref{8eom2d2}, 
\eqref{8eom2d3}, \eqref{8eom2d4} and is not there in the rest of the four equations in \eqref{8eom2d}.

Similarly, we can think of the equations \eqref{8eom2d5}, \eqref{8eom2d6}, 
\eqref{8eom2d7}, \eqref{8eom2d8} as being obtained by the action of $\xi\cd\partial$ on the second equation in \eqref{eomsch}. Also here the terms $\xi^{\mu} \xi^{\nu} \xi^{\beta} \nabla_{\mu} \nabla_{\nu} u_{\beta}$, $\xi^{\mu} \xi^{\nu} \nabla_{\mu} \partial_{\nu} \Sigma_i $ (which constitutes the 4 remaining second order quantities which are determined by the equation of motion) are generated by $\xi\cd\partial$ acting on the ${\cal L}^{(b)}_i$ terms. These dependent four second order quantities do not appear in the first four equations in \eqref{8eom2d}. This structure justifies the block diagonal form of the coefficient matrix in \eqref{coefmat}. 
%

%****************************************************************************
\section{Derivation of thermodynamic identities for normal fluids}
\label{na}
%****************************************************************************
%
In this appendix we present a derivation of the equations of motion  
\eqref{simpeom}. Our starting point is the observation that the scalar 
components of the equations of conservation of the stress tensor and current
reduce, at first order to  
\begin{equation}\label{stchcon1a}
\begin{split}
&u_\nu\nabla_\mu T^{\mu\nu} = u_\nu F^{\mu\nu} J_\nu = 0\\
&\Rightarrow (u\partial) \rho + (\rho + P) \Theta = 0\\
\end{split}
\end{equation}
\begin{equation}\label{stchcon1b}
\begin{split}
&\nabla_\mu J^\mu =0 \Rightarrow (u\cd\partial) q + q \Theta = 0\\
\end{split}
\end{equation}
where $\Theta = (\nabla\cd u)$

\cl{Derivation of the first equation in \eqref{simpeom}}\noindent
%^^^^^^^^^^^^^^^^^^^^^^^^^^^^^^^^^^^^^^^^^^^^^^^^^^^^^^^^^^^^^^^^^
%
Subtracting [ $\frac{\partial q}{\partial\nu} \times$ eq.\eqref{stchcon1b}] from  [$\frac{\partial \rho}{\partial\nu} \times$ eq.\eqref{stchcon1a}] we find the following relations: \begin{equation}\label{stchcon2}
\begin{split}
&\frac{\partial q}{\partial\nu} (u \cd \partial)\rho - \frac{\partial \rho}{\partial\nu}(u \cd \partial)q = -\left[(\rho + P)\frac{\partial q}{\partial\nu} - q\frac{\partial \rho}{\partial\nu}\right] \Theta\\
\Rightarrow &\left[\frac{\partial q}{\partial\nu}\frac{\partial \rho}{\partial T}-\frac{\partial \rho}{\partial\nu}\frac{\partial q}{\partial T}\right](u \cd \partial)T = -\left[(\rho + P)\frac{\partial q}{\partial\nu} - q\frac{\partial \rho}{\partial\nu}\right] \Theta\\
\Rightarrow &\left[\frac{\partial q}{\partial\nu}\frac{\partial P}{\partial T}-\frac{\partial P}{\partial\nu}\frac{\partial q}{\partial T}\right](u \cd \partial)T = -\left[\frac{\partial P}{\partial \rho}\right]_q\left[(\rho + P)\frac{\partial q}{\partial\nu} - q\frac{\partial \rho}{\partial\nu}\right] \Theta\\
\Rightarrow &\left[\left(\frac{\rho + P}{T}\right)\frac{\partial q}{\partial\nu}-Tq\frac{\partial q}{\partial T}\right](u \cd \partial)T = -\left[\frac{\partial P}{\partial \rho}\right]_q\left[(\rho + P)\frac{\partial q}{\partial\nu} - q\frac{\partial \rho}{\partial\nu}\right] \Theta\\
\Rightarrow &\frac{(u \cd \partial)T}{T} =-\left[\frac{\partial P}{\partial \rho}\right]_q \Theta
\end{split}
\end{equation}
In the third line we have used two  identities:
\begin{equation}\label{identet1}
\begin{split}
&\left[\frac{\partial P}{\partial \rho}\right]_q\left[\frac{\partial \rho}{\partial T}\right]_\nu = \left[\frac{\partial P}{\partial T}\right]_\nu -\left[\frac{\partial P}{\partial q}\right]_\rho \left[\frac{\partial q}{\partial T}\right]_\nu\\
&\left[\frac{\partial P}{\partial \rho}\right]_q\left[\frac{\partial \rho}{\partial \nu}\right]_\nu = \left[\frac{\partial P}{\partial\nu}\right]_\nu -\left[\frac{\partial P}{\partial q}\right]_\rho \left[\frac{\partial q}{\partial \nu}\right]_\nu\\
\end{split}
\end{equation}
In the last line of \eqref{stchcon2}
we have used the following three thermodynamic  identities.
\begin{equation}\label{identet2}
\begin{split}
&\left[\frac{\partial P}{\partial T}\right]_\nu = \frac{\rho + P}{T}\\
&\left[\frac{\partial P}{\partial \nu}\right]_T = T q\\
&\left[\frac{\partial \rho}{\partial \nu}\right]_T = T\frac{\partial^2P}{\partial T\partial\nu} - \frac{\partial P}{\partial\nu} = T^2\frac{\partial}{\partial T}\left[\frac{1}{T}\frac{\partial P}{\partial\nu}\right] = T^2\left[\frac{\partial q}{\partial T}\right]_\nu
\end{split}
\end{equation}

\cl{Derivation of the second equation in \eqref{simpeom}}\noindent
%^^^^^^^^^^^^^^^^^^^^^^^^^^^^^^^^^^^^^^^^^^^^^^^^^^^^^^^^^^^^^^^^^^^^
%
Adding[ $\frac{\partial P}{\partial q} \times$ eq.\eqref{stchcon1b}] and [$\frac{\partial P}{\partial \rho} \times$ eq.\eqref{stchcon1a}] we find
\begin{equation}\label{stchcon3}
\begin{split}
&\frac{\partial P}{\partial\rho} (u \cd \partial)\rho + \frac{\partial P}{\partial q}(u \cd \partial)q = -\left[(\rho + P)\frac{\partial P}{\partial\rho} + q\frac{\partial P}{\partial q}\right]\Theta\\
\Rightarrow & (u \cd \partial)P=  -\left[(\rho + P)\frac{\partial P}{\partial\rho} + q\frac{\partial P}{\partial q}\right]\Theta\\
\Rightarrow &\left[(\rho + P)\frac{(u \cd \partial)T}{T} +T q (u \cd \partial)\nu\right]=  -\left[(\rho + P)\frac{\partial P}{\partial\rho} + q\frac{\partial P}{\partial q}\right]\Theta\\
\Rightarrow & (u \cd \partial)\nu = -\frac{1}{T}\left[\frac{\partial P}{\partial q}\right]_\rho\Theta
\end{split}
\end{equation}
In the third line of \eqref{stchcon3} we have used the first law
\begin{equation}\label{firstlawnew0}
\begin{split}
&dP = \left(S + \nu q\right)dT + Tq d\nu 
= \left(\rho + P\right)\frac{dT}{T} + Tq d\nu \\
&\rho + P = T(S + \nu q)
\end{split}
\end{equation}
and in the last line of \eqref{stchcon3} we have used the first equation in \eqref{simpeom}.

\cl{Derivation of the third equation in \eqref{simpeom}}
%^^^^^^^^^^^^^^^^^^^^^^^^^^^^^^^^^^^^^^^^^^^^^^^^^^^^^^^^^
%
The third equation in \eqref{simpeom} follows from the vector component of the stress tensor conservation equation.
\begin{equation}\label{veceom4}
\begin{split}
&P_{\mu\theta}\nabla_\nu T^{\nu\theta} = P_{\mu\theta} F^{\theta\alpha} J_{\alpha} = q E_\mu\\
\Rightarrow & P_\mu^\theta\partial_\theta P +(\rho + P)(u.\nabla)u_\mu = q E_\mu\\
\Rightarrow & P_\mu^\theta\frac{\partial_\theta T}{T} + (u.\nabla)u_\mu = \frac{q}{\rho + P} V_{1\,\mu}
\end{split}
\end{equation}
In the last line of \eqref{veceom4} we have used the first law as written in \eqref{firstlawnew0}.

\section{Pullback ambiguity} \label{pullamb}

The parameter $c_0$ is essentially trivial and is related to a pullback ambiguity as we now explain.
It was pointed out in \cite{Bhattacharyya:2008xc} that the following set
of operations maps one positive divergence entropy current $J^\mu_S$ to
another
    \begin{enumerate}
    \item Dualize $J^\mu$ to a three-form.
    \item Shift this three-form by its Lie derivative with respect to any vector field $V^\mu$
    \item Dualize the resultant form back to a current.
    \end{enumerate}
The end result of this operation is a shift in the entropy current
given by (see eq. 6.6 in \cite{Bhattacharyya:2008xc})
\begin{equation}\label{pullback}
	\delta J_S^\mu = \nabla_\nu( J_S^\nu V^\mu-V^\nu J_S^\mu) + V^\mu \nabla_\nu J_S^\nu\,.
\end{equation}

In the current setup we are interested in first order corrections
to the entropy current. The right hand side of \eqref{pullback}
has an explicit derivative. Therefore, the entropy current on the right hand side should be
replaced by the perfect fluid entropy current $J_S^{\mu}  = s u^{\mu}$.
This implies that  the second term on the right hand side of \eqref{pullback} is zero
(recall that the perfect fluid entropy current is divergence free).
Moreover $V^\mu$ must be a derivative free vector field. 

In ordinary (non superfluid) fluid dynamics there is
a unique vector at the zero derivative order---the fluid velocity $u^\mu$. Since $J^\mu \propto u^{\mu}$ then $V^\mu \propto u^\mu$
implies that the first term on the right hand side of \eqref{pullback} also vanishes, and so
\eqref{pullback} leads to no ambiguity in the entropy current
at the first derivative order.

In superfluid dynamics there exist two zero derivative
vectors, $u^\mu$ and $\xi^\mu$. Consequently \eqref{pullback}
can be used to generate a shift in the current proportional to 
$\partial_{\nu} \left( c_0  {\cal Q}^{\nu \mu} \right)$ . We
conclude that the freedom to add the total derivative term 
$\partial_{\nu} \left( c_0  {\cal Q}^{\nu \mu} \right)$ is
precisely the `pullback ambiguity' freedom described in \cite{Bhattacharyya:2008xc}.
%**********************************************************************
\section{Details relating to parity violating superfluids}
\label{calc}
%**********************************************************************
%
In this Appendix we provide several computational details relating to 
section \ref{S:SuperfluidPodd}. 
%
%==================================================================
\subsection{All equations of motion for ideal superfluid}
%==================================================================
%
The equations of motion for a superfluid are listed in table \ref{tb:PESoneD}. There are four scalar equations, one pseudo scalar equation and 3 vector equations.
In this subsection, using thermodynamics, we simplify these equations so that they can be easily used to solve for the dependent fluid data in terms of the independent data.

First we note the identity
 \begin{equation}\label{step1}
 \begin{split}
 &\nabla_\mu\xi_\nu -\nabla_\nu \xi_\mu = F_{\mu\nu}\\
 \Rightarrow & -\xi \nabla_\mu\xi -(\xi\cd \nabla)\xi_\mu = F_{\mu\nu}\xi^\nu\,.
 \end{split}
 \end{equation}
In the second line we have contracted both sides of the equation with $\xi_\nu$.
Using \eqref{step1} and the first law we can simplify the stress tensor conservation equation projected in the direction perpendicular to $u^\mu$:
 \begin{equation}\label{step2}
 \begin{split}
 &P^\mu_{\theta}\nabla_\nu T^{\nu\theta} -P^{\mu\theta} F_{\theta\nu}J^\nu =0\\
 \Rightarrow & P^\mu_\theta\nabla_\nu \bigg[(\rho + P) u^\nu u^\theta + P \eta^{\nu\theta} 
+ f \xi^\nu \xi^\theta \bigg] - P^\mu_{\theta}F^{\theta\nu} \left(q u_\nu - f\xi_\nu\right)= 0\\
\Rightarrow &P^{\mu\theta}\left[\nabla_\theta P+ f (\xi\cd\nabla)\xi_\theta +f F_{\theta\nu}\xi^\nu\right]+(\rho + P) (u\cd\nabla)u^\mu
 +\zeta^\mu \nabla_\theta\left(f\xi^\theta\right)= q E^\mu\\
\Rightarrow &P^{\mu\theta}\left[\nabla_\theta P- f \xi\nabla_\theta\xi\right]+(\rho + P) (u\cd\nabla)u^\mu +\zeta^\mu \nabla_\theta\left(f\xi^\theta\right)= q E^\mu\\
\Rightarrow &P^{\mu\theta}\left[(\rho + P)\frac{\nabla_\theta T}{T} + Tq \nabla_\theta \nu\right]+(\rho + P) (u\cd\nabla)u^\mu +\zeta^\mu \nabla_\mu\left(f\xi^\mu\right)= q E^\mu\\
\Rightarrow & P^{\sigma\mu}\left(\frac{ \nabla_\sigma T }{T}\right)+  (u \cd \nabla ) u^\mu = 
\left[\frac{q T}{\rho + P}\right]\left( \frac{E^\mu}{T} -P^{\mu\theta}\nabla_\theta \nu\right)  - \zeta^\mu \left[\frac{\nabla_\theta\left(f\xi^\theta\right)}{\rho + P}\right]\\
 \end{split}
 \end{equation}
In the second step of \eqref{step2} we have used the identity \eqref{step1} and in the third step of \eqref{step2} we used the first law in the form:
 $$ dP = (\rho +P)\frac{dT}{T} + Tq~d\nu + f\xi ~d\xi\,.$$

Next, using \eqref{step2} we can simplify the  equation relating the curl of the phase velocity $\xi^\mu$ to the field strength (ie. $C_{\mu\nu} \equiv \nabla_\mu \xi_\nu - \nabla_\nu \xi_\mu - F_{\mu\nu} =0$).  
Consider the case where one of the two free indices of this equation is projected in the direction of $u^\mu$.
\begin{equation}\label{step3}
\begin{split}
&u^\beta \left(\nabla_\mu \xi_\beta - \nabla_\beta \xi_\mu -F_{\mu\beta}\right)=0\\
\Rightarrow & u^\beta\left[\nabla_\mu(-\mu~ u_\beta + \zeta_\beta) -\nabla_\beta( -\mu~ u_\mu + \zeta_\mu) \right] = E_\mu\\
\Rightarrow &\left[\nabla_\mu \mu + u^\mu (u\cd\nabla)\mu\right]  +\mu (u\cd\nabla)u_\mu -(u\cd\nabla)\zeta_\mu - \zeta^\theta\nabla_\mu u_\theta = E_\mu\\
\Rightarrow &\left[P^{\theta}_\mu\nabla_\theta \mu +\mu (u\cd\nabla)u_\mu\right] -(u\cd\nabla)\zeta_\mu - \zeta^\theta\nabla_\mu u_\theta = E_\mu\\
\Rightarrow & P^{\theta}_\mu\nabla_\theta \nu+\nu\left[P^{\theta}_\mu\nabla_\theta T +T (u\cd\nabla)u_\mu\right] -(u\cd\nabla)\zeta_\mu - \zeta^\theta\nabla_\mu u_\theta = E_\mu\\
\Rightarrow & (u\cd\nabla)\zeta_\mu = -T\left(1 - \frac{\mu q}{\rho + P}\right) \left( \frac{E_\mu}{T} -P_\mu^{\theta}\nabla_\theta \nu\right)  -\zeta^\theta\nabla_\mu u_\theta\\
\end{split}
\end{equation} 
In the last step we have used \eqref{step2}.
 
Next consider the case where both the indices of $C_{\mu\nu}$ are projected in the direction perpendicular to $u^\mu$. This can be simply analyzed by contracting the two free indices of $C_{\mu\nu}$ with $\epsilon ^{\mu\nu\lambda\sigma} u_\nu$
\begin{equation}\label{step4}
 \begin{split}
 &\epsilon ^{\mu\nu\lambda\sigma} u_\nu \left(\nabla_\lambda \xi_{\sigma} 
- \nabla_\sigma \xi_\lambda - F_{\lambda\sigma}\right)=0\\
 \Rightarrow & 2\epsilon ^{\mu\nu\lambda\sigma} u_\nu \nabla_\lambda \xi_{\sigma} 
- \epsilon ^{\mu\nu\lambda\sigma} u_\nu F_{\lambda\sigma}=0\\
 \Rightarrow &- 2\mu\epsilon ^{\mu\nu\lambda\sigma} u_\nu \nabla_\lambda u_{\sigma} 
+ 2\epsilon ^{\mu\nu\lambda\sigma} u_\nu \nabla_\lambda \zeta_{\sigma}- \epsilon ^{\mu\nu\lambda\sigma} u_\nu F_{\lambda\sigma}=0\\
  \Rightarrow & ~\Omega^\mu= \frac{B^\mu}{2} + \mu \omega^\mu\,.
 \end{split}
 \end{equation}
 where $\Omega^\mu$, $\omega^\mu$ and $B^\mu$ are defined as
 $$\Omega^\mu = \frac{1}{2}\epsilon^{\mu\nu\lambda\sigma}u_\nu \nabla_\lambda\zeta_\sigma,~~
\omega^\mu = \frac{1}{2}\epsilon^{\mu\nu\lambda\sigma}u_\nu \nabla_\lambda u_\sigma,~~
B^\mu = \frac{1}{2}\epsilon^{\mu\nu\lambda\sigma}u_\nu F_{\lambda\sigma}$$
If we project \eqref{step2}, \eqref{step3} and \eqref{step4} in the direction perpendicular to $\zeta^\mu$ , it will be a rewriting of the three vector equations as listed in Table \ref{tb:PESoneD}.
Contracting the free index with $\zeta^\mu$ we get  two of the scalar equations and one pseudo scalar equation from Table \ref{tb:PESoneD}, rewritten in our basis.\footnote{ The relevant equations are \begin{enumerate}
\item $\zeta_\mu\nabla_\nu T^{\mu\nu} - \zeta_\mu F^{\mu\nu}J_\nu = 0$
\item  $u^\mu\zeta^\nu C_{\mu\nu} =0$ 
\item $\epsilon^{\mu\nu\lambda\sigma}\zeta_\mu u_\nu C_{\lambda\sigma}=0$
\end{enumerate}
}

Next we rewrite the two remaining scalar equations of table \ref{tb:PESoneD}. First, using these two remaining scalar equations, we will show that the entropy current is conserved in equilibrium. It is most useful to use the basis in Table \ref{tb:fdpv}
\begin{equation}\label{step5}
\begin{split}
&u_\nu\nabla_\mu T^{\mu\nu} - u_\nu F^{\nu\mu} J_\mu = 0\\
\Rightarrow & u_\nu\nabla_\mu \bigg[(\rho + P) u^\mu u^\nu + P\eta ^{\mu\nu} + f \xi^\mu \xi^\nu \bigg] +fu_\nu F^{\nu\mu} \xi_\mu  =0\\
\Rightarrow & -(u\cd\nabla)\rho -(\rho + P)(\nabla\cd u) +\mu\nabla_\theta(f\xi^\theta) + f u_\nu \left[(\xi\cd \nabla)\xi^\nu + F^{\nu\mu}\xi_\mu\right] = 0\\
\Rightarrow &-\left[(u\cd\nabla) \rho + f\xi (u\cd\nabla)\xi\right]-(\rho + P)(\nabla\cd u) + \mu\nabla_\theta(f\xi^\theta)=0\\
\Rightarrow & -\left[T(u\cd\nabla) s + \mu (u\cd\nabla)q\right]-(Ts + \mu q)(\nabla\cd u) + \mu\nabla_\theta(f\xi^\theta)=0\\
\Rightarrow & - T\nabla_\mu (s u^\mu) -\mu \nabla_\mu (q u^\mu - f\xi^\mu) =0\\
\Rightarrow &\nabla_\mu (s u^\mu) = 0\,.
\end{split}
\end{equation}
In the fourth line we have used the identity \eqref{step1}. In fifth line we have used
$$d\rho = T~ds + \mu~dq - f\xi~d\xi$$
and
$$\rho + P = Ts + \mu q$$
In the last line we have used the fact that 
$$\nabla_\mu J^\mu = \nabla_\mu (q u^\mu - f\xi^\mu) = {\cal O}(\text{Two derivatives})\sim 0$$
Using $\nabla_\mu(su^\mu)=0$ and $\nabla_\mu (q u^\mu - f\xi^\mu)=0$ we find
\begin{equation}\label{step6}
\nabla_\mu[f\xi^\mu] = s (u\cd \nabla) \left[\frac{q}{s}\right]\,.
\end{equation}

To summarize, we list all the equations in simplified form. 
The four scalar and one pseudo scalar equation are given by
\begin{equation}\label{eomscalar}
\begin{split}
	(\nabla\cd u)\equiv\Theta = -\frac{(u\cd\partial)s}{s} &=\left[B_1 (u \cd \partial)\chi 
+ B_2 (u \cd \partial)\nu + B_3(u \cd \partial)T\right] \\
	\frac{\nabla_\theta(f \xi^\theta)}{(\rho + P)}\equiv  K &= \frac{s(u \cd \partial)
\left( \frac{q}{s}\right)}{(\rho + P)} = \left[K_1 (u\cd\partial)\chi + K_2 (u \cd \partial)\nu 
+ K_3(u \cd \partial)T\right] \\
	\frac{ (\zeta\cd\nabla) T }{T}+  \zeta^\theta(u\cd\nabla ) u_\theta &= 
R T( V_1\cd\zeta )- \zeta^2 K\\
	\zeta^\mu \zeta^\nu\sigma_{\mu\nu} &= T^2\left[A_1 (u \cd \partial)\chi + A_2 (u \cd \partial)\nu 
+ A_3(u \cd \partial)T\right] - T(1-\mu R) (V_1\cd\zeta)\\
	 \Omega\cd\zeta &= \frac{B\cd\zeta}{2} + \mu (\omega\cd\zeta)\\
\end{split}
\end{equation}
The three vector equations are given by
\begin{equation}\label{eomvector}
\begin{split}
\tilde P^{\sigma\mu}\left[\frac{ \nabla_\sigma T }{T}+  (u\cd\nabla ) u_\sigma\right] &= 
R T\tilde P_\sigma^{\mu} V_1^\sigma \\
 \tilde P^{\sigma\mu}\Omega_\sigma & = \tilde P^{\sigma\mu}\left[\frac{B_\sigma}{2} + \mu \omega_\sigma\right]\\
\tilde P^{\sigma\mu}(u\cd\nabla)\zeta_\sigma &= \tilde P_\sigma^{\mu}\left[-T(1 - \mu R)  V_{1}^\sigma
-\zeta^\nu \nabla_\sigma u_\nu\right]
\end{split}
\end{equation}
where we have defined
\begin{equation}\label{sevdef}
\begin{split}
\nu &= \frac{\mu}{T},\quad\chi = \frac{\zeta^2}{T^2},\quad K = \frac{\nabla_\theta [ f \xi^\theta]}{\rho + P},\quad R= \frac{q}{\rho + P}\\
B_1&=-\frac{\partial}{\partial\chi}[\log(s)],\quad B_2=-\frac{\partial}{\partial\nu}[\log(s)],\quad B_3=-\frac{\partial}{\partial T}[\log(s)]\\
K_1&=\frac{s}{\rho +P}\frac{\partial}{\partial\chi}\left[\frac{q}{s}\right],\quad K_2=\frac{s}{\rho +P}\frac{\partial}{\partial\nu}\left[\frac{q}{s}\right],\quad K_3=\frac{s}{\rho +P}\frac{\partial}{\partial T}\left[\frac{q}{s}\right]\\
A_1&=-\frac{1}{2}- \nu\chi(1 - \mu R) \left[\frac{\partial}{\partial\chi}\left(\frac{q}{s}\right)\right] + \frac{\chi}{3s}\frac{\partial s}{\partial\chi}\\
A_2 &= - \nu\chi(1 - \mu R) \left[\frac{\partial}{\partial\nu}\left(\frac{q}{s}\right)\right] + \frac{\chi}{3s}\frac{\partial s}{\partial\nu}\\
A_3 &= -\nu\chi(1 - \mu R) \left[\frac{\partial}{\partial T}\left(\frac{q}{s}\right)\right] + \frac{\chi}{3s}\left(\frac{\partial s}{\partial T}-\frac{3 s}{T}\right)\\
V_\mu &=\frac{E_\mu}{T} -  P^\sigma_\mu \nabla_\sigma
\left[\frac{\mu}{T}\right]\\
\Omega^\mu &= \frac{1}{2}\epsilon^{\mu\nu\lambda\sigma}u_\nu \nabla_\lambda\zeta_\sigma,~~
\omega^\mu = \frac{1}{2}\epsilon^{\mu\nu\lambda\sigma}u_\nu \nabla_\lambda u_\sigma,~~
B^\mu = \frac{1}{2}\epsilon^{\mu\nu\lambda\sigma}u_\nu F_{\lambda\sigma}\,.
\end{split}
\end{equation}
%

%==========================================================================================
\subsection{Showing the linear independence, of the first derivative scalar data}
%==========================================================================================
%
Using the scalar equations we can solve for four dependent scalars and one dependent pseudo scalar in terms of the independent ones as they appear in the list of ${\cal S}^c_i$ in table \ref{tb:fdpv}.
We choose the five dependent scalars to be:
\begin{equation}
\label{E:dependent}
	\Theta,~~K,~~\zeta\cd\sigma\cd\zeta,~~ \zeta^\nu (u\cd\nabla)u_\nu,~~\Omega\cd\zeta\,.
\end{equation}
Note that the last term is the pseudo scalar. 
The dependence of the scalars in \eqref{E:dependent} on the $\mathcal{S}_i^c$'s from table \ref{tb:fdpv} takes the following form
\begin{equation}\label{adishsol}
\begin{split}
&\Theta = B_3 {\cal S}^c_2 + B_2 {\cal S}^c_3 + B_1 {\cal S}^c_4\\
&K =K_3 {\cal S}^c_2 + K_2 {\cal S}^c_3 + K_1 {\cal S}^c_4\\
&\zeta\cd\sigma\cd\zeta = A_3 {\cal S}^c_2 + A_2 {\cal S}^c_3 + A_1 {\cal S}^c_4 - T(1 - \mu R)  {\cal S}^c_1\\
&\zeta^\nu (u \cd \nabla)u_\nu= RT {\cal S}^c_1 + \zeta^2\left[K_3 {\cal S}^c_2 + K_2 {\cal S}^c_3 + K_1 {\cal S}^c_4\right]- \frac{1}{T}{\cal S}^c_5\\
&\Omega.\zeta = \mu (\omega.\zeta) + \frac{(B.\zeta)}{2}
\end{split}
\end{equation}
where $B_i$, $K_i$ and $A_i$ are defined in equation \eqref{sevdef}.
Using the first three of the above equations one can form the $A^s$ matrix in \eqref{adef}.
%

%==========================================================================================
\subsection{Showing the linear independence, of the first derivative vector data}
%==========================================================================================
%
We wish to argue that the vectors $\mathcal{V}^c$ form a set of independent vectors. To do so we  solve for the three vectors that do not appear in the list of ${\cal V}^{c\mu}_i$ in terms of ${\cal V}^{c\mu}_i$ using the three vector equations of motion.
The three vectors that are not part of the set $\{\mathcal{V}^c_i\}$ are
\begin{equation}
	\tilde P^{\theta\mu}(u\cd\nabla) u_\theta \quad
	\tilde P^{\theta\mu}(u\cd\nabla) \zeta_\theta \quad
	\tilde P^{\theta\mu}(\zeta\cd\nabla) \zeta_\theta \quad
\end{equation}
From the first vector equation \eqref{step2} we can solve for $\tilde P^{\theta\mu}(u\cd\nabla) u_\theta$,
\begin{equation}\label{solvec1}
\begin{split}
\tilde P^{\theta\mu}(u\cd\nabla) u_\theta &=  \tilde P^{\theta\mu}\left(-\frac{\nabla_\theta T}{T} + R T{ V_1}_\theta\right)\\
&=RT~ {\cal V}_1^{c \mu}  - \frac{{\cal V}_3^{c \mu}}{T}\,.
\end{split}
\end{equation}
This equation has been used to form the matrix $A^v$ in \eqref{adef}.

Taking the star of the second vector equation \eqref{step4} we can solve for  $\tilde P^{\theta\mu}(\zeta.\nabla) \zeta_\theta$
\begin{equation}\label{solvec2}
\begin{split}
\tilde P^{\theta\mu}(\zeta\cd\nabla) \zeta_\theta&=  \tilde P^{\theta\mu}\left[\mu \zeta^\alpha (\partial_\alpha u_\theta - \partial_\theta u_\alpha) + \frac{T^2}{2}\partial_\theta \chi + T\chi \partial_\theta T - F_{\theta\lambda}\zeta^\lambda\right] \\
&=T\chi{\cal V}_3^{c \mu}+\frac{T^2}{2}{\cal V}_5^{c \mu}  +\mu\zeta^2{\cal V}_6^{c\mu}  +\zeta^2{\cal V}_7^{c\mu} \,.
\end{split}
\end{equation}

From the last vector equation \eqref{step3} we can solve for $\tilde P^{\theta\mu}(u \cd \nabla) \zeta_\theta$
\begin{equation}\label{solvec3}
\begin{split}
\tilde P^{\theta\mu}(u\cd\nabla) \zeta_\theta &=  \tilde P^{\theta\mu}\left[-  T(1 - \mu R) {V_1}_\theta -\zeta^\alpha\sigma_{\alpha\theta} -\frac{\zeta^\alpha}{2}(\partial_\theta u_\alpha - \partial_\alpha u_\theta)\right]\\
&=-  T(1 - \mu R) {\cal V}_1^{c \mu} -{\cal V}_2^{c \mu} -\zeta^2{\cal V}_6^{c \mu} \,.
\end{split}
\end{equation}
%

%==========================================================================================
\subsection{Showing the linear independence of the two derivative parity odd scalar data}
%==========================================================================================
%
Here we solve for the five two derivative parity odd scalars using the five parity odd two derivative equations of motion as listed in Table \ref{tb:potd}. The solution will be presented in terms of the three independent two derivative pseudo scalars as listed in the last column of Table \ref{tb:potd} and squares of single derivative terms.
The two derivative pseudo scalars that we shall solve for, are the following
\begin{align}
\notag
	&e^{\lambda\sigma}(u\cd\nabla)\nabla_\lambda u_\sigma&
	&\zeta^\mu u^\nu \nabla_\nu B_\mu&
	&\tilde P^{\mu\nu} \nabla_\mu B_\nu 
	\\
\notag
	&e^{\lambda\sigma}(u\cd\nabla)\nabla_\lambda\zeta_\sigma &
	&e^{\lambda\sigma}(\zeta\cd\nabla)\nabla_\lambda\zeta_\sigma\,.
\end{align}
In the solution below we shall write the dependence on the two derivative terms explicitly but will not write the terms which are squares of a single derivative.
\begin{equation}\label{twodscala1}
 \begin{split}
E_1\equiv &\left[\epsilon^{\mu\nu\lambda\sigma}u_\mu\zeta_\nu \right]\nabla_\lambda
(\nabla_\theta T^\theta _\sigma - F_{\sigma\theta}j^\theta)=0\\
\Rightarrow &\nabla_\lambda\bigg[\left(\epsilon^{\mu\nu\lambda\sigma}u_\mu\zeta_\nu \right)(\nabla_\theta T^\theta _\sigma - F_{\sigma\theta}j^\theta)\bigg] = \text{Products of one derivatives}\\
 \Rightarrow &\nabla_\lambda\bigg(\epsilon^{\mu\nu\lambda\sigma}u_\mu\zeta_\nu \left[(u \cd \nabla) u_\sigma + \frac{\partial_\sigma T}{T} - R  (E_\sigma - T\partial_\sigma\nu)\right]\bigg)= \text{Products of one derivatives}\\
 \Rightarrow &\left[\epsilon^{\mu\nu\lambda\sigma}u_\mu\zeta_\nu \right] 
(u^\theta \nabla_\theta\nabla_\lambda u_\sigma - R\nabla_\lambda E_\sigma)
 = \text{Products of one derivatives}\\
  \end{split}
 \end{equation}
Using $E_1$ we can solve $e^{\lambda\sigma}
(u \cd \nabla)\nabla_\lambda u_\sigma$ in terms of $e^{\lambda\sigma}
\nabla_\lambda E_\sigma$.
 \begin{equation}\label{twodscala2}
 \begin{split}
E_2\equiv  &\left[\epsilon^{\mu\nu\lambda\sigma}\zeta_\mu\right]
 \nabla_\nu F_{\lambda\sigma}=0\\
 \Rightarrow & \left[\epsilon^{\mu\nu\lambda\sigma}\zeta_\mu u_\nu\right](u \cd \nabla)F_{\lambda\sigma} 
+ 2 \left[\epsilon^{\mu\nu\lambda\sigma}\zeta_\mu u_\lambda\right]u^\theta \nabla_\nu F_{\theta\sigma}= 0\\
  \Rightarrow & \zeta_\mu(u \cd \nabla)\left[\epsilon^{\mu\nu\lambda\sigma} u_\nu F_{\lambda\sigma}\right] 
+ 2 \left[\epsilon^{\mu\nu\lambda\sigma}\zeta_\mu u_\lambda\right]\nabla_\nu \left(u^\theta F_{\theta\sigma}\right)
= \text{Products of one derivatives}\\
 \Rightarrow & \left[\epsilon^{\mu\nu\lambda\sigma}u_\mu \zeta_\nu u_\lambda\right]\nabla_\lambda E_\sigma 
- \zeta^\mu u^\nu \nabla_\nu B_\mu
 = \text{Products of one derivatives}\\
 \end{split}
 \end{equation}
Using $E_2$  we can solve $\zeta^\mu u^\nu \nabla_\nu B_\mu$ in terms of $e^{\lambda\sigma}
\nabla_\lambda E_\sigma$.
 \begin{equation}\label{twodscala3}
 \begin{split}
E_3\equiv &\left[\epsilon^{\mu\nu\lambda\sigma}u_\mu\right]
 \nabla_\nu F_{\lambda\sigma}=0\\
 \Rightarrow &\nabla_\mu \left[\epsilon^{\mu\nu\lambda\sigma}u_\nu
  F_{\lambda\sigma}\right]= \text{Products of one derivatives}\\
   \Rightarrow &\nabla_\mu B^\mu= \text{Products of one derivatives}\\
  \Rightarrow & \tilde P^{\mu\nu} \nabla_\mu B_\nu +
\frac{\zeta^\mu\zeta^\nu}{\zeta^2}\nabla_\mu B_\nu
 = \text{Product of one derivatives}\\
 \end{split}
 \end{equation}
Using $E_3$  we can solve $\tilde P^{\mu\nu} \nabla_\mu B_\nu $ in terms of $\frac{\zeta^\mu\zeta^\nu}{\zeta^2}\nabla_\mu B_\nu$.
\begin{equation}\label{twodscala4}
 \begin{split}
E_4\equiv&\left[\epsilon^{\mu\nu\lambda\sigma}u_\mu\zeta_\nu \right] 
u^\theta\nabla_\lambda C_{\theta\sigma}=0\\
\Rightarrow &\nabla_\lambda\bigg(\left[\epsilon^{\mu\nu\lambda\sigma}u_\mu\zeta_\nu \right] \left[(u \cd \nabla)\zeta_\sigma + T(1 - \mu R)V_\sigma + \zeta^\theta \nabla_\sigma u_\theta\right]\bigg)\\
&~~~~~~~~~~~~~~~~~~= \text{Product of one derivatives}\\
&\left[\epsilon^{\mu\nu\lambda\sigma}u_\mu\zeta_\nu \right] \bigg[(u \cd \nabla)\nabla_\lambda\zeta_\sigma + (1 - \mu R)\nabla_\lambda E_\sigma + \frac{\zeta^\theta}{2} [\nabla_\lambda,\nabla_\sigma] u_\theta \bigg]\\
&~~~~~~~~~~~~~~~~~~= \text{Product of one derivatives}\\
\end{split}
\end{equation}
Using $E_4$  we can solve $e^{\lambda\sigma}(u \cd \nabla)\nabla_\lambda\zeta_\sigma $ in terms of $e^{\lambda\sigma}\nabla_\lambda E_\sigma$.
\begin{equation}\label{twodscala5}
 \begin{split}
E_5\equiv &\left[\epsilon^{\mu\nu\lambda\sigma}u_\mu\zeta_\nu \right] 
(\zeta.\nabla)C_{\lambda\sigma}=0\\
\Rightarrow & \left[\epsilon^{\mu\nu\lambda\sigma}u_\mu\zeta_\nu \right] \bigg[(\zeta.\nabla)\nabla_\lambda\zeta_\sigma
 - \mu (\zeta.\nabla)\nabla_\lambda u_\sigma\bigg]
+ \frac{1}{2}\zeta^\mu\zeta^\nu\nabla_\mu B_\nu\\
&~~~~~~~~~~~~~~~~~~ = \text{Product of one derivatives} \\
\end{split}
\end{equation}
Using $E_5$  we can solve $e^{\lambda\sigma}(\zeta.\nabla)\nabla_\lambda\zeta_\sigma $ in terms of $e^{\lambda\sigma}(\zeta.\nabla)\nabla_\lambda u_\sigma$ and $\zeta^\mu\zeta^\nu\nabla_\mu B_\nu$. Thus, we can choose $e^{\lambda\sigma}\nabla_\lambda E_\sigma$, $e^{\lambda\sigma}(\zeta.\nabla)\nabla_\lambda u_\sigma$ and $\zeta^\mu\zeta^\nu\nabla_\mu B_\nu$ as the three independent two derivative parity odd scalar data.\footnote{Here $$C_{\mu\nu} \equiv \nabla_\mu \xi_\nu - \nabla_\nu \xi_\mu - F_{\mu\nu}$$ and $$e^{\lambda\sigma} =\epsilon^{\mu\nu\lambda\sigma}u_\mu\zeta_\nu $$}
%

%==========================================================================================
\subsection{Relating the two different entropy current}
%==========================================================================================
%
The relation between the variables in \eqref{parent} and \eqref{gencur}
can be obtained by using
\begin{equation}\label{id}
\begin{split}
       \frac{1}{2}\epsilon^{\mu\nu\lambda\sigma}\xi_\nu F_{\lambda\sigma}
       &= -\mu B^\mu + (B\cd\zeta) u^\mu -T \tilde{\cal V}^{c\mu}_1 -T \tilde{\cal V}_4^{c\mu}\\
       \frac{1}{2}\epsilon^{\mu\nu\lambda\sigma}\nabla_\nu [T u_\lambda
\zeta_\sigma]&=
       T\left[u^\mu (\omega\cd\zeta) + \frac{RT}{2}\tilde{\cal V}^{c\mu}_1\right] -
T\left(\mu \omega^\mu + \frac{B^\mu}{2}\right)\,
\end{split}
\end{equation}

Explicitly, we find 
\begin{align}
       \tilde  s_1^a &= T\sigma_1\\
       \tilde  s_2^a &= T\sigma_8\\
        \tilde  s_1^b &=\alpha_3 -\frac{T\mu}{\zeta^2}\sigma_1+\frac{T^2}{\zeta^2}\sigma_9\\
       \tilde  s_2^b &=\alpha_4 -\frac{T}{2\zeta^2}\sigma_1 -\frac{\mu}{\zeta^2}\sigma_8 + \frac{T}{\zeta^2}\sigma_{10}\\
       \tilde{v}_1 & = \alpha_1 + 
       \sigma_1\frac{R T^2}{2} - T\sigma_8\\
       \tilde{v}_2 & = \alpha_2\\
       \tilde{v}_3 & = T \sigma_3 + T \partial_{T} \sigma_1 \\
       \tilde{v}_4 & = T \sigma_4 + T \partial_{\nu} \sigma_1 - T\sigma_8\\
       \tilde{v}_5 & = T \sigma_5 + T \partial_{\chi} \sigma_1\\
       \tilde{v}_6 & = -T\mu\sigma_1 + T^2\sigma_9 \\
       \tilde{v}_7 & = -\frac{T}{2}\sigma_1 -\mu\sigma_8 + T\sigma_{10}
      \end{align}

%==========================================================================================
\subsection{Computation of the divergence of entropy current}
%==========================================================================================
%
$\nabla_\mu\Jsn$ can be calculated term by term.

\begin{equation}\label{divterm5}
\begin{split}
&\nabla_\mu \left(\epsilon^{\mu\nu\lambda\sigma}\nabla_\nu [T \sigma_1 u_\lambda\zeta_\sigma]\right) = 0
\end{split}
\end{equation}

\begin{multline}
\label{divtermw}
	\nabla_\mu [T^2 \sigma_9 \omega^\mu]
	=-2 T^2 \sigma_9 (\omega \cd\zeta) K + 2 R T^3 \sigma_9 (\omega \cd V) \\
	+ T^2  \left[\frac{\partial \sigma_9}{\partial T} (\omega \cd\nabla T)
+\frac{\partial \sigma_9}{\partial \nu} (\omega \cd\nabla \nu)
+\frac{\partial \sigma_9}{\partial\chi} (\omega \cd\nabla \chi)\right]
\end{multline}

\begin{multline}\label{divtermB}
\nabla_\mu [T \sigma_{10} B^\mu]
=- T \sigma_{10} (B.\zeta) K +  R T^3 \sigma_{10} (B \cd V)  - 2 T^2 \sigma_{10} (\omega  \cd V) \\
	- 2 T^2 \sigma_{10} (\omega \cd \nabla\nu)+ T \left[\frac{\partial \sigma_{10}}{\partial T} 
(B.\nabla T)+\frac{\partial \sigma_{10}}{\partial \nu} (B.\nabla \nu)
+\frac{\partial \sigma_{10}}{\partial\chi} (B.\nabla \chi)\right]
\end{multline}

\begin{multline}\label{divterm1}
\nabla_\mu ( \sigma_3 {\cal V}_3^{c\mu})
= -2\sigma_3 (\omega \cd\zeta)(u \cd \nabla)T-\frac{\partial \sigma_3}{\partial\nu}({\cal V}_4^{c\mu}.\nabla T)-\frac{\partial \sigma_3}{\partial\chi}({\cal V}_5^{c\mu}.\nabla T)\\
+\sigma_3\left[RT  (V \cd {\cal V}_3^{c\mu}) +  2\mu (\omega \cd\nabla T) + (B.\nabla T)\right]
\end{multline}

\begin{multline}\label{divterm2}
\nabla_\mu (T \sigma_4 {\cal V}_4^{c\mu})
= -2T\sigma_4 (\omega \cd\zeta)(u \cd \nabla)\nu+T\left[\frac{\partial \sigma_4}{\partial T}({\cal V}_4^{c\mu}.\nabla T)+\frac{\partial \sigma_4}{\partial\chi}({\cal V}_4^{c\mu}.\nabla \chi)\right]\\
+T\sigma_4\left[RT (V \cd {\cal V}_4^{c\mu}) +  2\mu (\omega \cd\nabla \nu) + (B.\nabla \nu)\right]
\end{multline}

\begin{multline}\label{divterm3}
\nabla_\mu (T \sigma_5{\cal V}_5^{c\mu})
= -2T\sigma_5 (\omega \cd\zeta)(u \cd \nabla)\chi +T\left[\frac{\partial \sigma_5}{\partial T}({\cal V}_5^{c\mu}.\nabla T)-\frac{\partial \sigma_5}{\partial\chi}({\cal V}_4^{c\mu}.\nabla \chi)\right]\\
+T\sigma_5\left[RT  (V \cd {\cal V}_5^{c\mu}) +  2\mu (\omega \cd\nabla \chi) + (B.\nabla \chi)\right]
\end{multline}

\begin{multline}\label{divterm4}
\nabla_\mu \left(\frac{\sigma_8}{2}\epsilon^{\mu\nu\lambda\sigma}\xi_\nu F_{\lambda\sigma}\right)
= (B \cd \zeta)(u \cd \nabla)\sigma_8 - \mu (B \cd \nabla)\sigma_8 - 2T \sigma_8 [(B \cd V) + B \cd \nabla\nu)] \\
+T\left[ \frac{\partial \sigma_8}{\partial T}(V \cd {\cal V}_3^{c\mu}) +\frac{\partial \sigma_8}{\partial\nu}(V \cd {\cal V}_4^{c\mu})+\frac{\partial \sigma_8}{\partial\chi}(V \cd {\cal V}_5^{c\mu})\right]\\
-T\left[\frac{\partial \sigma_8}{\partial T}({\cal V}_4^{c\mu}\cd \nabla T)+ \frac{\partial \sigma_8}{\partial \chi}({\cal V}_4^{c\mu}\cd \nabla \chi)\right]
\end{multline}
 
 In deriving these expressions we have used the following identities.
 \begin{equation}\label{identityty}
 \begin{split}
 \frac{1}{2}\epsilon^{\mu\nu\lambda\sigma}\xi_\nu F_{\lambda\sigma} &= -\mu B^\mu + (B \cd\zeta) u^\mu 
-{\cal V}_1^{c\mu} -{\cal V}_4^{c\mu}\\
 \frac{1}{2}\epsilon^{\mu\nu\lambda\sigma}\nabla_\nu [T u_\lambda \zeta_\sigma]&=
 T\left[u^\mu (\omega \cd\zeta) + \frac{RT}{2}{\cal V}_1^{c\mu}\right] - T\left(\mu \omega^\mu + \frac{B^\mu}{2}\right)
 \end{split}
 \end{equation}
%
%==========================================================================================
\subsection{Frame transformation formula}
%==========================================================================================
%
Under the frame transformation
\begin{equation}\label{frametran1}
\begin{split}
u_\mu' &= u_\mu - \delta u_\mu\\
T' &= T -\delta T\\
\nu' &= \nu - \delta\nu
\end{split}
\end{equation}
the stress tensor, current and $\mu_{diss}$ transform in the following way.
\begin{equation}\label{frametranstress}
\begin{split}
 T_{diss}^{\prime\mu\nu}-T_{diss}^{\mu\nu} &=\delta T_{diss}^{\mu\nu}
= \delta(\rho +P)~ u^\mu u^\nu + \delta P ~\eta^{\mu\nu} + \delta f ~\xi^\mu \xi^\nu
+ (\rho +P)(u^\mu \delta u^\nu + u^\nu \delta u^\mu)\\
J_{diss}^{\prime\mu} - J_{diss}^\mu &= \delta J^\mu_{diss} = \delta q~ u^\mu - \delta f ~\xi^\mu + q\delta u^\mu\\
\mu_{diss}'-\mu_{diss}&= \delta\mu_{diss} = \delta\mu -  \xi\cd\delta u
\end{split}
\end{equation}
where $\delta A$ for any scalar function $A$ denotes
$$\delta A = \frac{\partial A}{\partial T} \delta T + \frac{\partial A}{\partial \nu} \delta \nu$$ 

Using these expressions one can deduce how ${\mathbf s}_i$ transform under a change of frame
\begin{equation}\label{frametran3}
\begin{split}
\delta {\mathbf s}_1 &= 2\delta P\\
\delta {\mathbf s}_2 &= \delta P + \zeta^2\delta f\\
\delta {\mathbf s}_3 &= \delta \rho + \mu^2\delta f\\
\delta {\mathbf s}_4 &= \mu \zeta^2\delta f -(\rho +P)(\delta u \cd \zeta)\\
\delta {\mathbf s}_5 &= -\delta q - \mu \delta f\\
\delta {\mathbf s}_6 &= -\zeta^2\delta f + q (\delta u \cd \zeta)\\
\delta {\mathbf s}_7 &= -\delta\mu +(\delta u \cd \zeta)
\end{split}
\end{equation}
and also how the $\mathbf{v}_i$ transform under a change of frame.
\begin{equation}\label{frametran4}
\begin{split}
\delta {\mathbf v}_1^\mu &= -(\rho +P)(\tilde P^{\mu\nu}\delta u_\nu)\\
\delta {\mathbf v}_2^\mu &= 0\\
\delta {\mathbf v}_3^\mu &= q(\tilde P^{\mu\nu}\delta u_\nu)\\
\end{split}
\end{equation}
where `$\delta$' denotes a change under field redefinition.

Using these formula one can easily form the frame invariant combinations in the vector sector.
It turns out that ${\mathbf v}_2^\mu$ is frame invariant by itself .The other frame invariant combination is proportional to 
$$\left(\frac{q }{\rho + P}\right)~{\mathbf v}_1^\mu +  {\mathbf v}_3^\mu  = R~{\mathbf v}_1^\mu +  {\mathbf v}_3^\mu $$
These are exactly the combinations that appear in the left hand side of equation \eqref{E:decompositionV}.

Checking the frame invariance of the combinations appearing in the scalar sector is more involved and require many thermodynamic identities. Here we have used Mathematica (version 7) to impose all these identities by expressing $\rho$, $q$ and $f$ in terms of a single function $P(T,\nu,\xi)$ and its derivatives. Then we explicitly checked that 
the four combinations appearing in the left hand side of equation \eqref{sdis} are invariant under 
the transformations given by \eqref{frametran3}.
%

%%-------------Bibliography-------------------
%\bibliographystyle{utcaps_sl}
%\bibliographystyle{utphys}
\bibliographystyle{JHEP}
\bibliography{anom_finalDraft_v2}
\end{document}